# The impacts of alcohol taxes: A replication review

David Roodman[1]

Open Philanthropy Project

**Summary:** This paper reviews the research on the impacts of alcohol taxation outcomes such as heavy drinking and mortality. Where data availability permits, reviewed studies are replicated and reanalyzed. Despite weaknesses in the majority of studies, and despite seeming disagreements among the more credible ones—ones based on natural experiments—we can be reasonably confident that taxing alcohol reduces drinking in general and problem drinking in particular. The larger and cleaner the underlying natural experiment, the more apt a study is to detect impacts on drinking. Estimates from the highest-powered study settings, such as in Alaska in 2002 and Finland in 2004, suggest an elasticity of mortality with respect to price of −1 to −3. A 10% price increase in the US would, according to this estimate, save 2,000−6,000 lives and 48,000−130,000 years of life each year.

[1] I thank Alexander Berger for guidance; Phillip Cook, Gerhard Gmel, Jean-Luc Heeb, Pia Mäkelä, Bill Ponicki, and Alex Wagenaar for comments on earlier drafts; and Luke Muelhauser for a vigorous peer review.

# 1. Introduction

Heavy drinking is associated with many health and social problems, including liver disease, unsafe sex, domestic violence, homicide, and reckless driving. In 2012, 28,000 Americans died from alcohol-caused diseases. Another 10,000 lost their lives in alcohol-involved motor vehicle crashes, accounting for 31% of all motor vehicle deaths (Murphy et al. 2015, Table 10; NHTSA 2014, pp. 1−2). Worldwide in 2010, the death toll from alcohol-caused disease was 155,000 (calculated from WHO database). This is why the Open Philanthropy Project, or Open Phil, is exploring opportunities to influence public policy to reduce dangerous drinking.

Many policies affect drinking and related behaviors: criminal penalties for drunk driving, the minimum drinking age, state monopoly of retail, advertising rules, regulations on when bars can be open and who they can serve, outright prohibition, and more. In the US, alcohol taxes have hardly risen in a generation—indeed, have been drastically eroded by inflation (see figure below). In 1990, President Bush signed a deficit reduction bill that included an alcohol tax hike (visible below); thereby, Bush broke his "no new taxes" pledge, weakened his reelection bid, and helped make tax increases anathema in American politics. Conceivably, increasing taxation is now the low-hanging fruit in alcohol control policy.

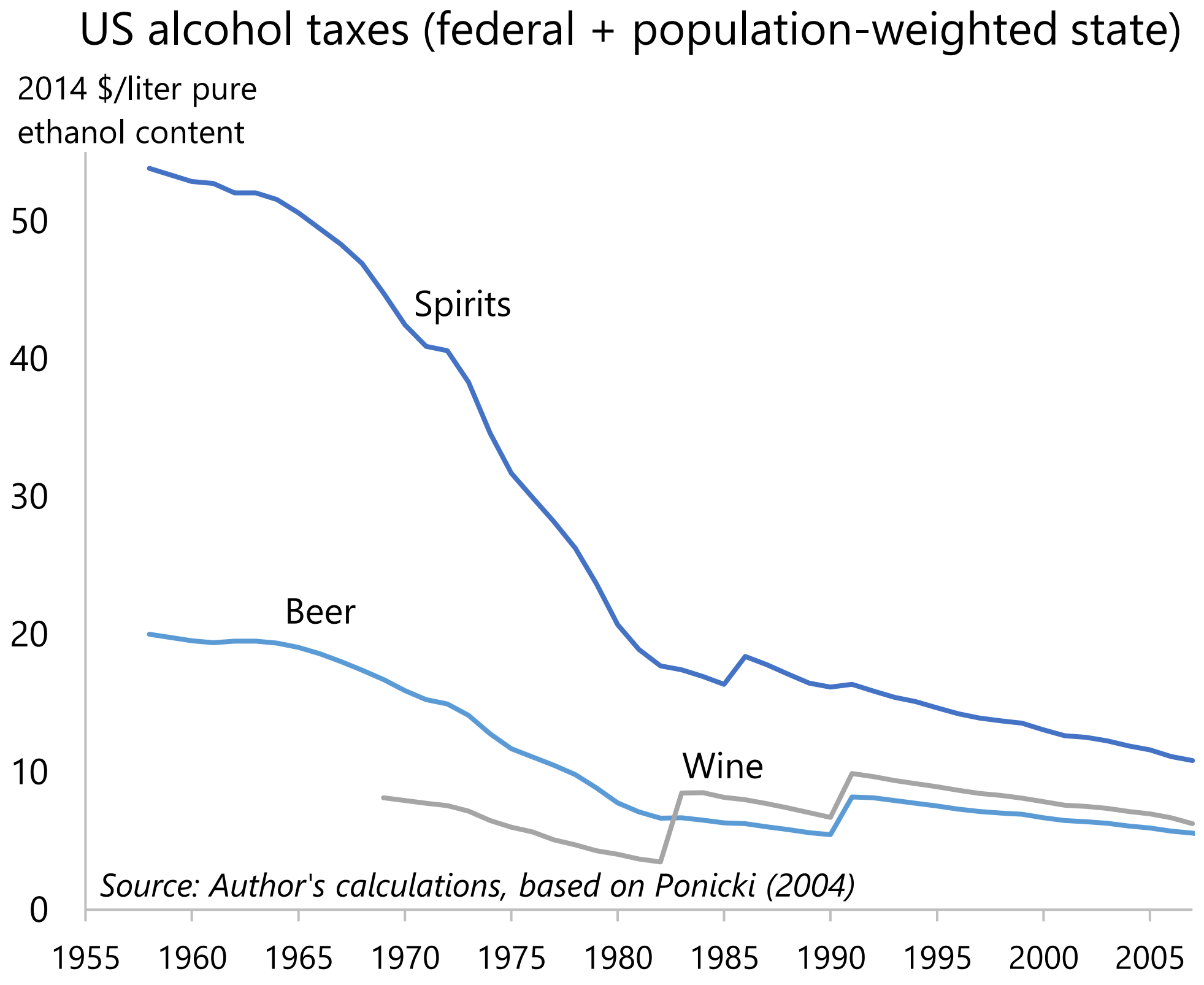

But if taxes are low-hanging fruit, how nutritious are they? How certain should we be that taxing alcohol reduces consumption in general and *problem* drinking in particular? How much illness, physical or social, would be averted?

Many studies examine the impacts of changes in alcohol taxes or prices. Nelson (2013, Table 1) finds 578. The literature is so big that it contains a sub-literature of systematic reviews (e.g., Wagenaar, Salois, and Komro 2009; Wagenaar, Tobler, and Komro 2010; Nelson 2013a, 2014a, 2014b).

Few of the underlying studies attain high-quality causal identification as that term is meant today in economics, exploiting randomized treatment or strong natural experiments. Here, I focus on the minority of studies that do use natural experiments—sudden changes in alcohol taxation in certain states or countries.

Superficially, the high-quality studies contradict each other. Alcohol tax *cuts* apparently did not increase problem drinking in Denmark or Hong Kong, for instance, but did in Finland and Switzerland.

Yet the overall pattern across the quasi-experiment studies is that the larger the experiment—the larger the price change—the clearer the effects. The 7% tax hike in Alaska on October 1, 2002, and the 18% cut in Finland on March 1, 2004, are leading examples.[2] The simplest and most plausible explanation for the "null" results in other contexts is that their natural experiments were too small to produce unambiguous consequences.

Overall, in my view, the preponderance of the evidence says that higher prices do correlate with less drinking and lower incidence of problems such as cirrhosis deaths. And, as I elaborate, I see little reason to doubt the obvious explanation: higher prices *cause* less drinking. A rough rule of thumb is that each 1% increase in alcohol price reduces drinking by 0.5% (Nelson 2013a, as discussed below). And, extrapolating from some of the most powerful studies, I estimate an even larger impact on the death rate from alcohol-caused diseases: 1−3% within months. By extension, a 10% price increase would cut the death rate 9−25%. For the US in 2010 (author's calculations, based on WHO), this represents 2,000−6,000 averted deaths/year.

How much a tax-induced price increase would affect violence and traffic deaths is harder to establish from the available studies. The clearest impacts in the literature have indeed been on the death rate from cirrhosis, in part because drinking is the primary cause, in part because heavy drinkers are presumably most sensitive to price, in part because the impact can be nearly immediate, making for easier statistical detection. (Although cirrhosis is a chronic disease, it is progressive, so that a sudden increase in drinking can speed death among those in whom the disease is most advanced. Seeley 1960.) Impacts on crime, suicides, and risky sexual behavior have been reported, but have not yet been demonstrated through strong natural-experiment−based studies.

And even the link to alcohol-caused diseases is less clear in the long term. It is difficult to pin down long-term impacts because tax changes mix with many other influences over time. This matters particularly for alcohol, because unlike with smoking, many studies find moderate drinking to be healthy. If the death increase from discouraging moderate drinking *only* surfaces after decades, it will be missed in all the studies reviewed.

However, my preliminary take on the epidemiological evidence is that the health benefits of moderate drinking are not certain—not as convincing, for instance, as the natural experiment−based tax impact studies featured in this review. Mendelian-randomized studies, which exploit genetic variation to construct natural experiments, have found no benefit (Holmes et al. 2014). Mendelian-randomized studies are not as reliable as conventional randomized trials (Thomas and Conti 2004). But they may supply the best evidence available since no randomized studies have been done on the question. So I think that on current evidence, Occam's Razor favors the simple theory that the harm of drinking rises steadily with quantity at all levels. (Notably, this implies that moderate intake of alcohol, like moderate intake of many things, does at most modest harm, so the point is not to warn people off drinking at all.) The bottom line for the present inquiry is that, on net, alcohol tax increases are likely to save lives in the long run too.

An important, if *en passant*, lesson from this review is that taxes are just one way to influence drinking. The lesson arises in the discussion below of the early-1980s Alaska and Florida tax hikes, whose effects are inscrutably intermingled with those of other, nearly simultaneous policy changes. Nothing in this review suggests that the most effective or politically realistic approach to alcohol control is to rely purely on taxes. The best approach may be the historical one, which is to pursue many policy reforms at once, however much that may befuddle future empiricists. Nevertheless, since many non-tax policies, such as the drinking age, have already been pushed to their limits, and since alcohol taxes are historically low, raising them may be a promising practical avenue to improving human welfare.

[2] Figures calculated below.

This document summarizes some recent systematic reviews of this alcohol price impact literature. It then examines the natural experiments in depth.

## 2. Process

Papers were included in this review through a two-stage process: search and filtration.

The search stage was informal and substantially ad hoc. In *systematic* reviews, discussed more just below, one typically searches databases of published papers—and perhaps the web as well, to catch "grey literature"—using carefully chosen and documented keywords. Here, in contrast, I found papers by chasing references (including ones in systematic reviews), conversing with authors of papers, and performing some web searches.

The filtration stage was more formal and exacting. Studies were retained only if they were designed to deliver *credible* identification of causal effects, in the sense of Angrist and Pischke (2010). They write:

> Design-based studies typically feature either real or natural experiments and are distinguished by their prima facie credibility and by the attention investigators devote to making the case for a causal interpretation of the findings their designs generate.

Since to my knowledge there have been no randomized experiments with alcohol taxes, all of the studies reviewed here attempt to exploit natural experiments, in which the tax level in a country or state suddenly changes and immediate consequences are tracked. Time series studies, reviewed first, look just at one jurisdiction at a time. Panel studies look at many at once—in our case, many US states at once, each of which enacted tax changes at different times.

Alexander Berger of GiveWell provided the starting point for the search: Ayyagari et al. (2009); Cook and Durance (2013); Elder et al. (2010); Nelson (2013a, 2013b, 2014a, 2014b); Saffer, Dave, and Grossman (2012); Wagenaar, Salois, and Komro (2008); Wagenaar, Tobler, and Komro (2010); Xuan et al. (2013). Of these, all but three—Ayyagari et al. (2009), Cook and Durance (2013), Saffer, Dave, and Grossman (2012)—are systematic reviews. Those by Nelson or by Wagenaar and colleagues are explored below as representative. However, since my goal was to focus on primary studies, I did not attempt to cover all relevant systematic reviews. Among the three primary papers, Ayyagari et al. (2009) and Saffer, Dave, and Grossman (2012) were set aside because they study not taxes but alcohol prices, which vary for many reasons other than the clean experiment of a sudden tax change. Cook and Durance (2013) focuses on the federal alcohol tax increase of 1991 and is reviewed below.

## 3. Systematic reviews

As their name suggests, systematic reviews survey and synthesize a set of studies all focusing on the same question, such as the effect of drinking on suicide. Sometimes systematic reviews perform their own statistical analysis, called *meta-analysis*: their input is not raw data about alcohol prices or suicide rates, but the characteristics and conclusions of the individual studies. They can check, for example, whether studies from a particular time period find larger impacts.

A common challenge in systematic reviews is expressing the results of underlying studies in a form that facilitates comparison. If one study finds that raising the beer tax by a nickel per bottle cuts drunk driving fatalities by 10 per year and another finds that every 10% rise in wine prices cuts deaths from liver disease by 1%, how are these two estimates to be compared or averaged? It requires re-expressing the results in more universal, abstract units.

The systematic reviews I read use two units. First, when linking prices to sales, they follow common practice in economics in speaking of *elasticities*, which are ratios between proportional changes in variables. An elasticity of −0.5 means that a 10% price rise (not tax rise) leads to a 5% sales drop. Second, many reviews use the correlation coefficient, which ranges between −1 and +1, with 0 indicating no correlation, +1 indicating perfect, positive correlation, and −1 indicating perfect negative correlation. A disadvantage of the correlation coefficient is precisely its abstractness: it tells you the sign and precision of a relationship, but not its real-world magnitude. It might be that for every $10 increase in the alcohol tax, drinking falls by exactly 1 glass/year/person, for a perfect correlation of −1, yet little real-world impact. But this disadvantage is not easily rectified, for it flows from the need to compare diversely denominated findings.

One important question in meta-analysis is how much weight to give each reviewed study (Borenstein et al. 2009, part 3). It can make sense to give a study with a sample of 1,000 people 10 times the weight as one with 100 people, since it seemingly contains 10 times as much information. And since a bigger sample normally manifests as a more precise estimate—narrower confidence intervals or lower *variance* in the estimates—it is common to generalize this thinking by weighting impact estimates in proportion to their precision (or inverse proportion to their variance). This is called the "fixed-effects" method of meta-analysis. It assumes that there is one, fixed value for the quantity of interest, such as the elasticity of wine purchases with respect to wine prices. Since all relevant studies are seen as estimating this one number, those with the most precision get corresponding weight.

A somewhat opposing approach is "random effects." It recognizes that the impact of, say, wine taxes, varies by place, time, demographic, and product. As a result, random-effects meta-analysis puts more weight than fixed effects does on small or otherwise imprecise studies, because they can illuminate the relationships of interest under less-studied conditions.

## 3.1. Wagenaar, Salois, and Komro (2009), "Effects of beverage alcohol price and tax levels on drinking: A meta-analysis of 1003 estimates from 112 studies," *Addiction*

Wagenaar, Salois, and Komro find 112 English-language studies of the impact of alcohol prices or taxes on drinking. Most contain several relevant statistical runs (regressions), so Wagenaar, Salois, and Komro gather a total of 1,003 distinct impact estimates.

Taking simple averages across studies, they find price elasticities of −0.46 for beer consumption, −0.69 for wine, and −0.80 for distilled spirits (Wagenaar, Salois, and Komro 2009, abstract). The −0.80 means, for instance, that a 1% price increase causes a 0.8% consumption decrease. Those estimates that do not break out by beverage type, being of total alcohol consumption, find an effect on the small end of this range, at −0.51 (their Table 1). This makes sense because a rise just in the price of spirits, say, could cause people to switch to beer or wine, making for a larger drop in liquor sales than overall alcohol sales; but an across-the-board alcohol price hike forces pushes people to cut back drinking per se, which meets more resistance.

Switching from elasticities to correlations for formal meta-analysis, and using random-effects weighting, Wagenaar, Salois, and Komro obtain average *correlations* between price and quantity sold of −0.17, −0.30, and −0.29 for beer, wine, and spirits. Though less intuitive, these more rigorously obtained numbers are all statistically significant and ratify the interpretation of the simple-average elasticities just mentioned as real-world negative associations.

Focusing on incidence of heavy drinking, which is the main public health concern, Wagenaar, Salois, and Komro (2009, Table 5) find an average overall *elasticity* of −0.28. Compared to the −0.51 elasticity for total alcohol consumption mentioned above, this suggests that the incidence of heavy drinking responds less to price increases than does drinking overall. If true, then studies of overall, population-level price elasticities may overestimate the benefits of alcohol tax increases, which arise from their impacts on problematic heavy drinking. However, most measurements of heavy drinking are based on self-reports on surveys, which are somewhat suspect. We will see that a more objective, if indirect, indicator of heavy drinking (Seeley 1960), deaths from alcohol-caused diseases, responds quickly and much more elastically.

## 3.2. Wagenaar, Tobler, and Komro (2010), "Effects of alcohol tax and price policies on morbidity and mortality: A systematic review," *American Journal of Public Health*

This study broadly resembles the previous one, but it assesses effects on health rather than drinking. One casualty of the switch is the relatively intuitive elasticity framework, which makes sense for beer sales, but not traffic deaths.

(One way to see this is to note that as the beer price goes to infinity, beer purchases must fall toward zero. Not so for traffic deaths, since they have causes other than drinking.) Now all results are expressed only as correlations. This table, based on the authors' Table 2, shows how they grouped studies by type of outcome, as well as the number of studies in each group, the average correlation, and the 95% confidence interval thereof:

| Outcome | Number of impact estimates[1] | Average correlation with taxes or prices | 95% confidence interval |
|---|---|---|---|
| Alcohol-related morbidity & mortality | 13 | −0.347 | [−0.457, −0.228] |
| Other morbidity and mortality | 2 | −0.076 | [−0.152, +0.001] |
| Violence | 10 | −0.022 | [−0.034, −0.010] |
| Suicide | 11 | −0.048 | [−0.102, +0.007] |
| Traffic | 34 | −0.112 | [−0.139, −0.085] |
| STDs and risky sexual behavior | 12 | −0.055 | [−0.078, −0.033] |
| Use of other drugs | 2 | −0.022 | [−0.043, 0.000] |
| Crime/misbehavior | 5 | −0.014 | [−0.023, −0.005] |
| **Overall** | 89 | −0.071 | [−0.082, −0.060] |

[1]Some studies count more than once because they estimate impacts on multiple outcomes.

For every outcome, the average impact is negative and the 95% confidence range either excludes 0 or only barely includes it. Again, there is a strong suggestion that raising alcohol prices reduces social ills.

I reviewed the abstracts of nearly all the underlying studies. My main discoveries were that:

- A minority of the underlying studies identify causation compellingly, as by exploiting a sharp natural experiment. The most promising examples are some of Finland, one of Alaska, and some US *panel* studies, meaning ones taking data across both states and time. I include these in my own review below.
- Most of the underlying studies draw data from U.S. states over a number of years. Since many social ills tend to be correlated, whether across states or over time, it is not clear that the studies are statistically independent observations of the impacts of alcohol price changes—that is, not as statistically independent as they are treated here in constructing the confidence intervals.

Despite these reasons for skepticism, the negative *association* between alcohol taxation and alcohol-related problems looks strong. The obvious and reasonable conclusion is that taxes are the cause and better health is the consequence. The key question is whether any competing theories should stay us from that conclusion.

And I struggle to come up with a strong alternative. Reverse causation is probably small: governments might raise alcohol taxes to compensate for declining alcohol tax revenue as people drink less; but usually total government revenue, to which alcohol taxes are a small contributor, is what matters for such decision-making. Also conceivable but probably secondary would be a legislative impulse to raise alcohol taxes *because* alcohol-related health problems are lessening and alcohol taxes are thus seen to be working well.

As for third variables that could be influencing both taxes and health, creating misleading correlations between them, the strongest candidate is income per capita. For example, poorer US states have worse health, at least as proxied by life expectancy. Perhaps, being politically conservative, they also have lower taxes, creating the negative correlation between drinking and alcohol taxes found in so many studies. But scatter plots of a cross-section of states bear out only the first of these hypothesized patterns. Across states, wealth and health do go together. But the relationship between income and alcohol taxes is, if anything, *positive*—though that is mostly because of the outliers Alaska and Washington:

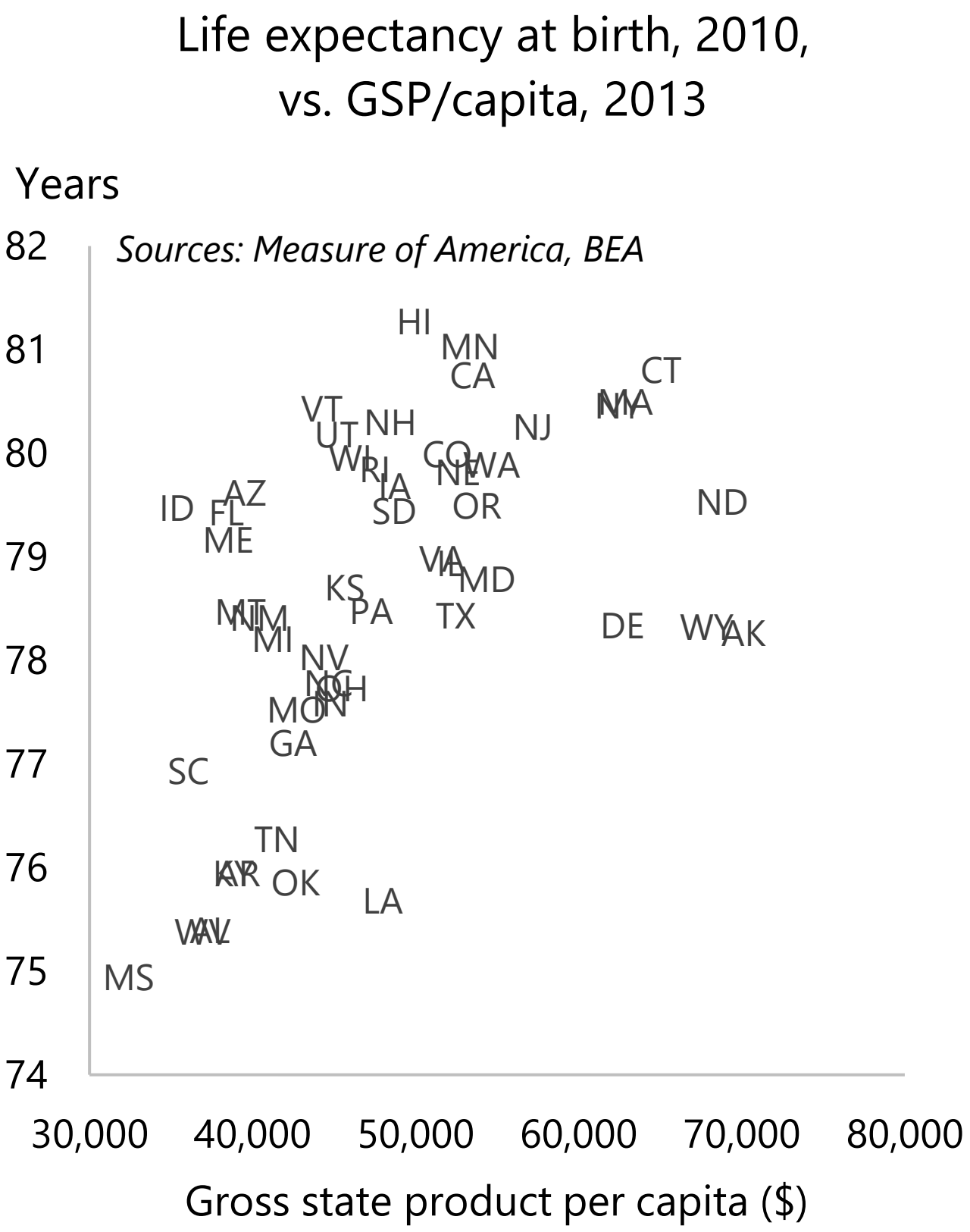

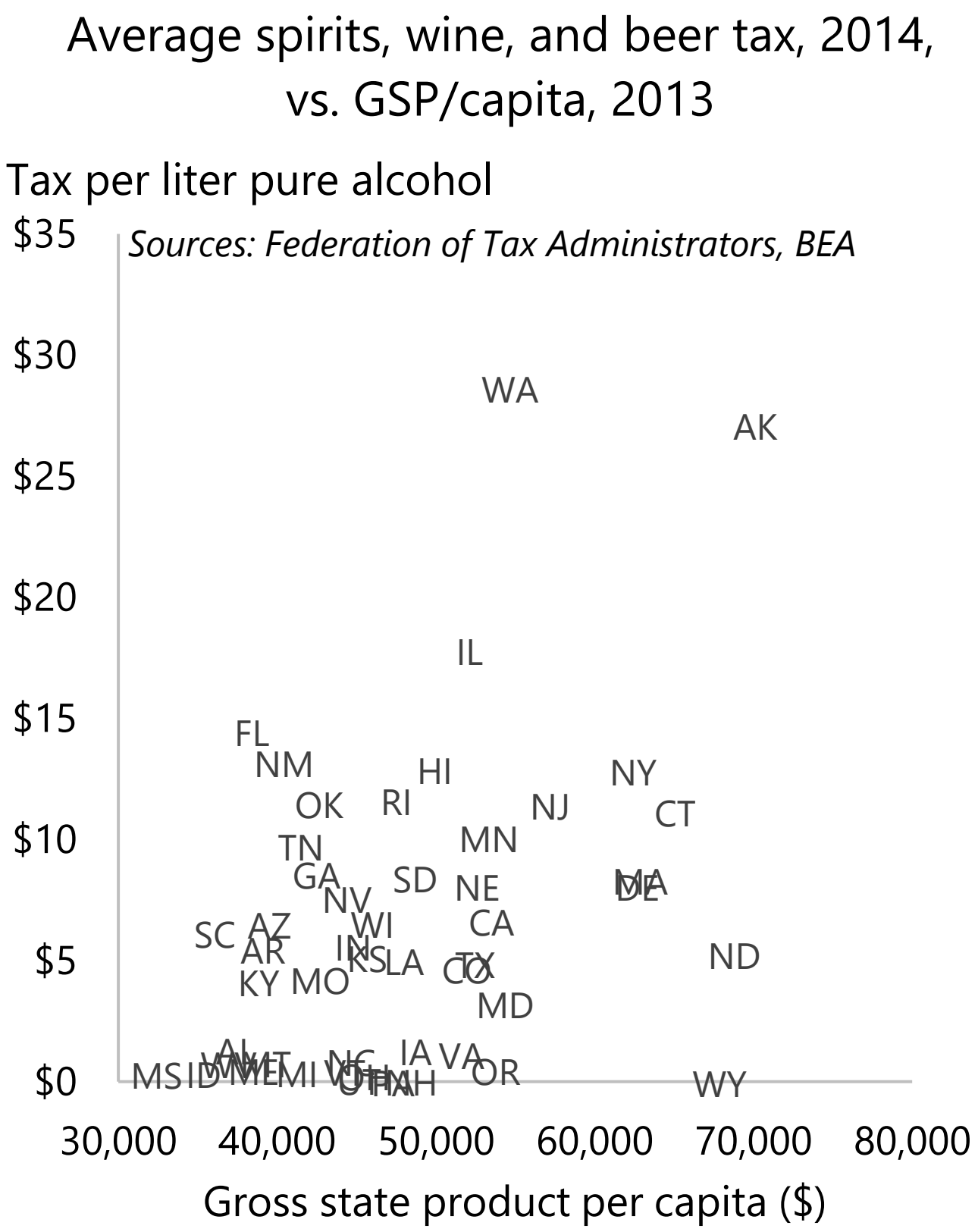

Since both of the graphed relationships are (arguably) positive, it seems hard to construct an alternative theory for the negative association between alcohol taxes and health.

## 3.3. Nelson (2013), "Meta-analysis of alcohol price and income elasticities—with corrections for publication bias," *Health Economics Review*

Jon Nelson, an emeritus economics professor at Penn State, recently published several systematic reviews of the effects of alcohol taxation. His research takes a more skeptical view of the claim that alcohol taxes affect behavior, especially among heavy drinkers. Although his work is funded by the International Center for Alcohol Policies, whose sponsors include Anheuser-Busch, Heineken, and other major alcohol producers (icap.org/AboutICAP/Sponsors), his critical perspective is useful.

Nelson (2013) returns to the territory of Wagenaar, Salois, and Komro (2009), reviewing the impact of taxes on sales. But Nelson departs from that review in several respects. Partly because he comes later, he unearths a lot more studies. He nets 578 at first, then reduces to 297 after applying several filters, such as requiring analysis of the impact of price changes, as distinct from just tax changes. (The elasticity with respect the tax rate is not very meaningful, since a 100% increase might only mean a doubling from one penny to two. What is conceptually coherent is the elasticity with respect to a change in after-tax price.) Nelson also applies some interesting techniques to detect and correct publication bias, which is the tendency of the publication process to select for certain kinds of results. Classic publication bias is the under-reporting of results that do not differ significantly from zero. It can arise from authors not writing up such unexciting results, or journal reviewers panning them, or editors passing over them. But publication bias can exhibit other patterns. In the alcohol impacts literature, for instance, researchers and journals may select for the expected negative correlations at the expense of unorthodox positive ones.

Nelson starts his treatment of publication bias with an informal graphical method called the funnel plot. The insight upon which it is based is that if there is one true average effect, and no publication bias, then the results should be distributed symmetrically around that average, more tightly so for studies with bigger samples. A scatter plot of all the studies' impact estimates versus sample size or some other measure of precision should be funnel-shaped—narrow at the precise-study end and wide at the imprecise end. Here are Nelson's funnel plots for the elasticities of beer, wine, and spirits consumption with respect to price. Each dot represents an estimate from a study. The vertical lines show the average estimate across all studies, when weighting by precision as in the fixed-effects meta-analysis approach explained earlier:

## Nelson (2013) funnel plots

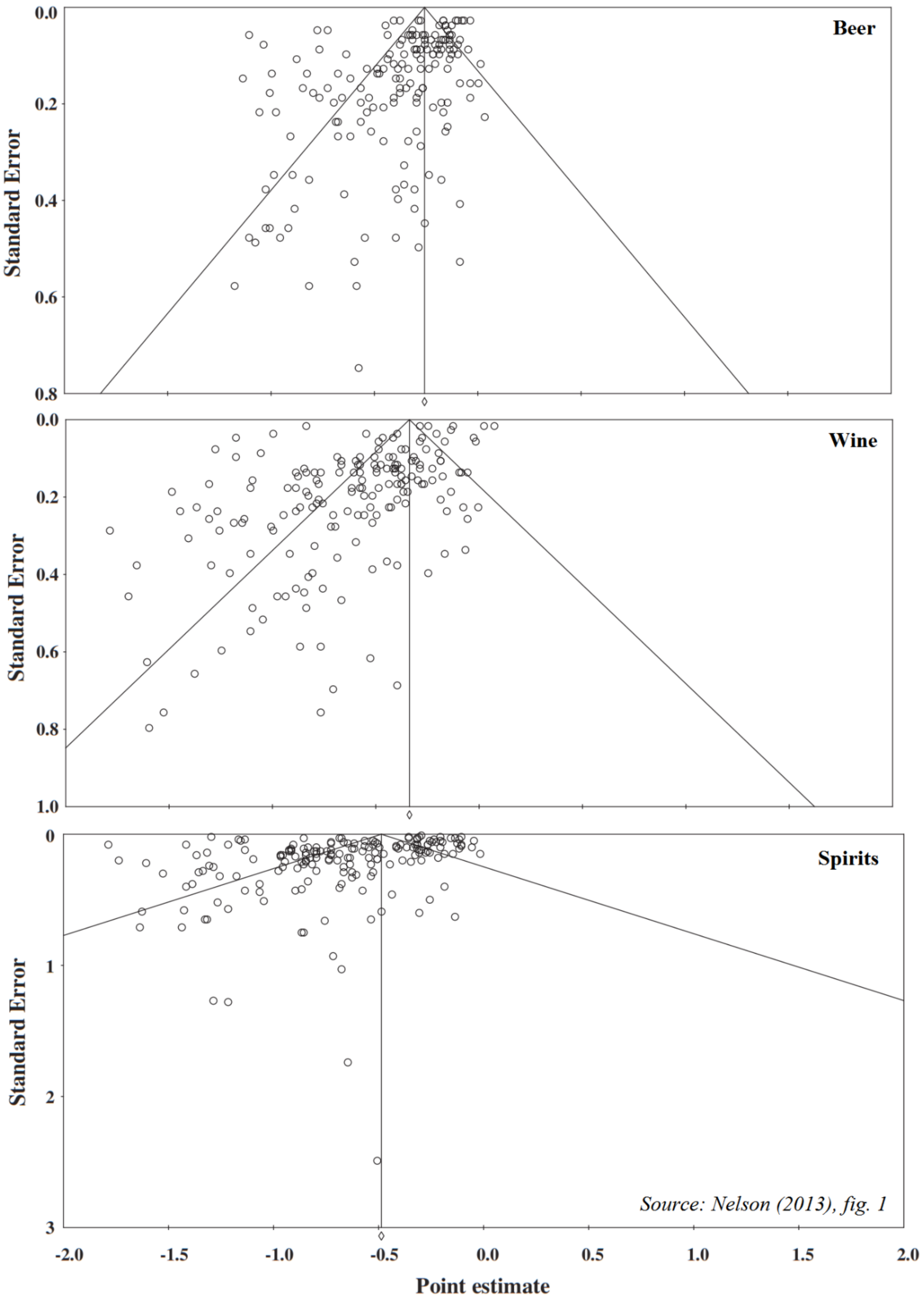

A formal statistical test confirms that all three funnel plots are skewed to the left: imprecise estimates, which are more scattered and feed more variation into the publication selection process, appear to be filtered toward reporting *negative* impacts of price increases. In other words, toward the bottom of each graph, most of the dots are on the left. This suggests that the literature on average overestimates the responsiveness of alcohol sales to price. To reduce this bias, Nelson drops the 50% of studies with the least precise estimates, the ones most susceptible to biased filtering (Nelson 2013a, Table 4).[3] The result, using random-effects weighting, is generally smaller estimates of alcohol price elasticities:

**Meta-analytic estimates of price elasticities**

| | Wagenaar, Salois, and Komro (2009) | Nelson (2013) |
|---|---|---|
| Beer | −0.46 | −0.29 |
| Wine | −0.69 | −0.46 |
| Spirits | −0.80 | −0.54 |
| Total alcohol | −0.51 | −0.49 |

I find Nelson's adjustments for publication bias persuasive, and so prefer his estimates as summaries of the literature. Note that despite his critical bent, Nelson agrees that higher prices lead to lower sales.

### 3.4. Nelson (2014a), "Estimating the price elasticity of beer: Meta-analysis of data with heterogeneity, dependence, and publication bias," *Journal of Health Economics*

Here, Nelson elaborates his analysis of the price elasticity of beer, "the drink of choice among youths" (Saffer and Grossman 1987a, pp. 353−54). He applies to the same collection of studies a broader set of methods for correcting publication bias. The various approaches yield estimates between −0.17 and the −0.30, the highest being essentially the same as the −0.29 in the table above. Nelson settles on −0.20 as a representative value (p. 186).

One source of variety among Nelson's methods is that in this study he performs fixed-effect as well and random-effect meta-analysis. The agnosticism of random effects—acknowledging that the impact of price changes varies by context—is intuitive and appealing. But what might be a compelling choice is complicated by publication bias. In one respect, random-effects is more vulnerable to publication bias: it puts more stock in the imprecise studies at the bottoms of those funnels, where the bias can thrive. On the other hand, researchers and journals can filter results based not only on whether the estimates are in a desired range but whether the apparent precision of those estimates is high. Given the choice between two estimates, researchers might report the one with the smaller standard errors, for example. This can push studies toward the tops of the funnels, where the fixed-effect approach especially will give them undue influence, since it weights only by reported precision.

Nelson (2014a) does not favor one approach over the other, and I am not able to either. In the event, the random-effects estimates of the responsiveness of sales to price are larger than the fixed-effect ones.

As a final step, Nelson performs *meta-regressions*. These take as inputs the output of the individual studies' regressions. The meta-regressions check, for example, whether being published in a journal, or using annual, country-level data, is associated with a lower or higher impact estimate. Altogether 17 traits are considered (Nelson 2014a, Table 4). One example of the results: elasticity estimates *not* appearing in journals are 0.1 larger in magnitude (more negative) on average (Nelson 2014a, Table 4, row 3). Nelson finds that studies with the characteristics he favors (including journal publication and use of annual data) put the elasticity of beer sales volume with respect to price at −0.17 (fixed effects) or −0.20 (random effects) (Table 4, row 1, columns 1 and 6). From these, it appears, he draws −0.20 as representative (p. 186).

[3] My colleague Luke Muehlauser points out that the 50% filtering is done on a sample that already excludes the 10% most extreme estimates, meaning the 2.5% highest point estimates, the 2.5 lowest, and the same for the standard errors. This double filtration does not minimally arbitrary, and so raises the question of whether the findings are sensitive to eliminating the first filtration.

Nelson's choices in performing meta-regressions look reasonable. But compared with the funnel plot analysis, the meta-regressions are more discretionary, creating their own opportunities for bias. A defender of a greater beer price elasticity might have chosen a different list of study traits, and expressed different preferences as to their best values. For example, it is not obvious to me that analyzing annual data produces more reliable results than analyzing quarterly or monthly data since high time resolution can reveal immediate impacts of sudden, tax-induced price increases, strengthening causal ascription. According to Nelson's (2014a, Table 4, row 4) using high-frequency data boost the magnitude of the elasticity by 0.2.

## 3.5. Nelson (2014b), "Binge drinking, alcohol prices, and alcohol taxes: A systematic review of results for youth, young adults, and adults from economic studies, natural experiments, and field studies," working paper

This review focuses on the price-responsiveness of *heavy* drinking. Unlike the other papers described so far, this one is a systematic review, but not a meta-analysis. It does not average numerical results across studies. Instead, it groups studies by type and counts how many in each group find significant impacts on heavy drinking.

From the abstract:

> **Results**: More than half of economic studies report insignificant results for prices or taxes (30 null of 56 studies), with mixed results in 13 studies and significant results in only 13 studies. Null results are equally distributed across age groups, but some mixed results reflect different outcomes by gender. Prices or taxes are insignificant for 11 of 16 samples for men and 7 of 14 samples for women. Four of five natural experiments report null results for country-level tax cuts. Six field studies examine a variety of pricing methods and drink specials, but results are mixed.
> **Conclusions**: A large body of evidence now indicates that binge drinkers are not highly-responsive to increased prices or taxes, and may not respond at all.

The counts of studies finding null results—lack of impact—are interesting but not rigorous. To understand why, imagine I have an unfair coin in my pocket. I commission 1000 studies of its fairness. In 999, the researchers flip the coin only once. They do their analysis properly, and so find no statistically significant evidence of unfairness. The last researcher flips the coin a million times and discovers its true nature. Working in Nelson's mold, we would conclude that the vast majority of studies find no evidence of unfairness. A proper meta-analysis, on the other hand, would reach the right conclusion by giving nearly all its interpretive weight to the one good study. The example is fanciful, but the lesson is practical: counting null results will lead one astray unless one also takes into account the *power* of each study to detect any impact. In our case, that depends on the size of the tax or price change in each case.

To appraise this review more directly, I followed up on the national-level natural experiments that Nelson refers to, which are the studies most likely to produce persuasive results. The natural experiments were sudden changes in the price or accessibility of alcohol in Finland, Sweden, Denmark, Switzerland, and Hong Kong. As discussed below, I corroborated Nelson's interpretations of these studies in all cases save Switzerland. Thus, three rather than four of the five natural experiments produce null results. And the three with null results—Hong Kong, Denmark, and Sweden—are the three where prices changed least (indeed, not at all in Sweden). So, as will be seen, a dose-response story best explains the evidence: the bigger the tax or price change, the clearer the impact.

In addition, any complete assessment of impacts on heavy drinking needs to embrace studies of outcomes to which heavy drinking has been tightly linked by medical science, such as cirrhosis deaths (Seeley 1960). If cirrhosis deaths plunge right after a tax rises, that should shift our beliefs about whether taxes affect heavy drinking. Nelson does not take that on, but this review does.

## 4. Times series studies

Having surveyed some recent surveys, we will now dive into some individual studies, ones that in my view had the most potential to produce credible evidence of causality, not just correlation. The studies are of three main kinds: time series studies, which as one would expect follow developments over time; cross-section studies, which compare countries or states at a given time, and panel studies with work across time and space at once.

The time series studies examined here are at their core "interrupted time series" (or "before-after") analyses. Within some jurisdiction, tax rates on alcohol change overnight. Researchers compare levels or time trends in drinking patterns and health outcomes before and after. The logic is intuitive. And the studies can be strong, especially when focusing on short-term impacts. If cirrhosis deaths plunge right after alcohol taxes spike, that can be compelling evidence. So I reviewed all the interrupted time series studies of sudden tax changes that I found.

But interrupted time series studies can also mislead. To be most compelling, they should:

- Be geared to detect short-term impacts. If a statistician finds that after taxes rose in April, deaths fell in May, that is far more persuasive than if she finds that deaths fell in 2010 after taxes rose in 2000. As Shadish, Cook, and Campbell (2002, p.173) write in their seminal text on impact evaluation, "With delayed effects, the longer the time period between the treatment and the first visible signs of a possible impact, the larger the number of plausible alternative interpretations."
- Strive to rule out competing explanations for any discontinuity found, such as simultaneous changes in non-tax policies.
- Perform a falsification test: demonstrate the absence of a discontinuity where we would not expect one, such as six months before or after an actual tax change.

The first three time series studies reviewed here track developments in individual states: Florida, Alaska, and Illinois. To set the stage, these graphs show how alcohol taxes evolved in those states, after adjusting for inflation, and expressing relative to gallons of pure alcohol content. The last graph is a composite of the first three, weighting by consumption of each alcohol category.[4] We again see the downdraft from inflation. Counteracting that trend are two increases each in the three states, signified by upward jumps. All but the first Illinois tax increase enters the studies below. The second Alaska increase emerges as easily the largest. Indeed, according to my calculations the tax increase on spirits was the largest of any state since 1971, after adjusting for inflation.

[4] Assumptions about pure ethanol are based the benchmark products used in the ACCRA price data: 4.5% for Bud/Miller Lite, 12% for Gallo Chablis, 43% for J&B scotch (Ponicki 2004, ACCRAdjust.xls). Weights are apparent consumption in gallons of pure ethanol terms, from LaVallee, Kim, and Yi (2014). typical weights are 50% beer, 20% wine, 30% spirits.

**Inflation-adjusted alcohol taxes in Florida, Alaska, Illinois (2014 $/gallon of pure alcohol)**

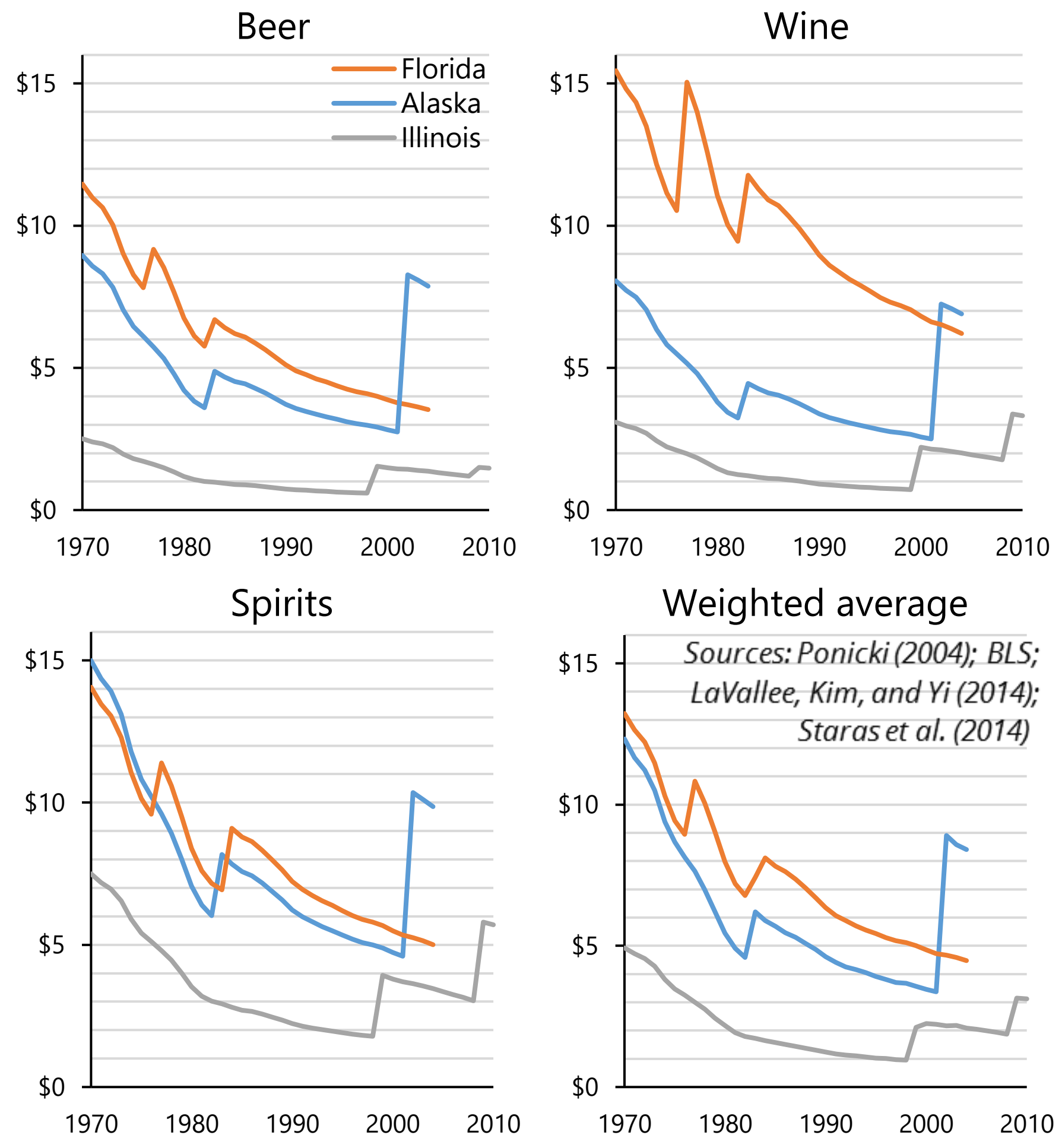

Alcohol taxes are transmitted to consumers, and thereby affect their behavior and health, through *prices*. Exactly how much and over what timeframe are unclear. Alcohol comes in many forms at many price points, so the relative price increase from a given per-gallon tax increase generally depends on the product. And retailers exercise discretion in how much and how quickly they raise sticker prices. The US Bureau of Labor Statistics maintains an alcohol prices index as part of its measurement of overall inflation, but does not break the index out by state. In a detailed study, Kenkel (2005, Table 1) found that shops and restaurants in Alaska typically passed on the 2002 tax increase *twice over*. This suggests that consumers did not aggressively compare prices before buying, limiting competitive pressures on sellers. Or perhaps retailers had resisted inflationary pressures for years, not fully passing on cost increases to their customers, until the tax increase broke the dam.[5] Looking across many states and the years 1982–97, Young and Bielinska-Kwapisz (2002) concur on the prevalance of 50–100% over-shifting, and find that it takes at most three months after a tax increase.

I have obtained state-level price data for products representing the three alcohol categories (Ponicki 2004, ACCRAdjust.xls): a six-pack of Bud Lite or Miller Lite, a 1.5 liter bottle of Gallo Chablis, and a 0.75 liter bottle of J&B scotch—much the same data as in Young and Bielinska-Kwapisz (2002). The data are available longest for the stand-in for spirits, 1976–2003 for most states, and are graphed here for the three states of interest. The first Alaska tax hike and the second Illinois one are outside the sample. The second Alaska one shows up strongly while the two Florida ones are striking for their near-invisibility. The federal tax increase of 1991 also shows up, with a modesty that happens to belie its true size because it applied less to spirits than to beer and wine, as is clear from the graph in the introduction to this review. It is important to keep these relative magnitudes in mind. If any of the state tax increases were big enough to send detectable ripples into the statistics on human health, that one should. On the flipside, lack of credible *evidence* of impact for the smaller increases could just as easily reflect lack of statistical power as it does lack of impact.

[5] Reviewer Bill Ponicki suggested this hypothesis.

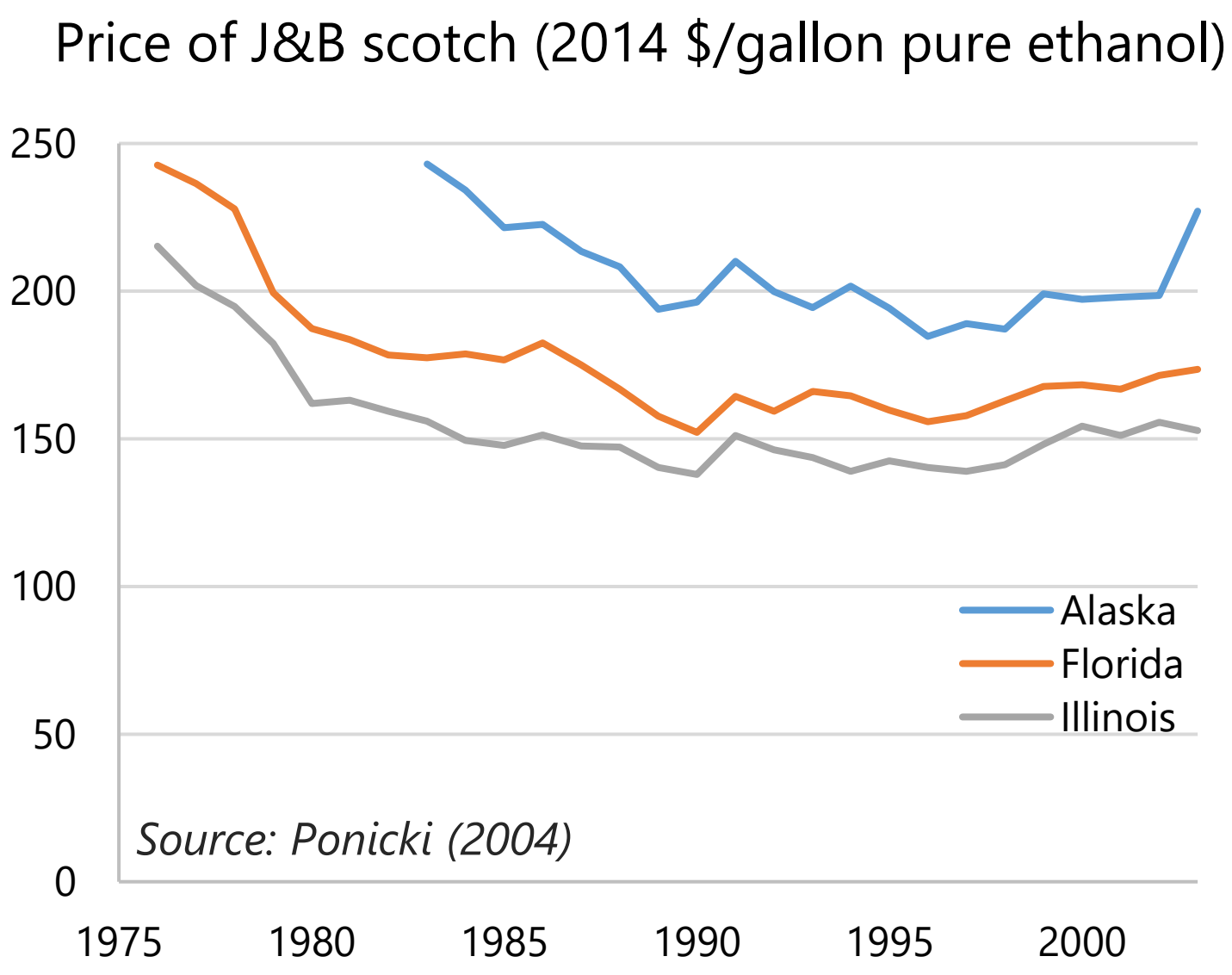

## 4.1. Maldonado-Molina and Wagenaar (2010), "Effects of alcohol taxes on alcohol-related mortality in Florida: Time-series analyses from 1969 to 2004," *Alcoholism: Clinical and Experimental Research*

Florida raised taxes on beer, wine, and liquor on July 1, 1977, and September 1, 1983 (DISCUS 1985, p. 45). Maldonado-Molina and Wagenaar (2010) looks for and finds effects on the rate of death from alcohol-related diseases in the state.

The "treatment" variable—Florida's inflation-adjusted tax rate—was graphed earlier. The next figure, which I computed using the primary data source (NCHS 1969−2004; population data from SEER 2014), shows Maldonado-Molina and Wagenaar's outcome variable[6]:

[6] This graph adopts the definition of "alcohol-related mortality" in Maldonado-Molina and Wagenaar (2010), Table 2; it filters by requiring that state of occurrence (not residence) is Florida and age of deceased ≥ 15. It corrects an apparent small error in the original study's 1969–78 figures by including ICD-8 code 303.9.

**Death rate from alcohol-linked diseases, Florida, 1969–2004**

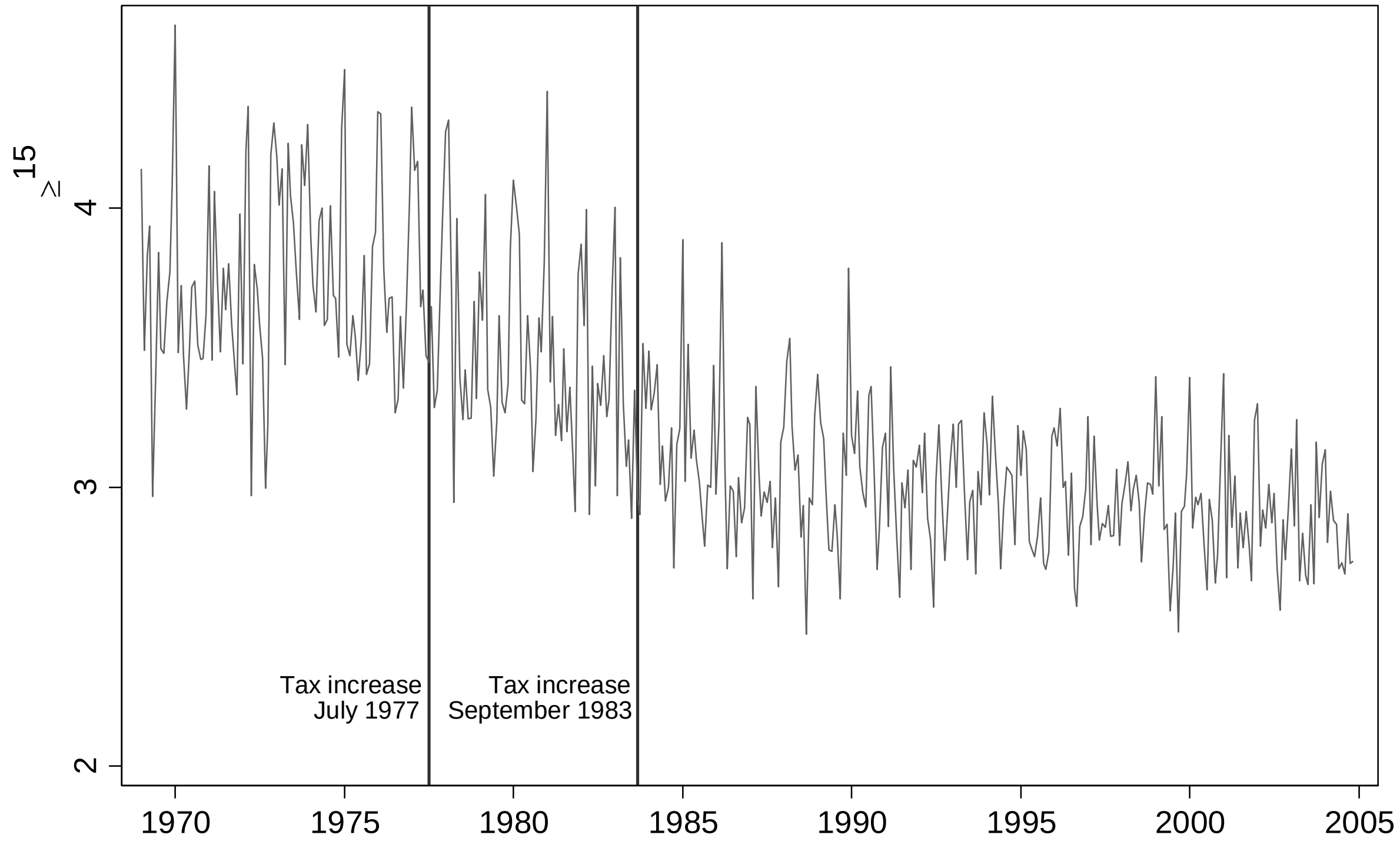

Maldonado-Molina and Wagenaar model the fluctuations in the outcome using the complex but common Autoregressive Integrated Moving Average (ARIMA) method. Then, in effect, they check whether the death rate fell in a way not predicted by that model in the months after each tax hike.

ARIMA models attempt to account for the distinctive traits of time series data. Previous values of the variable can affect future ones, making a variable "autoregressive," as when a rise in GDP leads to more investment, leading to more GDP. That is the “AR” in “ARIMA.” Meanwhile, sometimes a series is best seen not as a sequence of random numbers but as a running sum of such numbers. Here, the classic example is the random walk of a drunk, whose position at each moment is the cumulative sum of all random steps taken to that point. Technically, this trait makes a series "integrated," giving us the “I” in “ARIMA.” It tends to manifest as long-term but not necessarily straight-line trends; the series in the graph above, for instance, may be integrated because the death rate seems stable or rising in the early 1970s, declines for a stretch in the late 1970s and 1980s, and then flattens out. In addition, influences on the outcome that are themselves of secondary interest to the researcher can come and go on shorter time scales, creating a "moving average" process, signified by the “MA.” Finally, there can be seasonal fluctuations: for whatever reason, more people die in winter. By explicitly representing all these dynamics, ARIMA and seasonal ARIMA (SARIMA) models can help a researcher isolate changes *not* predicted by those dynamics, such as a sudden drop in the death rate.

One disadvantage of ARIMA models is that they have a lot of moving parts, which can be configured in a lot of ways: a model can be integrated or not, seasonal or not; transient moving average effects can last two periods or more; etc. And since real data are noisy and limited, often it is not obvious which ARIMA variant fits best. In an experiment run in the early 1980s with simulated data sets, 12 extensively trained graduate students were able to identify the correct ARIMA model only 28% of the time (Velicer and Harrop 1983).

Yet perhaps the slipperiness of the ARIMA fit is not a major concern for us. Unlike economists interested in how past inflation affects future inflation, we are only interested in understanding the time-series dynamics in order to remove them from the data; they are noise whose erasure, we hope, will more fully expose the signal of a tax's impact. If that impact is substantial enough, it should not matter much if the noise is imperfectly modelled and incompletely removed. (On the other hand, if the correctness of the ARIMA model is secondary, perhaps simpler methods such as a pure autoregressive model would serve.)

Maldonado-Molina and Wagenaar add controls to their ARIMA model to reduce the influence of statistical third factors that might have affected alcohol mortality around the time of the tax hikes: the alcohol-related mortality in the rest of the country, the *non*−alcohol-related mortality rate within Florida, and Florida's average personal income.[7] So for example, if abstemiousness rose nationwide around when Florida raised its taxes, then this would be picked up and removed by the variable representing the alcohol mortality rate in the rest of the country.[8]

And, crucially, their model also allows for sudden changes in death rates in July 1977 and September 1983, the two months that began with tax increases.[9] They conclude that those increases saved lives. If Florida brought inflation-adjusted tax rates back to the level reached after the 1983 increase, it would save 600−800 lives a year (Maldonado-Molina and Wagenaar 2010, p. 1920).

In assessing the credibility of these results, two questions seem paramount to me: Might other Florida-level factors have explained any reductions in alcohol-related deaths after the tax hikes? And would the results survive a falsification test? Neither is addressed in the paper.[10]

As for the first question, Florida, like nearly all states, tightened alcohol policies in the early 1980s. It made drunk driving *per se* a crime (even when no harm is done) as of January 1982. In mid-1985 it raised the minimum drinking age to 21 (NHTSA 2008, p. 19). So other forces were at work on drinking in Florida around the time of the second tax hike.

As for the question of falsification, I carried out the check myself after approximately replicating the Maldonado-Molina and Wagenaar regressions. The table below compares the authors' preferred regressions with my replications. Each coefficient measures the estimated impact of alcohol taxes, proxied by the beer tax, on deaths, with the death variable defined in three ways, as listed on the left edge of the table. For example, the value of −0.771 in the middle row implies that a $1 increase in the beer tax (in inflation-adjusted dollars of 2009) saved

[7] I had trouble determining precisely which ARIMA models the Florida and Alaska studies use. The Florida study states "First, we examined a seasonal ARIMA model with structure $(0,1,1)(0,1,1)_{12}$; and the final model is $(1-B^{12})Y_t = \alpha + \omega I_t + \beta Z_t + \boldsymbol{\psi}\mathbf{X}_t + (1-\Theta B^{12})u_t$," where $B^{12}$ is the 12-month lag operator, $Y_t$ is a death rate variable, $I_t$ is the tax rate, $Z_t$ is the alcohol-related death rate in the rest of the US, and $\mathbf{X}_t$ is the other controls. Going by this equation, the "final model" is purely seasonal; it would be denoted by $(0,0,1)_{12}$ and would be applied to the seasonally differenced $Y_t$. This seems a strange model: the seminal Box and Tiao (1975, eqs. 5.2, 5.4) study treats outcome variable and intervention dummy in parallel rather than seasonally differencing one but not the other. Moreover. Table 3 in the Florida study restores the $(0,1,1)(0,1,1)_{12}$ label in its title, and confirms that specification by reporting coefficients for moving average terms of orders 1 as well as 12. I also match Table 3's results better using that model, so this is what I use in text. My confusion carries over to the Alaska study, which presents a nearly identical structural equation (eq 1) and does not otherwise state the ARIMA parameters. In the face of uncertainty, and for consistency and conventionality, I use $(0,1,1)(0,1,1)_4$ in the Alaska study replication below. The match is reasonable, as will be shown.

[8] …at least to the extent that the Florida and national death rates are *linearly* related.

[9] Their preferred model (their Table 3) actually does not include dummies for the two tax increase dates, but a single variable, the inflation-adjusted (beer) tax rate. So technically, the preferred regressions are not an interrupted time series design. However, the authors report alternative regressions using ITS-style dummies; and about 75% of the variation in the tax rate variable can be explained by those dummies, so the distinction does not appear important.

[10] The first question constitutes the item 1 in the Cochrane quality checklist for interrupted time series designs (Cochrane EPOC 2002). The second is rarely raised in connection with time series, but is often mentioned in connection with regression discontinuity design (Imbens and Lemieux 2008; Lee and Lemieux 2010, p. 326), which is highly analogous with ITS (Shadish, Cook, and Campbell 2002, pp. 229–30).

0.771 lives/month per 100,000 Floridians aged 15 or older. The match between the original and the replication is not perfect, but is close enough to corroborate.

**Associations between inflation-adjusted beer tax rate and alcohol-related deaths, Florida, 1969–2004**

| Outcome | Original (Maldonado-Molina and Wagenaar 2010) | Replication | Replication, correcting 1969−78 death counts |
|---|---|---|---|
| Deaths/month | −69.280 | −70.781 | −62.358 |
| | (25.369)*** | (21.557)*** | (16.152)*** |
| Deaths/100,000/month | −0.771 | −0.750 | −0.823 |
| | (0.373)** | (0.285)*** | (0.233)*** |
| Log deaths/population/month | −0.271 | −0.219 | −0.268 |
| (elasticity model) | (0.115)* | (0.106)** | (0.099)*** |

Standard errors in parentheses. * $p<0.1$; ** $p<0.05$; *** $p<0.01$. Independent variable is state beer tax in 2009 dollars. All regressions control for Florida non-alcohol-related deaths, non-Florida U.S. alcohol-related deaths, and Florida personal income/capita in 2009 dollars. In replication, monthly values for population and personal income are interpolated. All replication regressions are ARIMA $(0,1,1)(0,1,1)_{12}$.
Source: Maldonado-Molina and Wagenaar (2010), Table 3; author's calculations.

However, it appears that for 1969−78, the years in which underlying causes of death were coded with ICD edition 8 (wolfbane.com/icd/icd8h.htm), the Florida and Alaska studies (the latter discussed next) inadvertently omit deaths with code 303.9, "Alcoholism: Other and unspecified."[11] Correcting this apparent error does not affect the Florida results much (last column of the table above).

Having matched well, I dropped the beer tax variable in favor of something more appropriate for interrupted times series studies, a "dummy" that represents the interruption, being 0 before and 1 after. By design, if this jump date coincides with a tax increase and the coefficient on the dummy is strong, that suggests an impact on mortality. To perform the falsification test, I repeatedly ran the regression corresponding to the bottom-right corner of the table above while varying the allowed jump date across all months between January 1970 and December 2004. This graph shows the coefficients on the dummy along with 95% confidence intervals:

[11] I exactly match the Alaska death counts (Wagenaar, Maldonado-Molina, and Wagenaar 2009, fig. 2), only if I exclude the 52 deaths so coded in 1976–78.

**Falsification test for Maldonado-Molina and Wagenaar (2010) model of log alcohol-related deaths/100,000, 1970–2004**

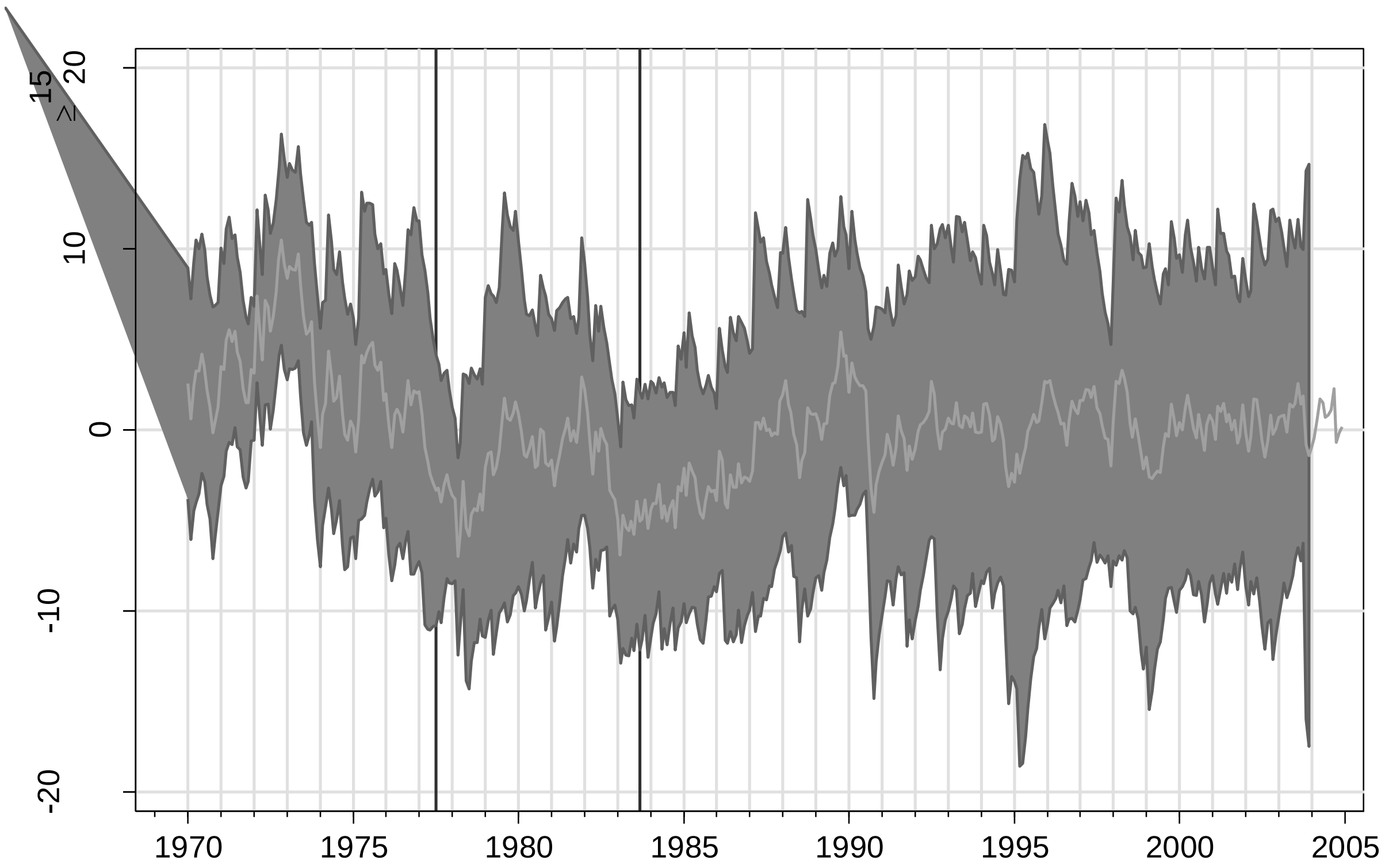

It bears emphasizing that this graph depicts not a mortality trend, but *month-to-month percent changes in mortality rates not otherwise predicted by the statistical model*. So, the high values in 1972 suggest that in those months, alcohol-related mortality rose more (or fell less) for several months in a row to a degree not otherwise predicted by the model. In the same sense, the death rate was low in the tax increase months, July 1977 and September 1983; the grey 95% confidence intervals there include zero, but only barely, so it is reasonable to view these numbers as non-zero. However, in both cases, the negative deviation from expected mortality rates appears *before* the tax increase—especially so for the 1983 one. Now, for technical reasons, this test can produce evidence of a decline one month earlier than it actually occurs.[12] But in the graph, the low values begin to appear at least 3 months early.

It seems most plausible to me that the drunk driving law entering into force in January 1982 had more to do with the early-1980s dip than the September 1983 tax increase did. The late-1970s dip, or at least its strong continuation after July 1977, can be ascribed more readily to a tax increase, but not with the certainty that a more superficial interpretation of the study might suggest.

## 4.2. Wagenaar, Maldonado-Molina, and Wagenaar (2009), "Effects of alcohol tax increases on alcohol-related disease mortality in Alaska: Time-series analyses," *American Journal of Public Health*

The Alaska study resembles the Florida one in subject, authors, method, and conclusion. It too finds that two tax hikes in recent decades cut deaths from alcohol-related diseases. And it too does not perform any time-shifting falsification tests, nor examine whether contemporaneous non-tax events could have explained the mortality drops.

[12] The study's model includes a first-order moving average term, and the unit of time is the month.

The study differs from the Florida one in some technical respects. The data series start in 1976 rather than 1969. Instead of regressing on the actual inflation-adjusted tax level, the study regresses on dummies for each tax increase date—August 1983 and October 2002—which holds truer to the spirit of interrupted time series analysis by focusing exclusively on the interruptions. The authors also drop two controls, state personal income/capita, and the state non-alcohol-related death rate, retaining only the alcohol-related death rate in the rest of the country. (In my tests, the excluded variables do lack statistical significance.) In addition, probably because Alaska's population is much smaller than Florida's, the study tallies deaths by quarter rather than month. Otherwise, noise in the data would especially dominate the signal because of Alaska's small population. In the first three months of 1976, for instance, Florida had 308, 295, and 264 alcohol-related deaths while Alaska had 7, 6, and 0, according to my calculations. The drop from 6 to 0 is a huge plunge, relatively speaking, but probably is mostly noise. It signifies much less than would a plunge from 264 to 0. Aggregating to the quarter removes some noise.[13]

The 2002 tax increase was especially large: from $0.35 to $1.07/gallon for beer, $0.85 to $2.50 for wine, and $5.60 to $12.80 for spirits (Ponicki 2004, BEERRATE.xls, WINERATE.xls, and SPIRRATE.xls). Relative to representative 2002 retail prices ($5.37 for a 6-pack of 12-ounce bottles of Bud or Miller lite, $7.40 for a 1.5 liter bottle or Gallo Chablis or equivalent, $25.71 for 0.75 liter bottle of J&B scotch or equivalent), the taxes represent price increases of 7.5%, 8.8%, and 5.5% respectively (Ponicki 2004, ACCRAdj.xls). Weighting by 2002 consumption in pure ethanol terms (LaVallee, Kim, and Yi 2014), the average tax increase was about 7%.[14]

As with the Florida study, despite some uncertainties about technicalities (see footnote 7), I match the Alaska results reasonably well. Compare the first two columns in this table:

**Associations between tax increases and alcohol-related deaths, Alaska, 1976–2004**

| Outcome | Original | Replication | Replication, correcting 1976–78 death counts |
|---|---|---|---|
| **Deaths/quarter** | | | |
| August 1983 tax increase | −5.65 | −2.70 | −1.01 |
| | (1.73)*** | (2.33) | (2.95) |
| October 2002 tax increase | −5.15 | −7.70 | −6.57 |
| | (2.11)** | (2.24)*** | (2.14)*** |
| **Deaths/100,000/quarter** | | | |
| August 1983 tax increase | −1.37 | −1.28 | 0.61 |
| | (0.50)*** | (0.57)** | (0.76) |
| October 2002 tax increase | −1.23 | −1.78 | −1.76 |
| | (0.57)** | (0.62)*** | (0.59)*** |
| **Deaths/100,000/quarter, controlling for alcohol-linked death rate in rest of US** | | | |
| August 1983 tax increase | −1.19 | −1.24 | −0.64 |
| | (0.52)** | (0.56)** | (0.72) |
| October 2002 tax increase | −1.36 | −1.70 | −1.49 |
| | (0.57)** | (0.68)** | (0.66)*** |
| **Log deaths/100,000/quarter, controlling for alcohol-linked death rate in rest of US** | | | |
| August 1983 tax increase | −0.19 | −0.21 | −0.12 |
| | (0.08)** | (0.09)** | (0.09) |
| October 2002 tax increase | −0.19 | −0.23 | −0.24 |
| | (0.09)** | (0.12)* | (0.11)** |

Standard errors in parentheses. * $p<0.1$; ** $p<0.05$; *** $p<0.01$. Independent variables are dummies for periods after the two tax increases. All regressions control for non-Alaska U.S. alcohol-related deaths. In replication, quarterly values for population are interpolated. All replication regressions are ARIMA (0,1,1)(0,1,1)4.
Source: Wagenaar, Maldonado-Molina and Wagenaar (2009), Table 1, panel 1; author's calculations.

[13] Technically, as the numbers fall, Poisson-type modeling becomes more necessary, but this is hard to combine with the ARIMA framework.

[14] The weights are 675360, 211560, and 537588 gallons ethanol.

As with the Florida study, I correct the apparent coding in the last column. This time the change greatly weakens the suggestion that one tax cut—that in August 1983—saved lives. However, the second, larger increase, in 2002, survives the fix.

As in Florida, the falsification graph (below, using data for 1969−2004) strongly suggests that any early-1980s decline in alcohol-related deaths started before the tax increase of that time. In contrast, negative trend deviations appear strongly only one quarter before the second tax hike. As mentioned, even if the true trend deviation occurred in the quarter of the tax hike, the statistical model can produce a large negative coefficient when the tax hike dummy is set for one period early, because the model assumes that large deviations persist across two time periods (according to an MA(1) process). Thus the falsification results for 2002 are consistent with the true trend break coinciding with the tax hike.

**Falsification test for Wagenaar, Maldonado-Molina, and Wagenaar (2009) model of log alcohol-related deaths/100,000, 1970–2004**

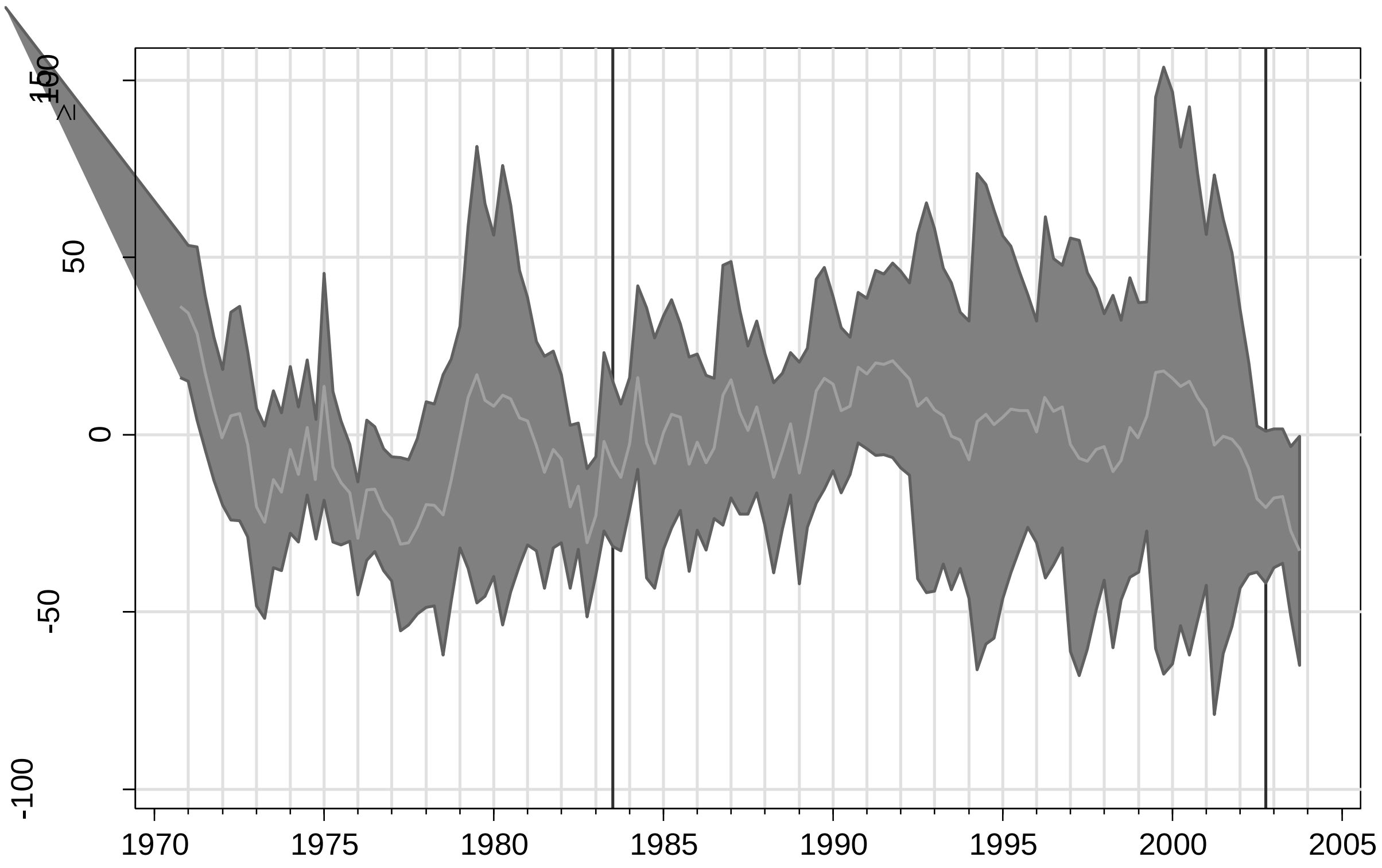

Like Florida, Alaska changed many policies in the early 1980s in order to discourage excessive drinking, as part of a nationwide movement spearheaded by groups such as Mothers Against Drunk Driving. These concomitant developments further muddy any statistical link between the August 1983 tax increase and mortality. Just as in Florida, Alaska's per se law went into force in January 1982 (NHTSA 2008, p. 19). Nine months later, refusing a blood alcohol test became punishable by loss of driver's license (NHTSA 2008, p. 19). Though aimed at traffic deaths, which are not included in the Wagenaar et al. analysis, the new rules may have discouraged heavy drinking generally. Also in October 1982, the minimum drinking age rose from 19 to 21. And on January 1, 1984, additional restrictions came into force, including on bars serving alcohol to intoxicated people. This government notice in the *Tundra Times* (State of Alaska 1983) portrays the sweep of the legal changes:

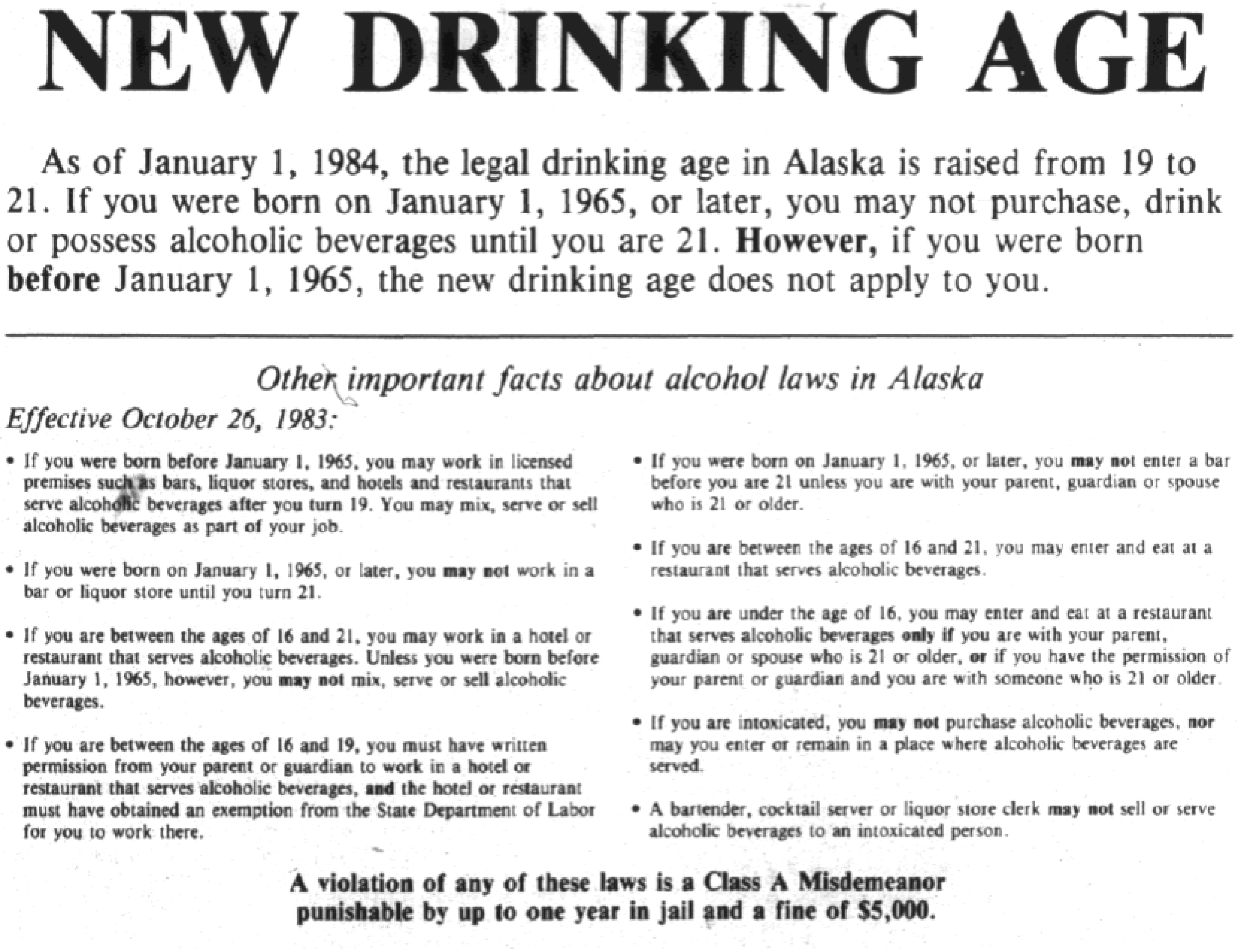

# NEW DRINKING AGE

As of January 1, 1984, the legal drinking age in Alaska is raised from 19 to 21. If you were born on January 1, 1965, or later, you may not purchase, drink or possess alcoholic beverages until you are 21. **However,** if you were born **before** January 1, 1965, the new drinking age does not apply to you.

*Other important facts about alcohol laws in Alaska*

*Effective October 26, 1983:*

- If you were born before January 1, 1965, you may work in licensed premises such as bars, liquor stores, and hotels and restaurants that serve alcoholic beverages after you turn 19. You may mix, serve or sell alcoholic beverages as part of your job.
- If you were born on January 1, 1965, or later, you **may not** work in a bar or liquor store until you turn 21.
- If you are between the ages of 16 and 21, you may work in a hotel or restaurant that serves alcoholic beverages. Unless you were born before January 1, 1965, however, you **may not** mix, serve or sell alcoholic beverages.
- If you are between the ages of 16 and 19, you must have written permission from your parent or guardian to work in a hotel or restaurant that serves alcoholic beverages, **and** the hotel or restaurant must have obtained an exemption from the State Department of Labor for you to work there.
- If you were born on January 1, 1965, or later, you **may not** enter a bar before you are 21 unless you are with your parent, guardian or spouse who is 21 or older.
- If you are between the ages of 16 and 21, you may enter and eat at a restaurant that serves alcoholic beverages.
- If you are under the age of 16, you may enter and eat at a restaurant that serves alcoholic beverages **only if** you are with your parent, guardian or spouse who is 21 or older, **or** if you have the permission of your parent or guardian and you are with someone who is 21 or older.
- If you are intoxicated, you **may not** purchase alcoholic beverages, **nor** may you enter or remain in a place where alcoholic beverages are served.
- A bartender, cocktail server or liquor store clerk **may not** sell or serve alcoholic beverages to an intoxicated person.

**A violation of any of these laws is a Class A Misdemeanor punishable by up to one year in jail and a fine of $5,000.**

Because of the fragility to correcting the death coding, the dissonance in the falsification test, and the competing explanators, I cannot confidently attribute any of the early-1980s drop in alcohol-related deaths to the tax hike.

The story in 2002 seems different, however. The statistical correlation is robust and the drop appears more sharply timed to the tax increase. And I have found no evidence of other relevant policy changes at that time. (See Hopkins 2013 for a journalist's account of the passage of the tax.)

Notably, as graphs at the top of the Florida discussion show, the second Alaska tax increase was the largest of the four so far discussed. Perhaps a dose-response story is at work: the larger the tax increase, the more clearly it shows up in the mortality data. Because this is a plausible synthesis of the evidence on the four tax increases in these two state studies—indeed, it is the simplest—I give disproportionate weight to the suggestive Alaska 2002 impact finding, of a 20−25% reduction in alcohol-related mortality (bottom row of table above).

## 4.3. Staras et al. (2014), "Heterogeneous population effects of an alcohol excise tax increase on sexually transmitted infections morbidity," *Addiction*

This paper carries forward the tradition of the Florida and Alaska studies. It looks for a link between Illinois's alcohol tax increase in September 2010, and the number of gonorrhea and chlamydia cases in the following months.

These graphs, inspired by ones in the paper, show incidence by race for the two diseases. Clearly, incidence is much higher among non-Hispanic blacks:

**STD cases per 100,000 population per month, by race, Illinois**

Chlamydia

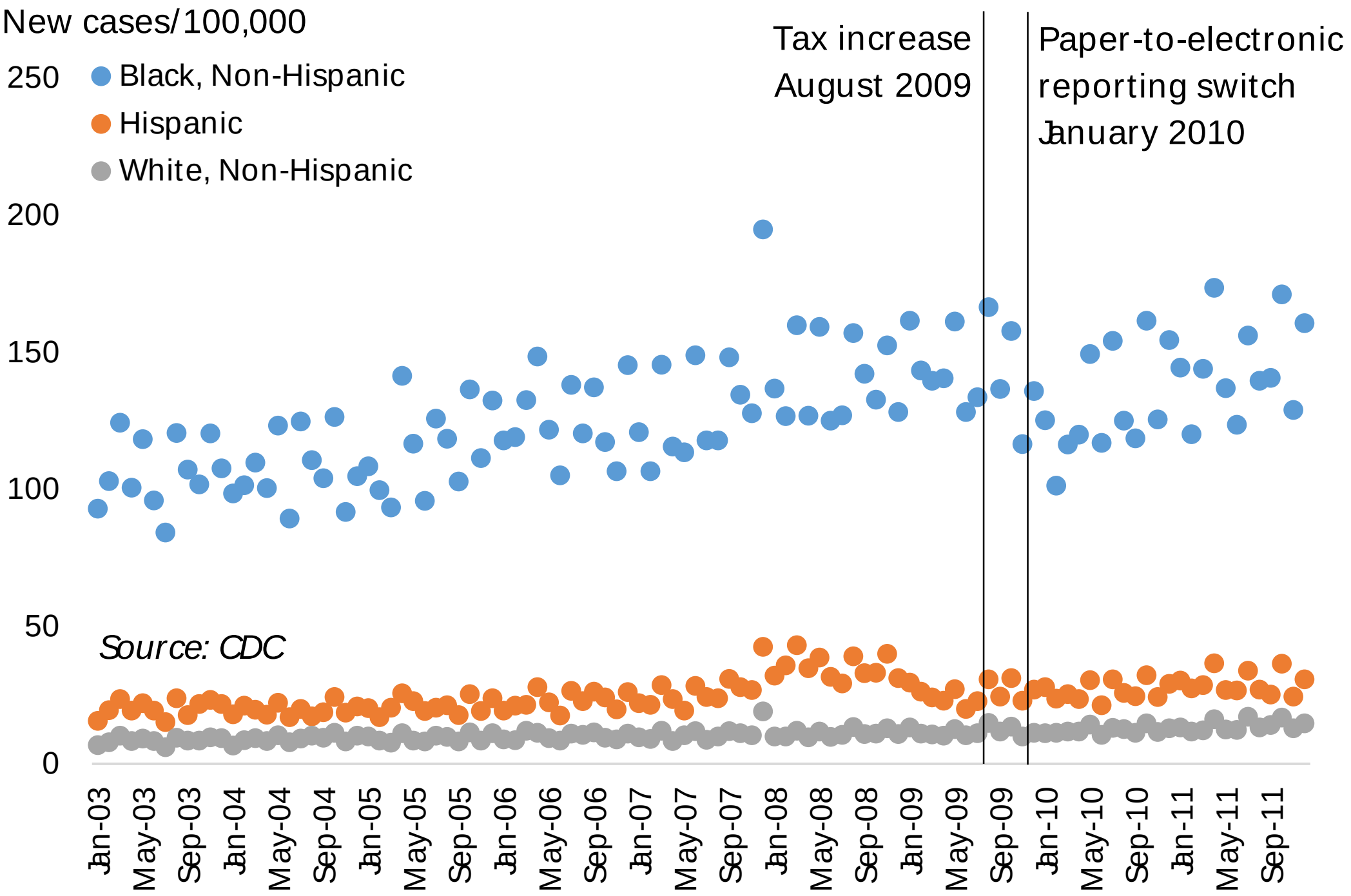

Gonorrhea

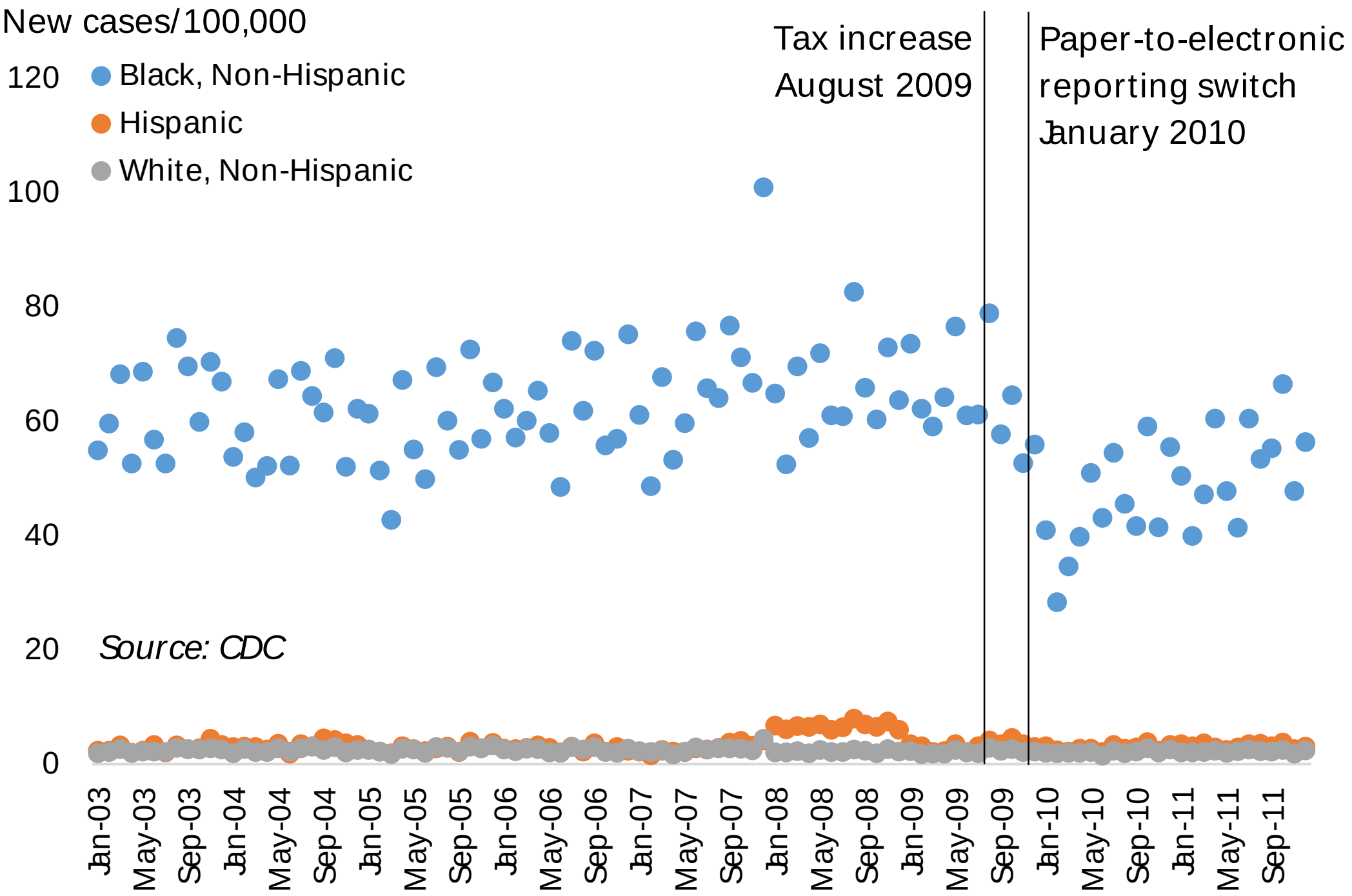

At first glance, the graphs might suggest that the tax increase had a strong impact on non-Hispanic blacks, for there are clear plunges after the tax increase. However, as the authors document, an unrelated event may explain most or all of the dramatic drops in gonorrhea and chlamydia in early 2010. In February 2010, the Illinois government, which collects STD reports and annually submits them to the federal Centers for Disease Control, switched from a paper to an electronic collection system. And the extremely low points for blacks a few months after the tax increase are both from February 2010. Staras et al. explain: "Technical problems and the additional time required for local health departments to report [STD] cases potentially caused underreporting as health departments adjusted to the electronic reporting system." My correspondence (Jan. 28, 2015) with the health official that they cite, Danucha Brikshavana of the Illinois Department of Public Health, confirms that the adjustment difficulties may have affected the January 2010 data points as well, since some January cases would have been entered into the system in February. Case reports for late 2009 were also taken well into 2010, but could be submitted on paper, so they may have been affected less, though whether they were completely immune to the problem is uncertain.

The dots between the vertical lines in the two graphs, which are for the fall of 2009 *may* be on the low side, but they do not lie outside the historical range of variation. Perhaps they too are slightly affected by the reporting problems.

The upshot is that 4−5 months after the tax increase, the STD data series appear to have been disrupted by a larger force. This makes it hard to attribute the obvious statistical declines among non-Hispanic blacks to the tax. The data problem is worsened by the paper's statistical approach. Instead of checking, as in the Florida and Alaska studies, whether month-to-month *changes* in outcomes were unusual right at the time of the tax increase, the regressions check whether incidence *levels* were lower on average in all the months afterward.[15] This allows the evident plunge from the paper-to-electronic switch in early 2010 to influence the apparent impact of the tax hike in mid-2009. Staras et al. report running ARIMA models as well, which may have addresses this concern (if they were integrated of order 1 as the Florida and Alaska models appear to be); however, Staras et al. (2014, pp. 3, 5) provides no details of the ARIMA models or results.

Staras et al. do address the statistical disruption of the paper-to-electronic switch in one way, by experimentally dropping data from Cook County, which includes Chicago; is home to a large share of cases; and is where under-reporting may have been particularly severe (citing state health official Brikshavana). Now the paper finds no clear impact on chlamydia, while the apparent impact on gonorrhea incidence falls from 3.2 to 1.9/100,000 people/month (95% confidence interval 1.3 to 2.5), which is about 1,750/year (Staras et al. 2014, p. 4).[16] But interpreting this number as a tax impact requires believing that while under-reporting was substantial in Cook Country it was minor beyond, a belief for which I do not see evidence.

I attempted to replicate Staras et al.'s (2014, Table 1) basic regressions—one each for gonorrhea and chlamydia, not breaking out by age or race. This table shows their results, my attempted replication, two variants, and a parallel regression for Alaska. The first variant cuts off the data after December 2009 because the disruption in the *data generating process* at the start of 2010 argues against using the same statistical model before and after that date.[17] The second variant stays within this sample while switching to the ARIMA approach of the Alaska and Florida papers.[18] Last is a regression that transfers the Staras et al. analysis to Alaska, whose tax increase in 2002, as we saw, was much bigger than Illinois's in 2009.[19]

[15] More technically, the regression model (p. 3) is static rather than dynamic in the way that an integrated ARIMA model is.
[16] The population of Illinois outside Cook County is about 12.5 million.
[17] The paucity of post–tax increase observations is undesirable but not necessarily catastrophic. The statistical question is whether those observations, however few, violate the pre-increase pattern, which is established from a larger sample.
[18] This monthly regression covers 2002–2012. In 2002, the CDC (2015) STD data set includes all states except Arizona. Before that, it is missing 10 or more states. This substantially impedes construction of a Staras et al. (2014, p. 2) control, the STD incidence rate in the 12 control states, which is highly significant.
[19] The same could not be done for Florida, because its last tax increase was in 1983, well before the STD data start.

**Associations between tax hike and new STD cases/100,000 population, Illinois, 2003–11**

| Disease | Original (Staras et al. 2014) | Replication | Replication, omitting post-2009 data | ARIMA, omitting post-2009 data | ARIMA, Alaska, 2002–12 |
|---|---|---|---|---|---|
| Gonorrhea | −3.2 | −1.8 | −0.3 | −2.1 | +2.3 |
| | [−4.2, −2.4] | [−3.3, −0.3] | [−1.2, +0.6] | [−5.2, +1.1] | [−2.1, +6.7] |
| Chlamydia | −4.8 | −2.1 | −2.0 | −4.0 | −0.1 |
| | [−7.8, −1.9] | [−3.9, −0.3] | [−3.9, −0.2] | [−9.9, +1.9] | [−10.4, +10.3] |
| Observations | 108 | 108 | 84 | 84 | 132 |

95% confidence intervals in brackets. First two replication regressions use Newey-West standard errors with a lag limit of 12 months to allow for seasonality. Independent variables are a dummy for the periods after the September 2009 tax increases (October 2002 in Alaska regression). All regressions control for real median household income, and average incidence in 12 states that allow private retail in alcohol, do not border Illinois, and did not increase alcohol taxes in the study period. ARIMA models are $(0,1,1)(0,1,1)_{12}$.
Source: Staras et al. (2014), Table 1; author's calculations.

The replication (column 2) is consistent with the original in the sense of pointing to statistically significant reductions in STD incidence. However the implied impacts are about half as big, which is a much larger discrepancy than in the earlier replications. Possibly underlying data have been revised, or there is some subtle design choice I have missed, or there is an error on one side or the other. At any rate, dropping the suspect post-2009 data weakens the apparent impact on gonorrhea, but not that on chlamydia (column 3).

More importantly, switching to an ARIMA model to zero in on whether the incidence trend changed immediately post-tax-hike produces much less precise estimates, whose confidence intervals embrace both positive and negative values (column 4). The widening makes sense since any measurement of the incidence *right* after the tax increase contains more noise than the average of measured incidence over a sequence of months beginning then. So the immediate drop needs to be larger in order to emerge as statistically significant in an ARIMA model fit.

Confidence intervals are also wide when transplanting the analysis to Alaska, despite the larger sample and bigger tax increase (column 5).

Because of the data problems and the apparent fragility of the results, I have low confidence that the tax hikes reduced STDs in Illinois or Alaska.

### 4.4. Chung et al. (2013), "The impact of cutting alcohol duties on drinking patterns in Hong Kong," *Alcohol and Alcoholism*

In 2008, the government of Hong Kong eliminated its 80% duty on wine and 40% duty on beer, as part of a successful effort to build the city into a regional wine-trading hub. The 100% duty on liquor remained. (Chung et al. 2013, p. 720.) Since the duties were calculated relative to import prices, retail prices fell much less, proportionally: 1.8% for beer and 14.3% for wine (Chung et al. 2013, p. 721). The authors surveyed Hong Kong residents before (2006) and after (2011−12) in order to assess the impact on drinking behavior. Surveyors asked such questions as whether respondents had had at least 5 drinks at a time in the last 30 days, the chosen definition of binge drinking. Analysis consisted of looking for changes between surveys.

The study has significant limitations. First, the price drop was small, being significant only for wine, which accounted for only 15% of Hong Kong alcohol consumption on a pure alcohol basis (Hong Kong DOH 2011, Table 3). Second, the global recession is itself a major potential confounder; a lot else happened in Hong Kong around 2008. This matters especially because the surveys did not take place in a tight window around the tax change—which would strengthen attribution of results to that event—but with a gap between surveys of 5−6 years. Changes in drinking over such a long and eventful period cannot be confidently attributed to a modest tax increase somewhere in the middle. By the same token, *absence* of clear change might merely reflect lack of statistical power rather than lack of impact.

Third, unlike the cause-of-death and crime data used in the studies discussed above, which come from governments and pertain to rather objectively observable events, drinking behavior is here self-reported—or not, if people refuse to talk to the surveyors. Heavy drinkers may systematically respond less or understate their troubles with alcohol.

Chung et al. find that the share of people who say they had ever had alcohol rose from 66.6% in 2006 to 82.0% in 2011 and 85.2% in 2012 (p. 720). The duty cuts may have contributed to that climb, but so might other societal factors. Surprisingly, the rate of binge drinking (≥5 drinks in one episode in the last month) *dropped*, from 9.0% to 7.3%. The authors conclude that cultural and biological factors limited heavy drinking. (As for biology, East Asians have less aldehyde dehydrogenase in their bodies, so alcohol accumulates more rapidly in the blood in its toxic form, and people get drunk faster.)

In light of the potential biases from self-reporting, I extracted data from the World Health Organization's Mortality Database on cirrhosis in Hong Kong. I reduced the influence of demographic trends, which could be steadily shifting the share of the population at prime risk for liver disease, by extracting the death rate for each age-sex group in each year, then estimating the number of deaths that would have occurred if the number of people in each of those groups had not changed since 2001. (Alcohol liver disease deaths are defined as those with an International Classification of Diseases (ICD)-10 code beginning with K70.[20]) These statistics corroborate the very low level of drinking-related disease—a mere 50 deaths/year—and the lack of any rise after 2008. (A graph for a broader family of alcohol-related diseases tells the same story.)

### Deaths from alcoholic liver disease, Hong Kong, 2001 age distribution

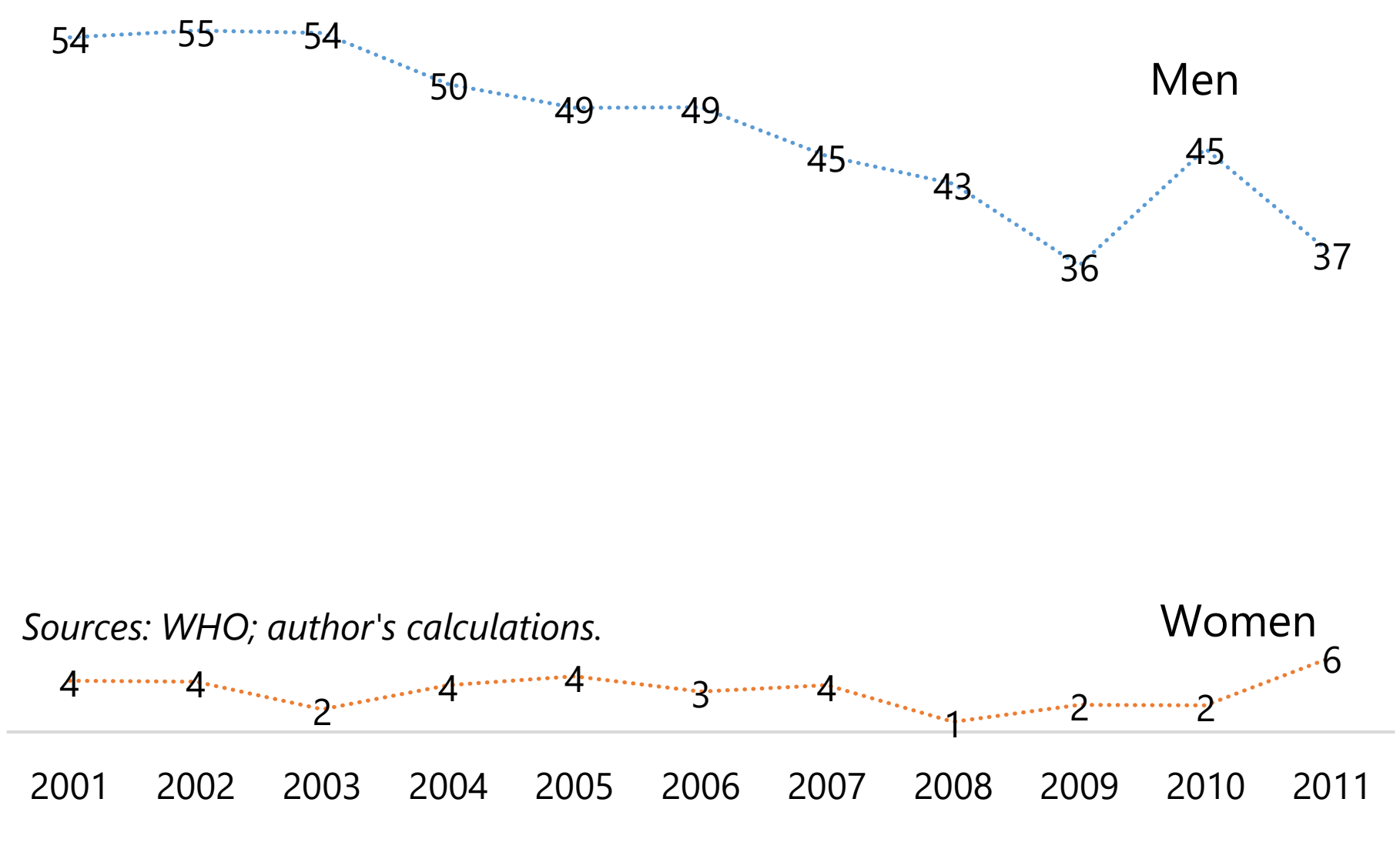

*Sources: WHO; author's calculations.*

Certainly there is little evidence here that the tax cut in Hong Kong increased drinking or its side effects. But with the price effect so small and the before-after study period so long, any other result would have been surprising. The conservative conclusion is that the null result indicates lack of statistical power rather than the impotence of taxation as a policy.

[20] http://apps.who.int/classifications/icd10/browse/2015/en#/K70-K77.

## 4.5. Heeb et al. (2003), "Changes in alcohol consumption following a reduction in the price of spirits: A natural experiment in Switzerland," *Addiction*<br>Kuo et al. (2003), "Does price matter? The effect of decreased price on spirits consumption in Switzerland," *Alcoholism: Clinical and Experimental Research*<br>Mohler-Kuo et al. (2004), "Decreased taxation, spirits consumption and alcohol-related problems in Switzerland," *Journal of Studies on Alcohol*<br>Gmel et al. (2007), "Estimating regression to the mean and true effects of an intervention in a four-wave panel study," *Addiction*

In July 1999, Switzerland cut import duties on foreign-made spirits in order to comply with World Trade Organization rules. This lowered domestic spirits prices 30−50% (Heeb et al., p. 1434). These four studies, all by essentially the same team, track the impacts.

To provide background, I tried to graph alcoholic liver disease deaths for Switzerland as I did for Hong Kong above. However, Switzerland's data uses a condensed coding system that does not break out such fine-grained categories as alcoholic liver disease, but only the broader categories of all liver disease. Since alcohol is blamed for the majority of liver disease deaths in neighboring countries (In France, about 67% for men and 59% for women in 2000; and in Germany 61% and 47%) the broader totals are still a relevant, if imperfect, indicator:

**Deaths from liver disease, Switzerland, 2001 age distribution**

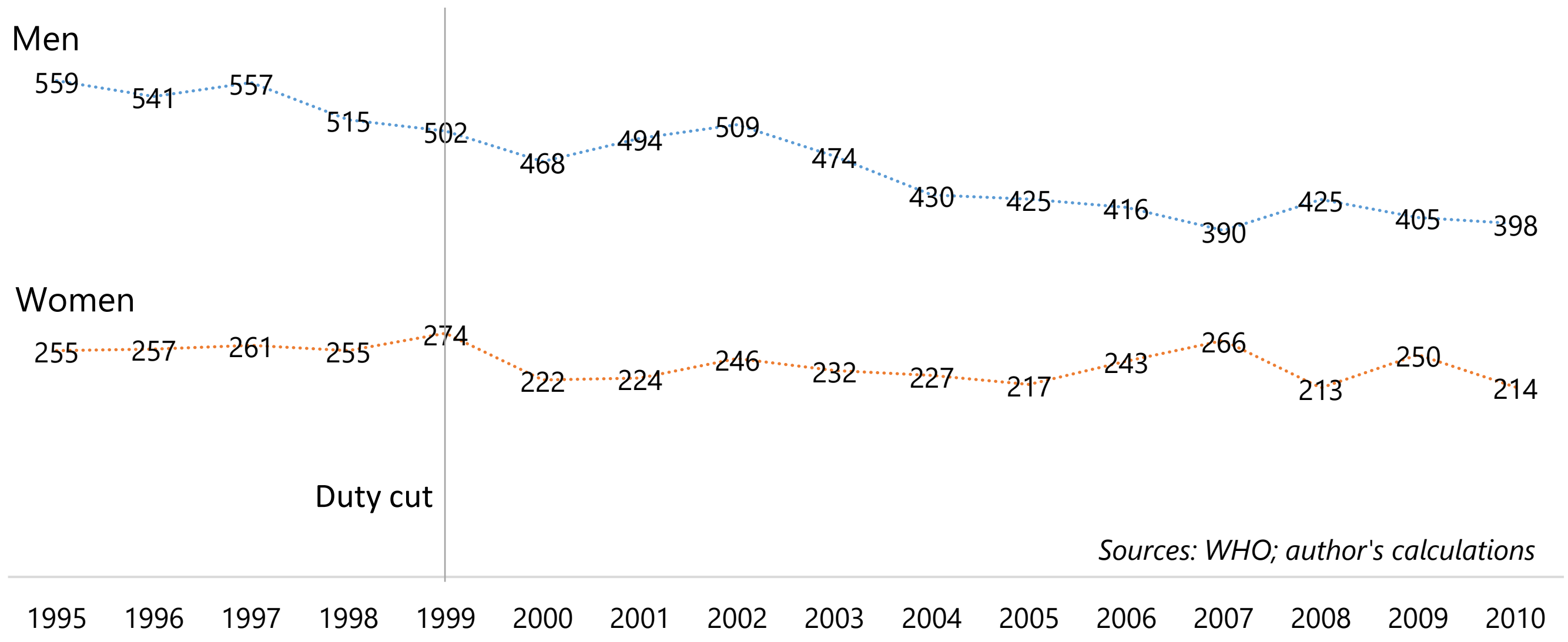

Surprisingly, the death rates *drop* in 2000, the first full year after the import duty was cut for spirits. This challenges the hypothesis that the duty cut increased drinking. But not definitively: Death from liver disease is not synonymous with death from alcoholic liver disease. More drinking is not synonymous with more alcohol-related liver disease deaths. Other factors may have reduced mortality, masking the contrary impulse from the price decline. Indeed, it is hard to fathom why a duty cut would have *caused* a mortality decline, which reminds us of our ignorance of all the forces at work. But it is not impossible: perhaps the duty cut shifted market prices in favor of liquors with lower alcohol content, enough to offset an overall increase in drinking. At any rate, the graphs put a perplexing cloud over the findings that follow.

Using before- and after-surveys, Heeb et al. (2003, p. 1433) report that in a six-month period bracketing the duty cut, spirits consumption rose 28.6% in Switzerland while wine and beer consumption remained unchanged. This combination of results looks very much like a fingerprint of the duty cut, since it too was specific to spirits.

Kuo et al. followed up with a survey two years later. They find the spirits consumption increase to have been sustained, at 38% cumulative. They break out the results by age, sex, and other traits. All subgroups except people over 60 reported increases:

### Changes in spirits consumption by subgroup, Switzerland, before and after July 1999 duty elimination

| | Consumption/day (grams pure alcohol) | | |
|---|---|---|---|
| Group | Spring 1999 | Fall 2001 | Change |
| All | 0.711 | 0.980 | +38% |
| Men | 0.936 | 1.217 | +30% |
| Women | 0.521 | 0.788 | +51% |
| Age <30 | 0.934 | 1.551 | +66% |
| Age 30−59 | 0.619 | 0.870 | +41% |
| Age ≥60 | 0.691 | 0.639 | −8%* |
| Working | 0.806 | 1.023 | +27% |
| Not working | 0.698 | 0.898 | +29% |
| High school degree only | 0.932 | 1.024 | +10% |
| Post-high school education | 0.695 | 0.959 | +38% |
| Smoker | 0.961 | 1.314 | +37% |
| Non-smoker | 0.587 | 0.827 | +41% |
| Averages ≥3 drinks/day | 1.337 | 1.563 | +17% |
| Averages <3 drinks/day | 0.452 | 0.737 | +63% |

*Only change not differing from zero at $p = 0.01$.
Source: Kuo et al. (2003, Table 1, right pane).

The statistics above are straightforward, but do not address how traits interact. It might be, for example, that the slightly greater rise among non-smokers owes entirely to non-smokers belonging disproportionately to the under-30 age group, which itself increased consumption more. Perhaps after controlling for age, the differential by smoking status evaporates.[21]

To check such possibilities, Kuo et al. (2003) run multivariate regressions. However, their methods are complex and cursorily described; and the column labels in the results table (their Table 2) confuse me: separate columns are labeled "Test of the effect *p*-value" and "*p*-value." This may help explain why Nelson's (2014b) systematic review on binge drinking appears to misinterpret the study. (Nelson 2014b, p. 47: "Any changes were not significant controlling for age, sex, and volume.") And, as is unfortunately the norm, data and code have not been posted. For myself, e-mail discussion with the lead author has not eliminated the confusion.[22] The interpretation that seems to fit best is that their Table 2 corroborates the bivariate results quoted above, showing increases in drinking among most subgroups, including self-reported heavy drinkers. Most of the reported *p* values are highly significant. It appears that those reporting in the "before" survey that they had ≥3 drinks/day increased alcohol consumption

[21] Whether Swiss non-smokers are young, I do not know. This example is purely for illustration.
[22] My guess is that the insignificant *p* value cited by Nelson (0.3778, in Kuo et al. Table 2) is for the hypothesis that whether a person responded to the follow-up survey was uncorrelated with his or her sprits consumption *before* the abolition of the duty—and that in the reference demographic, which is people aged 60 or more. The failure to reject that hypothesis only reduces the concern that attrition from the follow-up survey had a pattern that would bias the results. It does not support Nelson's interpretation.

0.58 grams/day after the tax cut. That was twice the national-average increase of 0.27 (from 0.711 to 0.980, in first row in table above).

Mohler-Kuo et al. (2004) is a companion to Kuo et al. (2003); in similar fashion, it analyzes answers to different survey questions. These ask not about the amount of drinking, but signs of problem drinking, such as "How often during the last 6 months have you been unable to remember what happened the night before because you had been drinking?" (p. 268). The authors fashion an index of problem drinking from the answers to such questions. The index rises on average after the tax cut, especially among young people. Put technically, a follow-up dummy (for the second survey round) receives statistically significant, positive coefficient in a regression of the index (Mohler-Kuo et al. 2004, Table 2, model 1). But then, when entered, a control for reported consumption of spirits outcompetes the dummy for explanatory power (models 2 & 5). The same does *not* happen when with beer or wine consumption, which did not have duties removed (models 3 & 4). This strongly suggests that the passage of time increased problem drinking via a rise in spirits consumption only. Again, the distinct pattern for spirits, versus beer and wine, helps convince.

Now, among the subgroup splits in the table above, the last—between heavy drinkers and the rest—is particularly susceptible to a kind of statistical bias. The numbers in the table say that those classed as heavy drinkers before the duty cut increased their drinking much less than the rest, by 17% instead of 63%. However, this difference may be exaggerated by *regression to the mean*. Most likely some people in the first survey, in spring 1999, happened to have drunk more recently than they normally did—whether in reality or imperfect recollection. Some would have been inappropriately classed as heavy drinkers, by clearing the 3 drinks/day threshold. Symmetrically, some would have drunk less than usual or underestimated their recent drinking, and been misclassified as non−heavy drinkers. If the tendency to mismeasure a given person is partly random over time, then we could expect that, like an average basketball player who happens to play one great game, the mismeasured drinkers would converge in the next survey toward population averages. Thus even if there were no true difference between heavy and light drinkers, the follow-up survey would show a smaller increase for the heavy drinkers. Turning that around, if regression to the mean could be purged from the data, then the table above could show an even larger relative rise among heavy drinkers.

Gmel et al. (2007) study this issue systematically.  They take advantage of the fact that two survey rounds were conducted after the tax cut. Since alcohol taxes did not change between these latter surveys, they can more confidently attribute any apparent convergence between heavy and non−heavy drinkers over those later rounds to the statistical mirage of regression to the mean. Then, having gauged the effect, they can then approximately remove it from results like those in the table above.

Gmel et al. (2007, p. 32) conclude that after the Swiss tax cut, heavy drinkers actually increased their drinking *more* than others—but only temporarily. By the two-year follow-up, the difference had disappeared. This suggests that in the long run, price changes affect heavy drinkers about as much as the general population.

The story most consistent with the evidence is a bit complicated: after the duty cut, problem drinking and drinking in general increased; yet deaths from liver disease, and thus probably from alcoholic liver disease, fell. As already conjectured, some of the foreign liquors that became more affordable may have contained less alcohol content than their Swiss competitors, causing people to drink more alcoholic beverages but take in less pure alcohol. This effect would have escaped detection in the surveys studied above, which asked people only about how many drinks they had or how often they drank. If the conjecture is correct, it would confirm the responsiveness of drinking behavior to taxation. And it would demonstrate the potential for unintentional side effects from tax changes that are not across the board and well calibrated to alcohol content.

## 4.6. Koski et al. (2007), "Alcohol tax cuts and increase in alcohol-positive sudden deaths: A time-series intervention analysis," *Addiction*

Mäkelä et al. (2007), "Changes in volume of drinking after changes in alcohol taxes and travellers' allowances: Results from a panel study," *Addiction*
Mäkelä and Österberg (2009), "Weakening of one more alcohol control pillar: A review of the effects of the alcohol tax cuts in Finland in 2004," *Addiction*
Herttua, Mäkelä, and Marikainen (2008), "Changes in alcohol-related mortality and its socioeconomic differences after a large reduction in alcohol prices: A natural experiment based on register data," *American Journal of Epidemiology*
Herttua et al. (2008), "The impact of a large reduction in the price of alcohol on area differences in interpersonal violence: A natural experiment based on aggregate data," *Journal of Epidemiology and Community Health*
Bloomfield et al. (2010), "Changes in alcohol-related problems after alcohol policy changes in Denmark, Finland, and Sweden," *Journal of Studies on Alcohol and Drugs*
Helakorpi, Mäkelä, and Uutela (2010), "Alcohol consumption before and after a significant reduction of alcohol prices in 2004 in Finland: Were the effects different across population subgroups?", *Alcohol and Alcoholism*
Gustafsson (2010), "Alcohol consumption in southern Sweden after major decreases in Danish spirits taxes and increases in Swedish travellers' quotas," *European Addiction Research*

In integrating with the European Union, Denmark, Finland, and Sweden changed several alcohol policies in 2003 and 2004. In January 2004, the countries raised their limits on how much alcohol travelers could carry when entering from other EU member states. Then in May, Estonia joined the EU, making it easier for people to buy alcohol in Estonia and bring it into nearby Finland. Anticipating competition for domestic retailers through this import channel, Denmark and Finland cut domestic taxes. In March, Finland reduced taxes on spirits (by 44%), fortified wine (40%), table wine (10%), and beer (32%). Denmark had cut its spirits tax 45% the previous October. For its part, Sweden did not cut alcohol taxes, but at least in the southern part of the country, alcohol became more available because of the higher traveler's allowances and lower retail prices in nearby Denmark. (Bloomfield et al. 2007, pp. 182–83.)

These graphs show how the policy changes translated into prices.[23] In Sweden, where no taxes were cut and the government held a retail alcohol monopoly, prices did not change. In Finland, where taxes fell the most and the government also holds a retail monopoly, the drop was dramatic. As expressed here, the index fell 18%, from 107.5 in February 2004 to 88.1 in March. An immediate price impact is also visible in Denmark, but it is smaller, presumably because only the tax on spirits fell. Perhaps as well, the subsequent downward trend reflects delayed market adjustment to the duty cut; in Denmark alcohol sales are the province of the private sector. (Market structures from Örnberg and Ólafsdóttir 2008.)

[23] Data are from Eurostat via FRED: FRED data series CP0210DKM086NEST, CP0210FIM086NEST, CP0210SEM086NEST.

**Alcohol consumer price indexes (January 2000 = 100)**

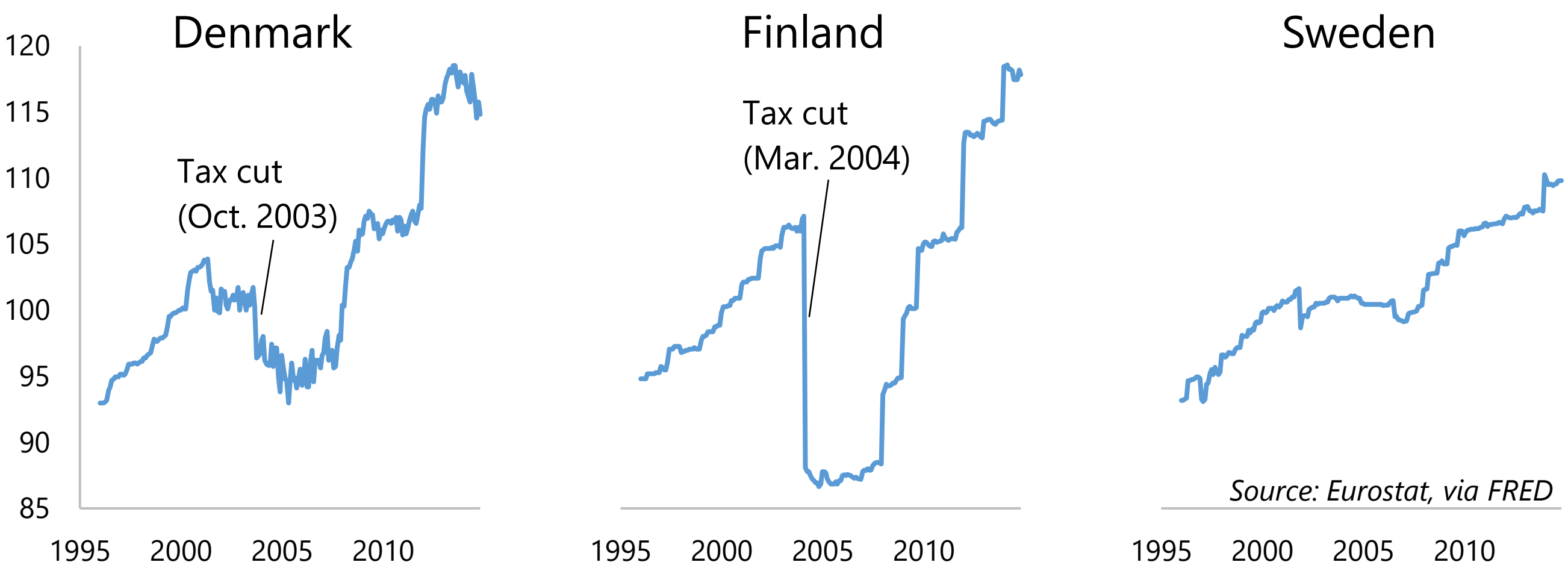

A substantial literature has tracked these natural experiments. The findings are reassuringly consistent across studies—and arguably coherent with the price graphs above, in the sense that impacts are found only where the natural experiment was largest, in Finland. There, several studies show, people in some demographics drank more, and died more from alcohol-linked diseases such as cirrhosis. By and large, Denmark and Sweden did not experience such changes. That said, despite my hint (after the fact) that the results were about what one should expect, the lack of impact in Denmark and Sweden surprised authors of the studies reviewed here. After all, it came in spite of the increased ease of import in both countries and the price drop in Denmark. (Mäkelä et al. 2008, p. 181: "The results did not confirm expectations." Bloomfield et al. (2010), p. 38: "…surprising findings on self-reported alcohol problems and alcohol consumption.")

The thrust of the results from the various studies is conveyed in three more graphs, which are analogous to my earlier one for Hong Kong. In Finland, the numbers of men and women dying from alcoholic liver disease jumped between 2003 and 2004 and kept climbing through 2007, until the alcohol price began to recover. No such developments appear clearly in Denmark and Sweden.

**Deaths from alcoholic liver disease, 2001 age demographics**

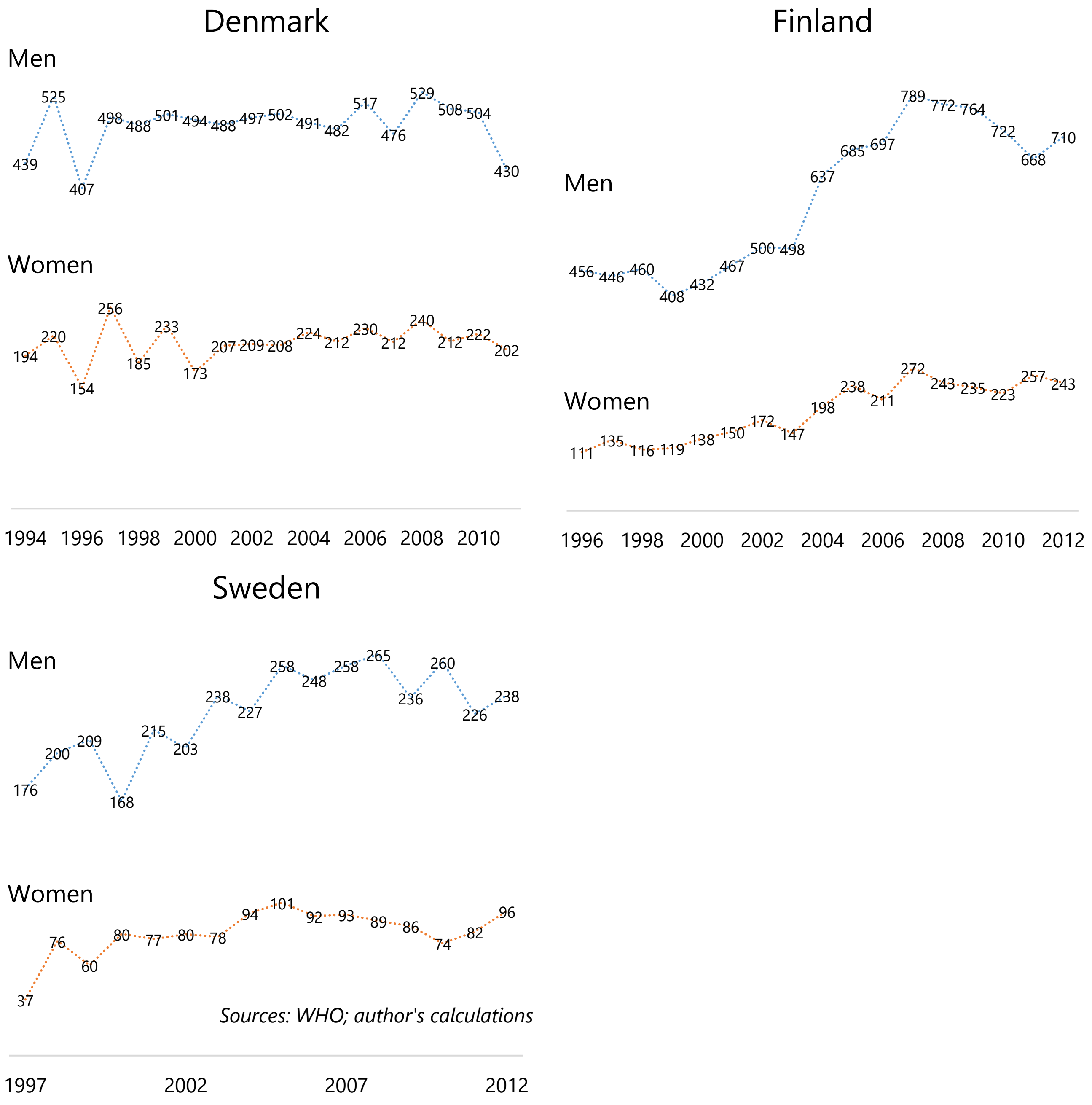

Mäkelä and Österberg (2009, fig. 2, copied below) confirm that in Finland, alcohol sales rose after the tax cut. The grey line in their graph shows the percentage increase in 2004 sales relative to the same week in 2003. The black line smooths those numbers with a 5-week moving average. Before week 10, when the tax fell, the year-on-year change hovered around 0. After, it jumped to 20−40% and then settled into a long-term 10% rise. The other major spikes in the graph should be ignored as artifacts of major holidays falling in different weeks in the two years,

**Alcohol sales by week in Finland, 2004, relative to 2003**

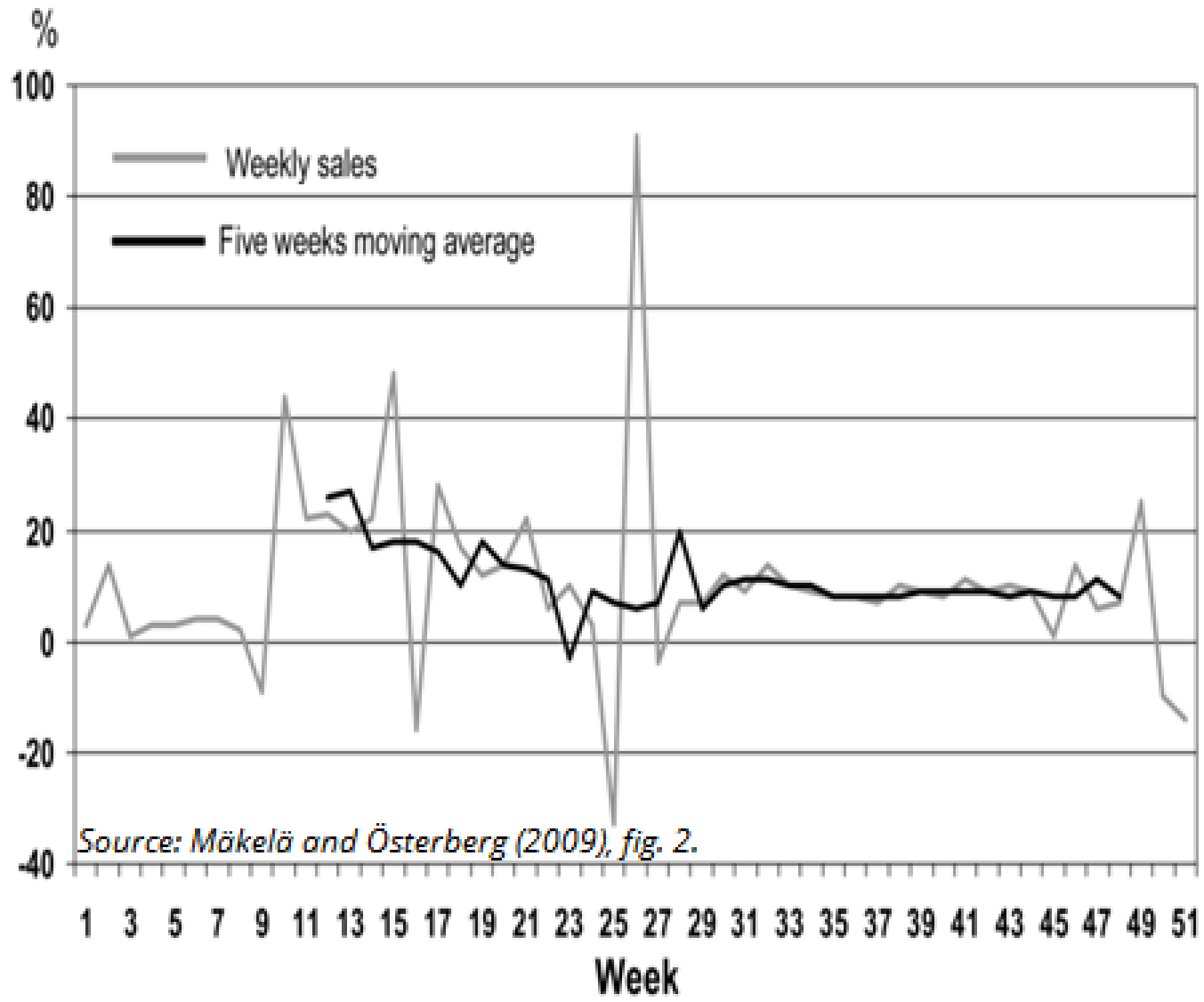

Herttua, Mäkelä, and Martikainen (2008) find that Finland's worse-off suffered the largest increases in deaths from alcohol-related diseases—notably the unemployed and those in the lowest three income quintiles. (However, the increase was not quite as large in the lowest quintile, perhaps because alcohol was still expensive for this group.) To wit, this table excerpts their results:

**Change in death rate from alcohol-related diseases between 2001–03 and 2004–05, by subgroup, Finland**

| | Men | | Women | |
|---|---|---|---|---|
| | % change | 95% confidence interval | % change | 95% confidence interval |
| **People aged 30+** | | | | |
| Age* | | | | |
| 30–39 | +13.5 | [+0.8, +27.8] | +1.8 | [−23.0, +34.5] |
| 40–49 | +6.3 | [+0.0, +14.1] | +7.3 | [−8.0, +25.1] |
| 50–59 | +23.9 | [+17.4, +30.8] | +52.7 | [+36.0, +71.5] |
| 60–69 | +15.4 | [+7.8, +23.5] | +30.6 | [+13.1, +50.8] |
| 70–79 | +6.0 | [−4.3, +17.4] | +23.8 | [+0.0, +55.1] |
| 80+ | +16.9 | [−9.9, +51.7] | −16.1 | [−50.0, +40.9] |
| Education | | | | |
| Upper tertiary | +7.8 | [−11.3, +31.1] | +56.0 | [+0.4, +142.3] |
| Lower tertiary | +16.0 | [+4.7, +28.5] | +11.5 | [−9.9, +38.0] |
| Secondary | +20.3 | [+13.9, +27.0] | +39.8 | [+23.5, +58.3] |
| Basic | +16.1 | [+11.0, +21.5] | +34.2 | [+22.0, +47.6] |
| Social class | | | | |
| Upper white-collar | +8.4 | [−3.7, +22.0] | +31.9 | [+2.3, +70.1] |
| Lower white-collar | +14.2 | [+4.4, +24.9] | +34.9 | [+19.5, +52.2] |
| Skilled worker | +17.0 | [+10.4, +24.0] | +10.6 | [−9.0, +34.5] |
| Unskilled worker | +18.2 | [+11.6, +25.1] | +33.5 | [+18.0, +51.1] |
| Self-employed | +14.9 | [+1.8, +29.6] | +17.6 | [−12.9, +58.7] |
| Other | +11.1 | [−0.5, +23.9] | +38.4 | [+7.4, +78.1] |
| All | +17.1 | [+13.4, +21.0] | +30.6 | [+21.7, +40.0] |
| | | | | |
| **People aged 30–59** | | | | |
| Economic activity* | | | | |
| Employed | +2.9 | [−4.7, +11.1] | +8.0 | [−9.4, +28.7] |
| Unemployed > 2 years | +21.2 | [+11.7, +31.5] | +49.9 | [+24.8, +79.9] |
| Unemployed ≤2 years | +30.1 | [+14.2, +48.1] | +81.3 | [+39.8, +135.1] |
| Pensioned | +27.1 | [+18.0, +36.9] | +36.6 | [+16.3, +60.5] |
| Other | +21.5 | [+6.3, +38.8] | +54.0 | [+18.3, +100.4] |
| Income quintile* | | | | |
| 1 (highest) | +17.2 | [+5.6, +30.1] | +17.8 | [−6.8, +48.9] |
| 2 | +17.5 | [+5.3, +31.0] | +61.2 | [+28.3, +102.5] |
| 3 | +78.8 | [+61.7, +97.7] | +81.5 | [+46.2, +125.3] |
| 4 | +64.0 | [+50.6, +78.6] | +78.2 | [+49.1, +112.9] |
| 5 (lowest) | +19.4 | [+10.4, +29.2] | +48.4 | [+25.1, +76.1] |
| All | +19.2 | [+14.6, +24.1] | +33.6 | [+22.4, +45.8] |

*Hypothesis of same change in all subgroups is rejected at $p = 0.1$ for age, economic activity, and income quintile, for both men and women.
Source: Herttua, Mäkelä, and Marikainen (2008, tables 1 & 2).

The various studies of these natural experiments are summarized as follows:

- *Koski et al. (2007)* performs time series analysis like that in the Florida and Alaska studies, using Finnish government data on deaths by cause and week for 1990−2004. They find that alcohol-related deaths jumped in 2004—by about 8 deaths per week, or 400 per year. This is about twice the increase seen in my Finland graph above, partly because mine is restricted to liver disease. (A version of my graph embracing all alcohol-related diseases shows a rise from 1243 to 1505 a year for men and from 360 to 423 for women.) The study gains strength from de-facto falsification tests. Since the traveler's allowance increase, tax cut, and accession of Estonia took place in January, March, and May of 2004 respectively, the regressions allow for trend breaks at all three points. Only the break corresponding to the tax cut is statistically significant. In addition, rates of death from causes *not* linked to alcohol show no changes at any of the dates (Koski et al. 2007, table 1).
- *Mäkelä et al. (2007)* is based on surveys fielded in Denmark, Finland, and southern Sweden. Self-reported drinking did not increase in any of the regions. In all three, it tended to converge across age groups, with older people drinking less and younger drinking more. The *lack* of apparent overall increase in Finland contradicts the sales data from the Finnish retail monopoly graphed above, not to mention the jump in deaths, suggesting that surveys based on self-reports underestimate drinking changes.
- *Bloomfield et al. (2010)*: Like Mäkelä et al. (2007), and involving most of the same authors, this study analyzes results from surveys fielded in Denmark, Finland, and Sweden in 2003−06. Northern Sweden is taken as the control region. This study focuses on self-reported signs of problem drinking, as Mohler-Kuo et al. (2004) does for Switzerland. If anything, prevalence of alcohol problems *decreased* in Denmark, Finland, and southern Sweden relative to the control region.
- *Helakorpi, Mäkelä, and Uutela (2010)* also works from a repeated survey, but one performed only in Finland. Perhaps because the sample is larger—79,100 received the survey by mail and 72% responded—this survey-based analysis, unlike Mäkelä et al. (2007), detects changes quite in line with what Herttua, Mäkelä, and Marikainen (2008) find in government data. Moderate and heavy drinking became more common, especially among those who were 45−64 years old or less educated.
- *Mäkelä and Österberg (2009)* documents alcohol-related trends in Finland including the sales data graphed earlier. In addition, they show how annual statistics point to tax-cut-attributable increases in deaths from alcohol-related diseases and drunk driving (the latter temporary), but not to a link with violent crime:

Assaults and drunk driving cases per 100,000 population, Finland

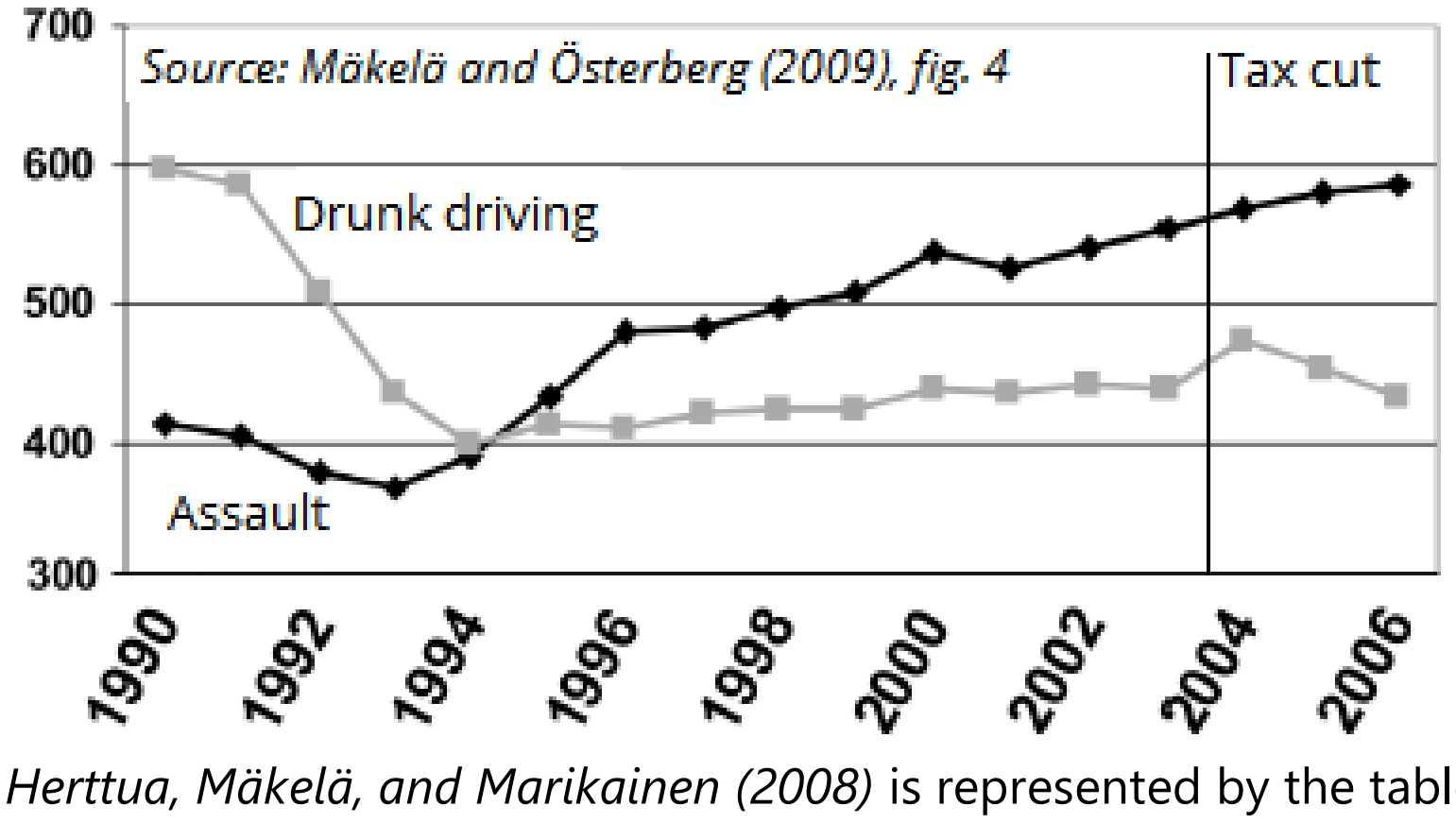

- *Herttua, Mäkelä, and Marikainen (2008)* is represented by the table above.
- *Herttua et al. (2008)* looks across 86 administrative tracts in Helsinki, and finds no rise in interpersonal violence from 2002−03 (pre-tax-cut) to 2004−05 (post-tax-cut).
- *Gustafsson (2010)* surveyed people in Denmark and southern Sweden by telephone during 2003−06 about how much they drank, and how often they engaged in heavy drinking. Again, no sudden changes were found in these countries despite the policy changes of 2003−04.

# 5. Cross-section studies

## 5.1. Cook and Durance (2013), "The virtuous tax: Lifesaving and crime-prevention effects of the 1991 federal alcohol-tax increase," *Journal of Health Economics*

In 1990, President Bush and the US Congress agreed to a deficit reduction package that was famously taken to violate Bush's "no new taxes" pledge. Among other things, it raised taxes on alcohol: from $0.29 to $0.58/gallon for beer, $0.17 to $1.07/gallon for wine, and $12.50 to $13.50/gallon for spirits (the last expressed in volume of pure ethanol content). As a result, consumer alcohol prices rose 10% in January 1991:

**U.S. alcohol consumer price index (March 1983 = 100)**

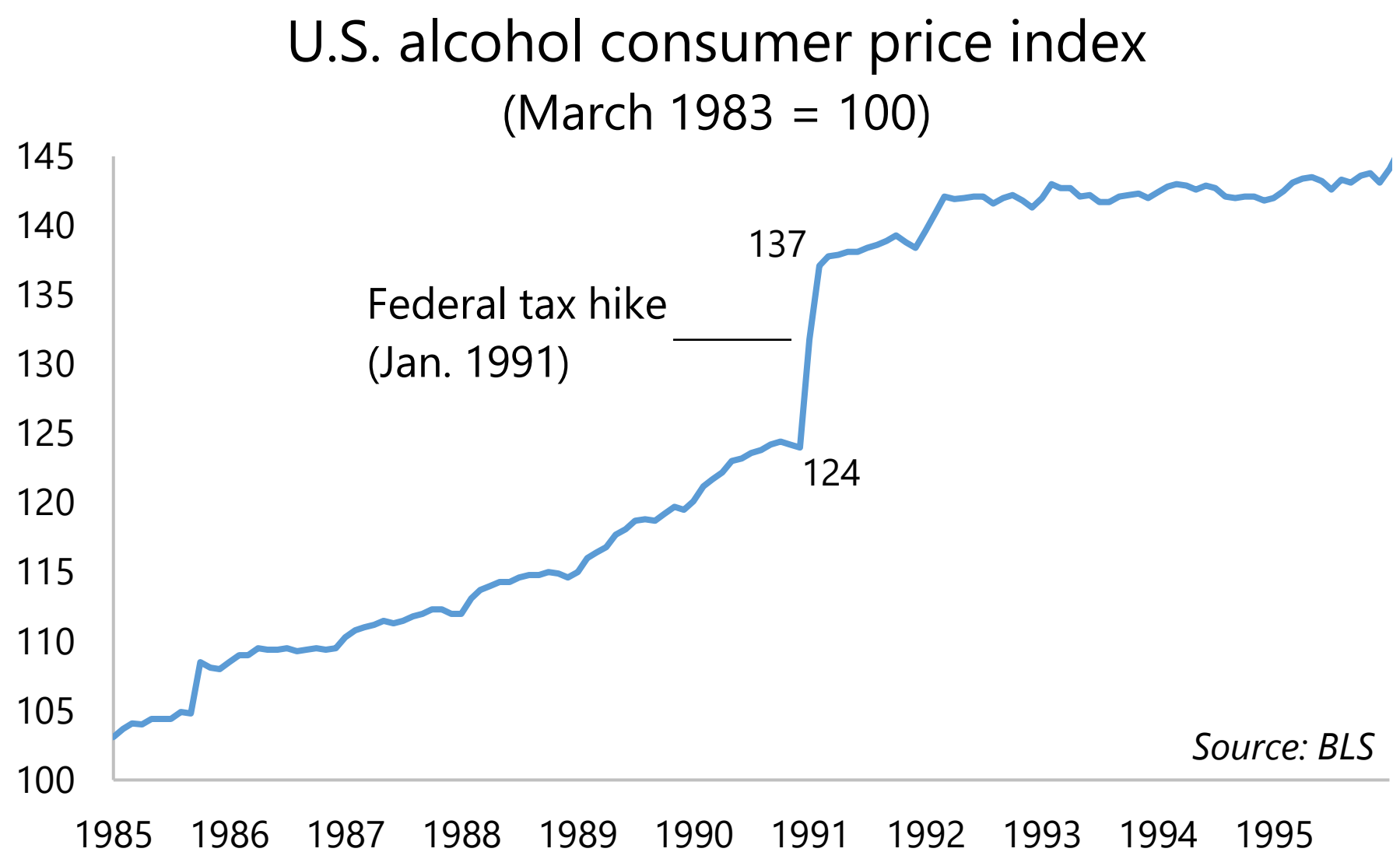

The paper of Cook and Durance works to exploit this natural experiment in a way quite different from the studies above. In fact, technically, it is not a time series study, nor really a quasi-experimental one. Instead of testing whether drinking or its side effects dropped nationwide right after the tax rose, Cook and Durance compare trends across states. They hypothesize that drinking fell more in states where it had farther to fall: people in "wetter" states felt the added tax more than those in "drier" states, and cut back more in absolute terms. This in turn may have manifested as larger drops (or smaller increases) in violent crime, traffic deaths, and suicides in the wetter states. Cook and Durance (Tables 2 & 3) check for this cross-state pattern and find it for suicide, assault, and robbery.

This study too must clear additional hurdles before it can convince. One challenge is that it wants to attribute convergence to a tax increase in a context in which convergence was already occurring. Even before the tax increase, with the exception of Utah, alcohol sales were falling most where they were highest to start with:

**Alcohol sales by state, 1980 & 1990**

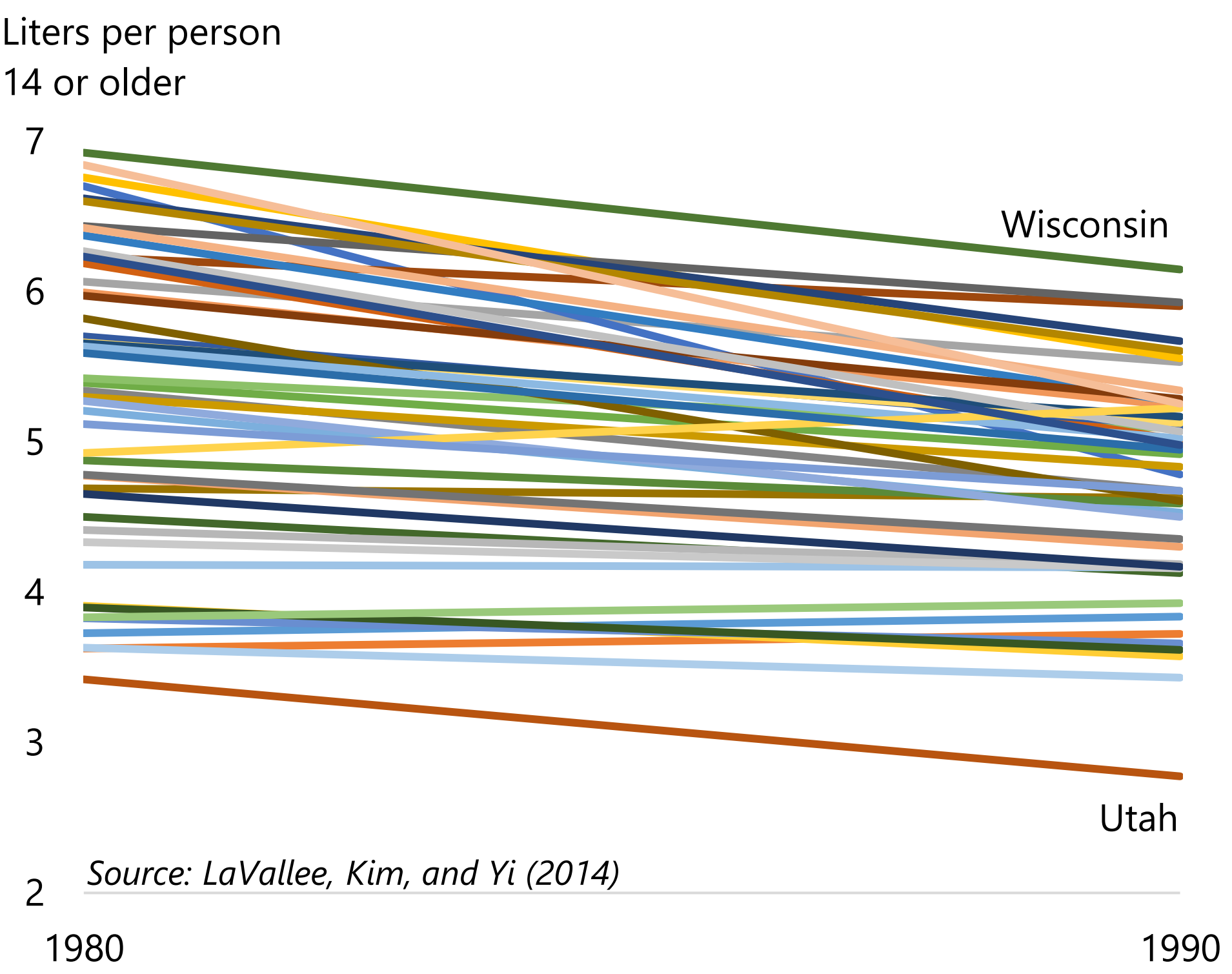

If the Cook and Durance theory holds, long-term convergence in drinking is a force for convergence in crime, traffic deaths, and suicides regardless of any tax changes. It therefore does not suffice to demonstrate that states converged on these outcomes right after the alcohol tax rose: that would be expected anyway. Cook and Durance need to show that convergence *temporarily accelerated* in 1991. Understanding this, the authors run regressions that control for state-specific long-term trends.

A second hurdle is the standard concern about the exogeneity of "treatment." The clever Cook and Durance strategy requires that a state's per-capita alcohol consumption just before the tax correlated with subsequent trend deviations in crime, traffic deaths, and suicides *only* through the causal chain they sketch. But low- and high-consumption states appear to have differed systematically in many respects. The 10 lowest-consumption states in 1989 (Utah, West Virginia, Oklahoma, Kentucky, Kansas, Arkansas, Alabama, Tennessee, Iowa, Mississippi) were relatively southern, rural, and poor. The 10 highest-consumption ones (Colorado, Rhode Island, Massachusetts, Hawaii, California, Vermont, Arizona, Florida, Delaware, Wisconsin) were on average more urban and wealthy, and perhaps healthier and better educated too. For reasons separate from alcohol taxes, the two groups may have exhibited systematic differences in levels and trends in crime, traffic deaths, and suicide. Perhaps, for example, the poor, dry states suffered more from the 1990–91 recession, causing their crime rates to climb relative to the wet states. This could produce the sort of result Cook and Durance find—improvement in crime in wet states relative to dry states—without involving alcohol taxes.

The third concern is the lack of falsification tests. If the Cook and Durance theory is right, shifting their analysis backward or forward a few years should produce null results.

To better understand the data, in particular to apply that falsification test, I replicated Cook and Durance, starting from the same public data sources (LaVallee, Kim, and Yi (2014) for the latest update on alcohol consumption; FBI Uniform Crime Reporting Statistics for crime data; CDC for fatal injury reports). One subtlety needs explaining. The arrival of the tax on January 1, 1991 appears to have distorted alcohol buying in both 1990 and 1991, making both years' sales data unreliable proxies for actual drinking. Evidently, people rushed to buy alcohol before the price rose, shifting sales from early 1991 to late 1990:

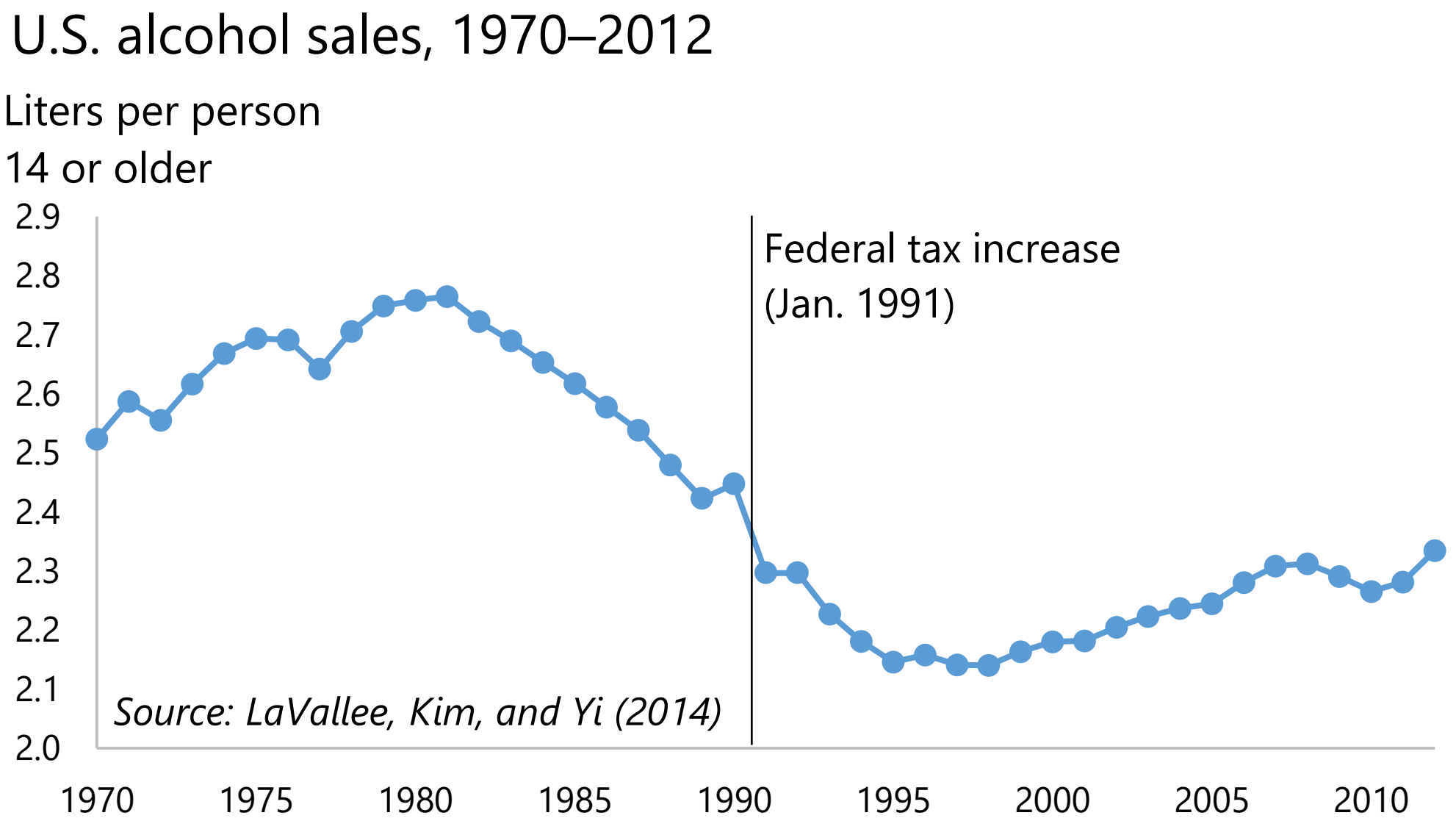

This is why Cook and Durance use 1989 rather 1990 sales data to measure how "wet" a state was.

For each outcome of interest (homicide, traffic deaths, etc.), Cook and Durance perform three kinds of regressions. Their results and my replications are in the table below. The matches are not perfect, but again good enough to head off any doubts about implementation errors or fragility to data revision. Read this table less to digest individual results than to confirm the consistency between the original results, on the left, and the new ones, on the right.

The first type of Cook and Durance regression, in the first and fourth columns, applies to a cross-section: it checks whether, across states, per-capita alcohol sales in 1989 correlated with the subsequent change in each outcome, between 1990 and 1991. As the first column shows, all of the correlations are negative, which is consistent with the theory that where people drank the most, drinking and its side effects fell more after the January 1991 tax increase. And many of those negative correlations are statistically significant.

However, these regressions do not address the possibility highlighted above that long-term trends driven by factors other than the tax increase can fully explain convergence. The regressions reported in the remaining columns take on this concern by incorporating annual data for each state for 1985−95, and allowing for fixed state and year effects. They ask whether the 1990−91 deviation in each outcome, relative to each state's 1985−95 straight-line trend, and after removing nationwide ups and downs, was correlated with 1989 alcohol sales. The "Panel II" regressions in the last column are more conservative still in abstracting from long-term convergence in a second way, by controlling for the recent alcohol sales level (1 or 2 years back). In other words, to the extent that a greater decline in crime, traffic deaths, or suicide could be predicted by a higher drinking level in the previous year, this effect too is saved from accidental attribution to the tax hike. In the event, this conservative adjustment does not affect results much: "Panel I" and "Panel II" results nearly match.

**Cross-state correlations between 1989 per-capita alcohol sales and 1990–91 changes in death and crime rates**

| | Original (Cook and Durance 2013) | | | Replication | | |
|---|---|---|---|---|---|---|
| | Cross−section | Panel I | Panel II | Cross−section | Panel I | Panel II |
| **Alcohol sales (liters/person 14 or older, change, 1989−92)** | | | | | | |
| | | | | −0.203 | −0.128 | −0.127 |
| | | | | (0.040)*** | (0.041)*** | (0.038)*** |
| **Deaths/1000 people (change, 1990−91)** | | | | | | |
| All injury types | −4.266 | −4.481 | −4.501 | −4.381 | −4.248 | −4.341 |
| | (1.715)** | (1.852)** | (1.837)** | (1.720)** | (1.889)** | (1.871)** |
| Homicide | −2.153 | −3.641 | −3.657 | −2.190 | −4.972 | −5.271 |
| | (6.003) | (6.429) | (6.429) | (6.004) | (6.054) | (6.016) |
| Suicide | −6.559 | −6.712 | −6.703 | −6.557 | −6.585 | −6.584 |
| | (3.158)** | (3.476)* | (3.476)* | (3.175)** | (3.665)* | (3.673)* |
| Traffic | −5.371 | 4.368 | 4.410 | −5.369 | −4.878 | −5.050 |
| | (2.909)* | (3.175) | (3.175) | (2.915)* | (3.145) | (3.157) |
| **Crimes/100,000 people (change, 1990−91)** | | | | | | |
| All violent crime | −8.952 | −7.773 | −7.773 | −8.994 | −7.872 | −7.883 |
| | (2.679)*** | (2.651)*** | (2.654)*** | (2.684)*** | (2.576)*** | (2.601)*** |
| Murder | −6.054 | −4.078 | −4.006 | −6.034 | −4.245 | −4.239 |
| | (6.163) | (7.369) | (7.402) | (6.172) | (6.880) | (6.949) |
| Rape | −4.149 | −2.086 | −2.118 | −4.138 | −0.783 | −0.902 |
| | −4.026 | (3.948) | (3.970) | (4.043) | (4.130) | (4.161) |
| Assault | −10.080 | 9.367 | −9.361 | −10.133 | −10.037 | −10.069 |
| | (3.284)*** | (3.458)*** | (3.461)*** | (3.294)*** | (3.513)*** | (3.555)*** |
| Robbery | −11.236 | −9.692 | −9.690 | −11.264 | −9.571 | −9.496 |
| | (5.069)** | (4.761)** | (4.771)** | (5.071)** | (4.807)* | (4.830)* |
| Property | | | | −5.296 | −3.547 | −3.465 |
| | | | | (2.016)** | (2.052)* | (2.067) |
| Burglary | | | | −5.815 | −3.603 | −3.564 |
| | | | | (3.046)* | (2.932) | (2.939) |
| Larceny | | | | −4.275 | −2.257 | −2.211 |
| | | | | (1.983)** | (2.069) | (2.073) |
| Vehicle theft | | | | −13.097 | −14.697 | −14.395 |
| | | | | (4.109)*** | (4.444)*** | (4.471)*** |

Heteroskedasticity-robust standard errors in parentheses. *$p<0.1$; **$p<0.05$; ***$p<0.01$. AK, DC, NH, NV excluded. Panel regressions cover 1985−95 and include state and year dummies. "Panel II" regressions control for lagged alcohol sales, with the lag of 1 year in the original and 2 in the replication to avoid distorted 1990 data.
Source: Cook and Durance (2013), Tables 2 and 3; author's calculations.

My results, in the right half of the table, also add rows for outcomes not found in the published Cook and Durance study. In particular, the first row checks for whether alcohol consumption indeed fell more in "wetter" states during 1989−92 as Cook and Durance hypothesize; it does seem to have, even after controlling for long-term convergence (upper-right corner). Because of the sales data distortion, the results here pertain to the changes over 1989−92 rather than 1990−91 as in the other regressions. In addition, to guard against data mining, the bottom rows of my results restore four outcomes examined in the Cook and Durance (2011) working paper but dropped from the 2013 journal publication: property crime, burglary, larceny, and vehicle theft.

If we set aside “All injury” and “All violent crime” as aggregates that include, and thus are not independent of, the other rows’ outcomes, then 4 of the 11 distinct death and crime rate outcomes in the table exhibit negative deviations from long-term trends in 1990−91 that are statistically significant at $p = 0.1$ (last column of table). This is much more than the 10% that would be expected from pure chance if Cook and Durance’s model were completely wrong.

To apply a falsification test, I reran the regressions in the last column while varying the focus period across the 1985−95 range of the data set. In addition to checking for a correlation between 1989 sales and 1990−91 trends, I checked for a correlation between 1988 sales and 1989−90, trends, etc. Because of the sales data distortions just mentioned, I omitted regressions involving 1990 and 1991 sales (and thus 1991−92 and 1992−93 outcome changes). I expanded the samples as much as data availability permitted, to 1967−2004 for injury deaths and 1967−2012 for crime deaths.[24]

The figure below displays the results. Red dots show the best-fit estimates and blue lines show 95% confidence intervals. The results in the last column of table above correspond to the data points for 1991 in each plot. For instance, in the “All injury” graph in the top left we see that any drops in the death rate from injuries in 1990−91, over and above what would be expected from broader trends, is significantly, negatively related to a state’s pretax ”wetness,” as proxied by 1989 alcohol sales. To repeat, this fits the Cook and Durance hypothesis that wetter states reduced drinking and its side effects more.

[24] Wine consumption data are unavailable before 1967. Post-2004 geocoded NCHS mortality data have not yet been obtained. For precision and consistency outcome variables were recalculated using population denominators from SEER (2014).

## Falsification test of Cook & Durance "Panel II" regressions

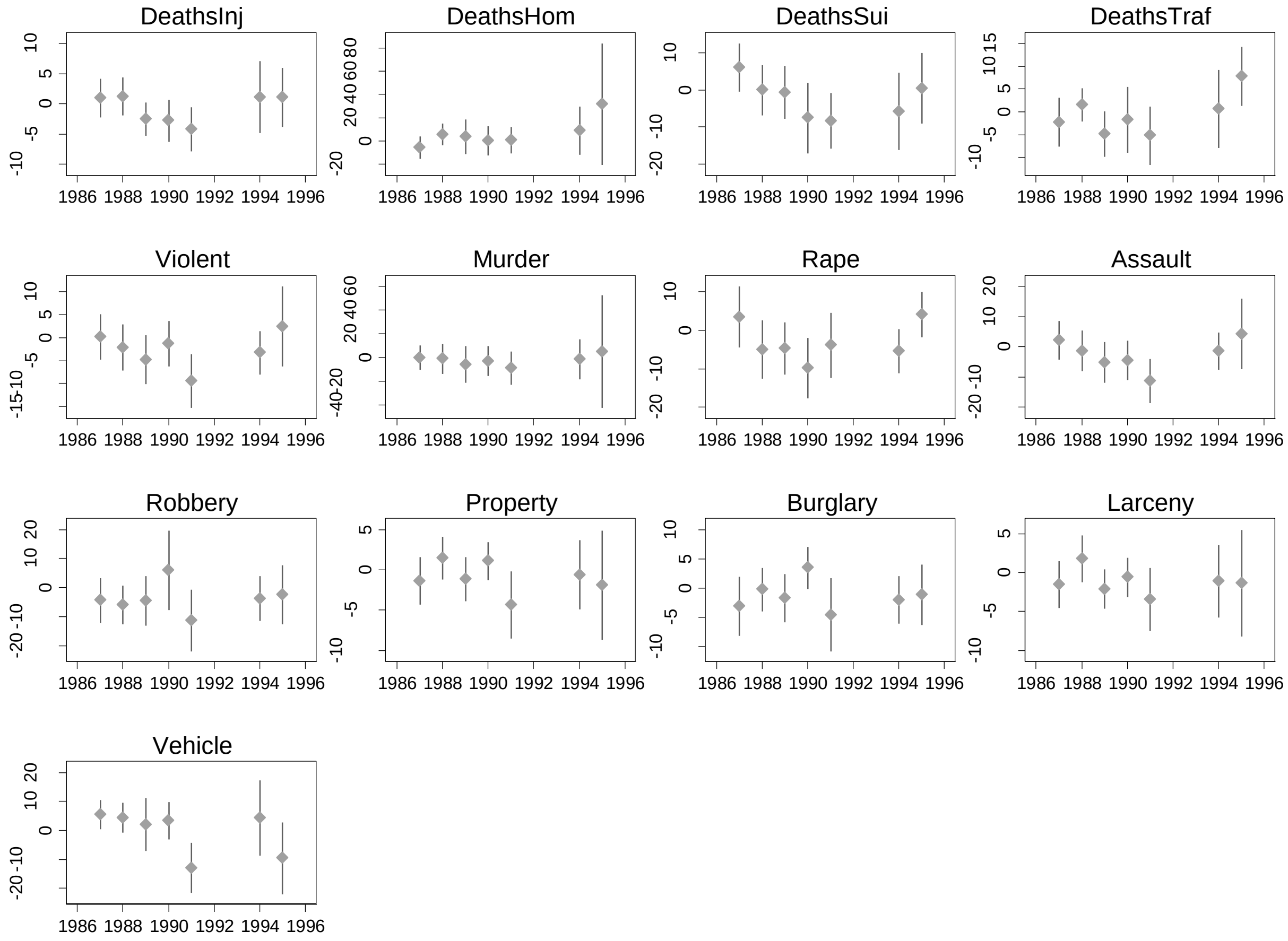

Perusing the plots, the associations between initial alcohol sales and subsequent trend deviations seems weaker for causes of death (first row of graphs) than violent crime (remaining graphs). For almost every crime category, 1990−1991 (plotted at 1991) is the period with the strongest negative association between how wet a state was and how much the crime rate dropped soon after: 1991's red dot is usually the lowest. In contrast, 1991 does not stand out as much in the first row's plots relating to causes of death.

So Cook and Durance appear to have found a real statistical pattern consistent with theory in the case of crime—other than murder—but not injury deaths. The next question is what else might explain this matter. While this study centers on a natural experiment—a sudden, national tax rise—the statistical comparisons made (technically, the source of identification) are cross-state. And cross-state comparisons are not generally seen as quasi-experimental, since states differ systematically in many correlated ways. Other forces, such as the 1990−91 recession might also explain the drops in crime in 1990−91 in wet states relative to dry ones, if the recession raised crime more in drier states. Ruling out such competing explanations would require delving more into the drivers of temporal and geographic crime patterns, which Cook and Durance do not do. Such a counterargument *cannot* be raised against the time series studies looked at earlier, because they compare purely across stretches of time during which taxes suddenly changed, not across geographies differentiated by complexes of correlated traits. On balance, the Cook and Durance hypothesis that the alcohol tax hike reduced crime looks plausible, but not certain.

# 6. Panel studies

Except for Cook and Durance (2013), the studies discussed so far track events in a single state or country over time. We will next look at papers that combine time series data from many geographies into a single analysis. These are called panel studies because they follow a panel of observational units—in our case, US states—over time.[25]

At their best, panel studies increase statistical power—that is, the ability to detect impacts—by studying many natural experiments at once. They can also reduce publication bias to the extent that they prevent researchers from focusing, however unconsciously, on the places where correlations happen to manifest as hoped. However, some panel studies relating to alcohol policy cannot lay such strong claims to causal identification. A panel study relating, say, drinking levels to drunk driving deaths would not qualify as quasi-experimental because both variables exhibit variation over time and space that may make them correlated for reasons other than the direct causal link of interest. To be more compelling, I think a panel study should:

- As with times series studies, focus on the impacts of sudden changes in policy.
- As with times series studies, be geared to detect short-term impacts.
- As with times series studies, use additional controls or knowledge outside the statistical analysis to rule out competing explanations for any correlation found.
- Control for any effects of variables outside the analysis that are fixed over time, within states; and likewise control for effects fixed across states, within each time period. (For more on fixed effects, see Roodman (2014).) This is *not* possible with time series studies; the possibility with panels is one of their strengths.
- Perform falsification tests. This idea too carries over from time series studies, but the application needs to change. Varying "the" assumed date of a tax change makes less sense when dozens occur within the sample. But other falsification tests can be run, for example of whether an alcohol tax change affects *non*−alcohol-related mortality.

These methods are powerful in combination. For example, among US states, inflation-adjusted alcohol tax increases were larger and more frequent in the 1960s than the 1990s (calculated from Ponicki 2004 tax rate files), and yet the US cirrhosis death rate rose in the 1960s and fell in the 1990s (Yoon, Chen, and Yi 2014, fig. 1). A study that only met my first criterion could conclude that raising taxes increases cirrhosis rates. But if time and state fixed effects were also controlled for, satisfying the third criterion, then such coincidentally correlated time trends would be expunged from the analysis, at least to the extent they were nationwide.

To survey the panel literature, I reviewed abstracts of studies included in the Wagenaar, Salois, and Komro (2009) systematic review. I read in full those that focused on taxes rather than prices, and that included fixed effects. The review that follows is not exhaustive but should be representative.

Apparently the first to study several alcohol tax natural experiments at once was Julian Simon (1966); in so doing, he coined the term "quasi-experiment," which means nearly the same thing as "natural experiment" (the ambiguous distinction having to do with whether a tax change is "natural" or whether the before-after framing around such an event is an imposed construct of the researcher). But we will start our selective review in 1982.

The panel literature splits rather cleanly into two strands: one that takes cirrhosis as the outcome and emphasizes liquor taxation; and one that takes traffic fatalities, typically among young people, as the outcome and focuses on beer.

## 6.1. Liquor taxes and cirrhosis

### 6.1.1. Cook and Tauchen (1982), "The effect of liquor taxes on heavy drinking," *Bell Journal of Economics*

[25] The Saffer (1991) data set takes countries rather than states as the unit of observation, but has information on prices only, not taxes.

Cook and Tauchen (1982) brought panel econometric methods to alcohol taxes. Their sample consists of the 30 states where governments license private companies to sell hard liquor. (Where state governments monopolize retail the concept of "tax" becomes fuzzy since the state sets prices and retains all profits: raising a tax without raising prices would reduce profits, for zero net effect.) Alaska, Hawaii, and the District of Columbia also license private retail but were excluded, apparently because they are seen as unrepresentative. The time dimension runs from 1962 to 1977 with annual observations. Cook and Tauchen analyze these data in a way that meets all my criteria. State and year fixed effects are included; the time frame for impact is short, at 1−2 years; and the variation driving the natural experiments is discrete, occasional changes in the inflation-adjusted liquor tax rate.

Converting to dollars of 2014, Cook and Tauchen find that on average, a tax increase of $1/liter of pure alcohol was associated with declines of 1.7% in liquor consumption and 1.6% in cirrhosis mortality in the same calendar year (Cook and Tauchen 1982, Table 2).[26,27] For reference, that tax increase would be $0.32 if imposed on a 0.75 liter bottle of 86-proof Scotch.[28]

Cook and Tauchen follow up with Granger and Sims causality tests, which are fairly rigorous checks on whether tax changes are typically followed by changes in liquor consumption and cirrhosis mortality. The Granger test regressions differs from the one just cited in also controlling for previous-year values of a given outcome variable. The idea is this: With standard regression methods, a computer can derive from the history of, for example, the unemployment rate, a formula for predicting each month's rate based on past months' rates. The formula may work well most of the time but fail when some unanticipated event, such as a financial crisis, occurs. In that case, adding to the forecasting formula an indicator of systemic financial duress can significantly improve predictions. If it does, then the financial indicator is said to *Granger-cause* unemployment, for it suggests that financial duress today does not merely *predict* unemployment in the future, but truly *affects* it. The Sims test is similar in spirit and equivalent in theory (Sims 1972), if backwards in structure. It asks whether a predictions of one variable based only upon its own past values can be improved by including *future* values of another, just as knowing that US unemployment spiked in January 2009 makes it more likely, in some retrospective sense, that the financial system seized up months earlier.

In the case at hand, the tests confirm that liquor taxes Granger-cause liquor purchases and deaths from cirrhosis, but not vice versa (Cook and Tauchen 1982, p. 387).

One concern about this study is that tax changes might merely be proxying for other alcohol policy changes typically occurring at about the same time. However, this seems unlikely because in the United States alcohol control policies generally weakened from the end of federal Prohibition until the mid-1970s, when the modern movement for stricter alcohol controls gained steam—and that was after this study's timeframe. So it is hard to see how non-tax policies could have mimicked alcohol tax hikes in effect. Between 1969 and 1976, no state for which I have data raised the minimum drinking age, while most lowered it, at least for beer and wine (Ponicki 2004, minage.xls). No state outlawed drunk driving per se until 1983 ("drunk" meaning having a blood-alcohol level above 0.08; Ponicki 2004, 08BACLAW.SAV). And the number of people living in counties that were "dry for liquor," i.e., where sale was illegal, fell steadily through this period (Ponicki 2004, wet.sav and BeerDry.sav).

---

[26] The Cook and Tauchen tax variable is in 1967 dollars per proof gallon, which is a gallon of liquor that is 100 proof, meaning 50% ethanol by volume. Liquor excise taxes are normally calibrated to ethanol content, so lower- and higher-proof liquors are proportionally taxed. The Consumer Price Index averaged 33.4 in 1967 and 236.736 in 2014 (download.bls.gov/pub/time.series/cu/cu.data.1.AllItems). The coefficients on the contemporary tax variable are –6.3% and –6.0 %/(1967 $/proof gallon) (Cook and Tauchen 1982, Table 2). Multiplying those by 33.4 / 236.736 / 2 proof gallons/gallon ethanol × 3.7854 liters/gallon = 0.267 gives the figures in text.

[27] Cirrhosis mortality is age-adjusted to 1970 demographics, much as in my earlier graphs for Hong Kong, Denmark, Finland, and Sweden.

[28] 86 proof means 43% ethanol. $1 × 0.75 × 43% = $0.3225.

To assess the robustness of these results I approximately replicated them and introduced variations. The major apparent discrepancy between my data set in the original is in the construction of the mortality variable. Cook and Tauchen use cirrhosis mortality among those at least 30 years old. For mundane reasons having to do with data availability and the need to construct data sets to support replications of studies with diverse variable definitions, my outcome variable is cirrhosis deaths among those 15 and older divided by an approximation of the population at least 30 years old. This discrepancy can be expected to affect the magnitude of the coefficients, less so their statistical significance or meaning.

Once again, the replication is reasonably good (first two columns of table below). The coefficient on the current-year alcohol tax is similar (−0.055 instead of −0.063 proportional reduction from a $1 tax increase in 1967 dollars) and remains statistically strong (first row). The coefficient on the previous-year spirits tax (second row) remains close to zero. The sums of these pairs of coefficients (third row) match well too.

To check robustness I expanded the sample from 1962−77 to 1960−2004 (third column below). This only strengthened the results, doubling the sum of the current- and previous-year coefficients. Next, for comparability with earlier studies, I expanded the outcome from just death from cirrhosis to include alcohol poisoning and terminal damage to brain, heart, stomach, and pancreas (definition from Maldonado-Molina and Wagenaar 2010, Table 2). This further increases coefficients (fourth column). Finally, as a falsification test, I switched from cirrhosis deaths to non−alcohol-related deaths. Here, we expect and find no impact (last column). Finally, I ran Granger tests for the expanded-sample regressions. These pointed clearly to causality from the spirits tax rate to cirrhosis and alcohol-related mortality more generally ($p$ = 0.01, 0.0000) but hardly at reverse causality ($p$ = 0.20 and 0.79).[29]

[29] The Granger regressions include state and year dummies, two lags of the tax, mortality, and personal income. Two lags are used, just as in Cook and Tauchen (1982), because that is what is favored by Akaike and Bayesian information criteria. Standard errors are clustered two-way by year and state. In the death rate Granger regressions, observations are weighted by $\sqrt{np/(1-p)}$, where $n$ is the population in a state-year and $p$ is the crude probability of death (Maddala 1983, p. 29).

**Associations between spirits tax and death rates, US states**

| | Cook & Tauchen original (cirrhosis, 1962−77) | Replication (cirrhosis, 1962−77) | Expansion to 1960−2004 | Expansion to 1960−2004, alcohol-caused deaths | Expansion to 1960−2004, non−alcohol-related deaths |
|---|---|---|---|---|---|
| Spirits tax, current year (1967 $/gal) | −0.063<br>(0.015)*** | −0.055<br>(0.027)**<br>[−0.107,−0.003] | −0.080<br>(0.022)***<br>[−0.123,−0.037] | −0.092<br>(0.029)***<br>[−0.149,−0.036] | −0.007<br>(0.004)<br>[−0.015,0.002] |
| Spirits tax, previous year (1967 $/gal) | −0.009<br>(0.015) | −0.007<br>(0.026)<br>[−0.059,0.044] | −0.059<br>(0.021)***<br>[−0.100,−0.017] | −0.074<br>(0.028)***<br>[−0.129,−0.019] | 0.000<br>(0.004)<br>[−0.008,0.008] |
| Sum, current & previous | −0.072<br>(0.017)*** | −0.062<br>(2.28)**<br>[−0.115,−0.009] | −0.139<br>(5.35)***<br>[−0.190,−0.088] | −0.167<br>(4.74)***<br>[−0.236,−0.098] | −0.006<br>(1.08)<br>[−0.018,0.005] |
| Observations | 480 | 480 | 1320 | 1320 | 1320 |

Standard errors in parentheses. 95% confidence intervals in brackets. *$p$<0.1; **$p$<0.05; ***$p$<0.01. All regressions use generalized least squares to correct for AR(1); the replication regressions use the Durbin-Watson correlation estimate. Coefficients on current and lagged per-capita personal income and state and year dummies not shown. Taxes are for liquid gallons of beverage. Dependent variable in original (first column) is cirrhosis death rate for people age≥30. In replications, it is deaths among people age≥15 divided by an approximation of population≥30. In replications, coding of cirrhosis deaths is from Yoon, Chen, and Yi (2014, p. 5) and those of alcohol-caused and non-alcohol-related deaths from Maldonado-Molina and Wagenaar (2010, table 2).
Sources: Cook and Tauchen (1982, table 2); author's calculations, from Ponicki (2004); NCHS; BEA (State personal income data); SEER (2014).

The rigorous design, the Granger-Sims tests, the robustness, the coherent results on the falsification test, and the difficulty of constructing an alternative explanation for the findings all make this study persuasive.

#### 6.1.2. Ponicki and Gruenewald (2006), "The Impact of Alcohol Taxation on Liver Cirrhosis Mortality," *Journal of Studies on Alcohol*

This study updates and augments Cook and Tauchen (1982), shifting the timeframe from 1962−77 to 1971−98 and adding beer and wine taxes. Unsurprisingly in light of my own update, Ponicki and Gruenewald largely support Cook and Tauchen: a liquor tax increase of $1/liter pure alcohol (in 2014 dollars) is quickly followed by a 1.2% drop in cirrhosis deaths, which is comparable to the 1.6% estimated by Cook and Tauchen.[30] Beer and wine taxes, interestingly, do not seem to have much effect (Ponicki and Gruenewald 2006, Table 1, model 3). This suggests that alcoholics get more of their alcohol from liquor, making them more sensitive to its price. Possibly the effect is confined to those of limited means, since they may be more price-sensitive.

The main methodological innovation is the use of "random effects" instead of fixed effects. Ponicki and Gruenewald start with fixed effects, like Cook and Tauchen, meaning that they enter a dummy variable for each state. In the statistical model, this allows each state's cirrhosis mortality rates to be a fixed percentage above or below the national average. For example, New York's rate could be 10% above the national average in every year of the study, all else equal. Pennsylvania's could be 5% lower. And so on. In initial, unpublished regressions,

[30] The Ponicki and Gruenewald tax variable is in 2000 dollars per liter of ethanol. The Consumer Price Index averaged 172.2 in 2000 and 236.736 in 2014 (download.bls.gov/pub/time.series/cu/cu.data.1.AllItems). The coefficient on the contemporary tax variable is –1.7 %/(2000 $/liter ethanol) (Ponicki and Gruenewald 2006, Table 1, model 3). Multiplying that by 172.2 / 236.736 gives the figure in text.

Ponicki and Gruenewald observe that these fixed effects for the 30 states are distributed like a bell (normal) curve.[31] Then, in the published regressions, they impose the *assumption* that the state effects are distributed normally and estimate only the center and spread (standard deviation) of this distribution. This is the random effects model. By estimating two numbers instead of the 30 individual state effects, they conserve statistical power.[32] Since the data sample—the repository of information—is finite, the more parameters a regression estimates, the less power it has per parameter to estimate precisely. So the reduction in moving to random effects strengthens the model's ability to detect tax impacts, at the price of a stronger and therefore more debatable assumption about cross-state cirrhosis mortality patterns. Ponicki and Gruenewald counterbalance this shift to a stronger assumption by entering controls for a rich set of state traits, including composition by race, age, and religion.

One weakness of the random effects approach is that if the initial fixed-effects regression is so swamped by parameters that its results are imprecise—marked by wide confidence intervals—then the inability to *reject* the needed assumption of normal distribution is faint endorsement for the random-effects model.

But since my regressions for 1960−2004 (above) embrace the Ponicki and Gruenewald timeframe while sticking with the more conservative fixed effects model, it seems likely that this concern does not matter much in practice. The three sets of regressions—Cook and Tauchen, Ponicki and Gruenewald, and my own—tell a coherent story. The short-term statistical link from liquor taxation to cirrhosis mortality looks strong.

## 6.2. Beer taxes and traffic deaths

### 6.2.1. Saffer and Grossman (1987a), "Beer taxes, the legal drinking age, and youth motor vehicle fatalities," *Journal of Legal Studies*[33]

In 1984, President Reagan signed the Federal Uniform Drinking Age Act, which reduced federal highway funding to states that had a minimum drinking age below 21. This reinforced the nationwide trend toward that minimum, which had begun in 1976. By 1988 all states had moved to 21:

---

[31] They perform a Hausman test.

[32] Included in the 2 and the 30 is the constant term.

[33] Saffer and Grossman (1987b) elaborate their analysis in one way, instrumenting the drinking age with all the assumed-exogenous controls in a first-stage ordered-probit equation. There are no excluded instruments. Identification comes rather from the nonlinearity of the impact of the exogenous variables on the outcomes via the quantization of the drinking age into integer values. The identification assumptions required here are strong, because the nonlinear model for drinking age can pick up any other nonlinearities between the exogenous variables and the outcomes (Roodman 2011, p. 180). At any rate, the results for the impact of beer taxes are similar.

## Minimum drinking age by state

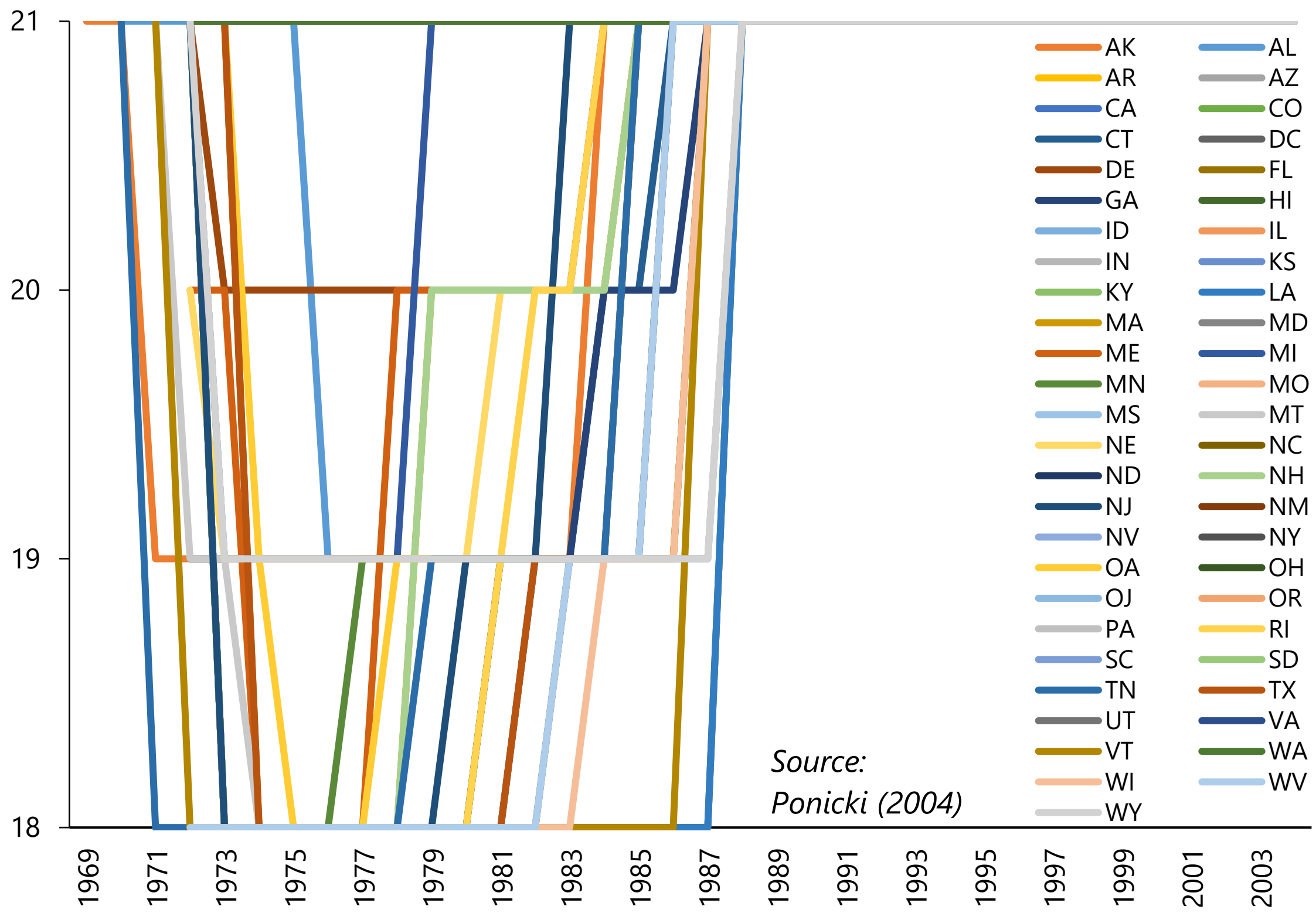

This policy evolution attracted academic interest. Saffer and Grossman (1987a) use a state panel for 1975−81 to study the impact of rises in the drinking age on motor vehicle fatalities. Along the way, they estimate impacts of alcohol taxes too—not those on liquor, as in Cook and Tauchen, but on beer, which is "the drink of choice among youths" (Saffer and Grossman 1987a, pp. 353−54). Thus while the study resembles Cook and Tauchen analytically, being based on a panel of US states, it differs scientifically, addressing a different health problem, within a demographic defined by youth rather than (effectively) by a long history of chemical addiction.

Saffer and Grossman do run regressions that control for all state fixed effects, as I prefer. However, the brevity of the timeframe, just seven years, makes for a low ratio of data points to variables since each state has a fixed-effect dummy and just seven observations. This appears to be the cause of some strange results such as that a higher drinking age increases deaths for 18-to-20-year-olds. "These results suggest that a model with state dummies is overdetermined and plagued by multicollinearity. The implausible nature of the estimates that emerge from this specification provides a justification for not emphasizing it" (Saffer and Grossman 1987a, pp. 369−70). Rather like Ponicki and Gruenewald, Saffer and Grossman prefer a smaller set of controls for factors that might affect alcohol policy and/or traffic deaths: whether states inspect vehicles annually, per-capita income, miles driven per driver, the religious composition of the state, and the fraction of people living in "wet" counties, where alcohol could be sold.

Saffer and Grossman run separate regressions for three age groups: 15−17, 18−20, and 21−24. Raising the drinking age to 21 registers its largest impacts for the middle group, which makes sense and adds credibility to the entire study (Saffer and Grossman 1987a, Table 2, row 1). (Youths in the other groups could still benefit from a drinking age rise, since 18−20-year-olds can share alcohol with their younger siblings and friends, and drive cars with passengers of other ages.)

In contrast, a beer tax increase appears to affect mortality about the same within the two older groups. This too makes sense since taxes do not discriminate by age. The tax impact does appear lower for the youngest group, which is also plausible since 15-17-year-olds presumably buy alcohol much less. In particular, according to the more conservative of the non-fixed-effects regressions, a $1/six-pack tax increase in today's dollars cut annual traffic deaths by 10% among 15−17-years, 18.5% among 18−20's and 21−24's.[34] Fixed-effects regressions produced similar results. (Saffer and Grossman 1987a, Table 2) When this paper was written, the federal beer tax had last been raised in 1951, to $9/barrel ($0.33/six-pack); the authors estimate that if the federal government had continually raised the tax to offset inflation, this would have saved the lives of 1,022 18−20-year-olds in the late 1970s (Saffer and Grossman 1987a, p. 374).

As Dee (1999, p. 304) points out, the apparent tax impact looks unrealistically large when you consider that only about half of traffic fatalities involve alcohol. An 18.5% reduction in fatalities among 18−20's and 21−24's means twice that among *alcohol-related* fatalities; a 40% reduction seems extraordinary for a $1 tax increase on a six-pack that goes for some $7.50. On the other hand, the subtleties just noted in the pattern of the results across age groups do cohere with the interpretation of these results as tax impacts. It is hard to be certain what this study's results mean, beyond that they are suggestive of impact.

#### 6.2.2. Dee (1999), "State alcohol policies, teen drinking and traffic fatalities," *Journal of Public Economics*

Dee (1999) is one of several contemporaneous papers to challenge earlier findings that beer taxes prevent traffic deaths.[35] At least one in this new crop, Young and Likens (2000), was supported by beer industry funding.[36]

Dee lodges several criticisms. He starts by working with a novel outcome variable, self-reported drinking behavior from nationwide surveys of high school seniors conducted by a group called Monitoring the Future. Examining data for 1977−92, he finds that when a state raised its minimum drinking age from 18, heavy drinking among seniors fell 8.4%. But beer tax hikes made little difference. (Dee 1999, p. 291.) Dee argues that the first finding shows that the surveys produces more than random noise, which makes the second finding more credible. However, I wonder whether the results merely show that raising the drinking age had more impact on teenagers' conceptions of the "right" answers to questions about their drinking. It would be surprising if outlawing drinking for 18-year-olds did not cause self-censoring. A beer tax hike would not do that. This theory can also explain Dee's survey-based results.

Dee then switches to a more common outcome variable, government-tabulated traffic deaths. He shows that drinking age increases are associated more reliably with lower traffic deaths than is raising the beer tax. His most rigorous regressions control not only for fixed effects for each state but straight-line trends. Thus, if some unidentified factor is driving both alcohol policy and traffic deaths, creating a false appearance of a direct causal link between them, this will be removed to the extent that its effects *change at a constant rate* over time—rather than, as with fixed effects, being merely *constant* over a time. (Constant rate of change of effect is more general than constant effect because the constant rate could just be zero, leading to constant effect.) This controlling for state-specific trends undoes the beer tax impact result, even flipping the tax-fatality association to positive. But raising the minimum drinking age to 21 continues to reduce death rates. (Dee 1999, Table 4, last column.)

Dee acknowledges that this constant-rate-of-change test is perhaps too tough in that it leaves the beer tax little variation with which to distinguish itself from all the controls. But then he administers what is perhaps the *coup de grâce*. He disaggregates the results by time of day of crash. Since, he states, the share of crashes that involve alcohol is 3.4−7.5 times higher at night (Dee 1999, p. 311), alcohol policies should affect *total* nighttime traffic fatalities much more than proportionally. For the policy change of raising the drinking age to 21, this proves so, with the impact coefficient 2.3 times higher by night. For the beer tax, the ratio is only 1.4. (Table 5, middle column; pp. 311–13.) Dee concludes that something other than causality from taxes to traffic deaths is driving the beer tax results, because otherwise they make little sense.

[34] Coefficients in Saffer and Grossman (187a, Table 2) are semi-elasticities with respect to taxes in 1967 dollars per case (24 12-ounce bottles). Conversion to 2014 dollars per six-pack multiples by 33.4 / 236.736 × 4 = 0.564. See footnote 28.

[35] Earlier studies also finding strong effects include Chaloupka, Saffer, and Grossman (1993) and Ruhm (1996). Other critical studies of Dee's (1999) vintage are Mast, Benson, and Rasmussen (1999) and Young and Likens (2000).

[36] Conversation with Douglas Young, April 8, 2015.

In my view, Dee succeeds in casting some doubt on earlier findings. Yet the criticisms also feel a bit strained. Raising the minimum drinking age to 21 *also* has less impact at night than Dee's argument suggests it should: 2.3 is outside the 3.4–7.5 range. And controlling for state-specific trends with a time dimension of 16 years is extremely demanding.

#### 6.2.3. Young and Bielinska-Kwapisz (2006), "Alcohol prices, consumption, and traffic fatalities," *Southern Economic Journal*

This paper adds an unexpected coda to the beer tax–traffic death subliterature. Sharing a coauthor with one of the skeptical circa-2000 studies (Young and Likens 2000), it concludes more hopefully on the benefits of alcohol taxation. It also introduces a smart methodological innovation. Yet the authors conclude with a caution about the reliability of the entire subliterature.

Young and Bielinska-Kwapisz (2006) technically violates one of my quality criteria, by looking at the impact of prices rather than taxes—but there is no violation in spirit. For they draw the causal chain,

Alcohol taxes → Alcohol prices → Alcohol sales → Traffic deaths,

and then use taxes to *instrument* prices. (On this concept, see [my review of the mortality-fertility link](my review of the mortality-fertility link).) They do the same with alcohol sales. Their instrumented regressions can be depicted as

Alcohol taxes → Alcohol prices → Traffic deaths

Alcohol taxes → Alcohol sales → Traffic deaths,

where the rightmost arrows are of primary interest. Each of these gives insight into the full causal chain. The assumption required to interpret Young and Bielinska-Kwapisz's correlations as causation is the same as needed in studies that regress outcomes directly on taxes: namely, that alcohol tax changes are exogenous after conditioning on controls.[37]

Young and Bielinska-Kwapisz (2006, Table 2) start by demonstrating the importance their instrumentation. Non-instrumented regressions of traffic fatality rates on an index of alcohol prices (not just for price of beer, the beverage of interest in other traffic fatality studies) produce positive coefficients, about half of which are statistically significant. The naïve interpretation: a higher price for booze causes *more* accidents. But instrumenting prices with taxes flips these results even as it puts them on firmer ground. After all, it seems, higher prices *reduce* accidents. Perhaps when there is no instrumentation, reverse causality obscures this connection, with higher local demand for alcohol raising local prices. Or perhaps for some other reason, alcohol costs more in states where traffic accidents happen more.

Young and Bielinska-Kwapisz (2006, Tables 3 & 4) find that a tax-driven 10% alcohol price increase cuts the accident rate by 6.9% on weekend nights and by 3.9% at other times. Among teens (actually, 16–20-year-olds), the numbers are reversed, at 3.5% and 9.3%. A tax-driven 10% alcohol *sales* reduction lowers the accident rate by 10.8% on weekend nights and 11.1% otherwise, and 10.2% and 14.1% just among teens. Since these four numbers for the impact of a 10% sales reduction are roughly twice their counterparts for a 10% *price* increase, this suggests that a 10% price increase saves about the same number of lives as a 5% sales decrease; from there, it is a short step to conclude that a 5% sales drop happens when there is a 10% price increase. Working more precisely, Young and Bielinska-Kwapisz (2006, p. 698) compute an alcohol price-sales elasticity of −0.51, which is in line with the meta-analytic estimates discussed early in this review.

[37] Young and Bielinska-Kwapisz (2006, Table A1) document very high first-stage $R^2$s, so bias from instrument weakness seems unlikely.

Young and Bielinska-Kwapisz close by doubting their results. Perhaps their control set is too rich for the sample size, so that near-collinearity among some variables leads to odd results. For when they drop controls relating to religion and the importance of tourism in a state's economy—whose coefficients are either insignificant or surprising in sign—the impacts of a price change fall by about half, and by about three-quarters among teens (their Table 5, first row), even if they remain statistically distinct from zero.

The authors may be too hard on their work. Several of their criticisms, such as that the impacts appear implausibly large (pp. 700−01), refer to the regressions run *before* those suspect controls are dropped. Nevertheless, taken in the context of an internally contradictory subliterature, it is hard to end our survey highly confident that alcohol taxes have significantly cut traffic deaths.

# 7. On long-term effects

The literature persuades me that raising alcohol taxes reduces deaths from alcohol-related diseases, notably cirrhosis. However, key studies such as Wagenaar, Maldonado-Molina, and Wagenaar (2009) on Alaska, Cook and Tauchen (1982) on US states generally, and Koski et al. (2007) on Finland persuade precisely by showing *quick* impacts. And their focus on the short-term leaves the long-term impacts less clear. This is not a criticism: as I have argued, long-term impacts are harder to prove (Shadish, Cook, and Campbell 2002, p.173).

In the absence of certainty, what is the most plausible prior? One the on hand, the benefits of an alcohol tax increase almost surely swell with time. Cirrhosis, for example, progresses over decades. If a higher price for alcohol slows the progression from near-death to death, which is what the studies are detecting, then it ought also to slow the progression at all stages. On the other hand, many studies suggest that moderate drinking, as compared to abstention, reduces coronary heart disease and other ailments (Fekjær 2013, p. 2015). Alcohol tax increases may well impede this apparently healthy moderation. The harm per person might be much smaller than the benefit per person among heavy drinkers of a tax hike, but the population harmed could be far larger. And this effect might play out purely in the long-term, so that it is missed in the studies favored here. In principle, this leaves the sign of net long-term impact ambiguous.

A full review of the relevant epidemiological literature on moderate drinking is beyond the scope of this document. However, an initial scan leaves me reasonably confident that tax increases save lives in the long run too. The reasons:

- No randomized trials have checked on the benefits of moderate drinking (Holmes et al. 2014, p. 4).
- The belief in such benefits derives from non-experimental, observational studies, whose claims to causal identification make them akin to the alcohol tax studies *passed over* for low credibility in this review. In epidemiology, as in economics, observational studies have come under a shadow in the last decade or so, as randomized trials have upended established doctrines such as the belief that hormone replacement therapy reduces heart disease and cancer (Writing Group for the Women's Health Initiative Investigators 2002).
- A new generation of *Mendelian randomization* studies challenges the conventional wisdom on moderate drinking, albeit not with the overwhelming force that conventional randomized trials would (Davey Smith and Ebrahim 2003, p. 10). Mendelian randomization studies strive to exploit the randomness inherent in sexual reproduction as well as the now-low cost of gene sequencing, in order to fashion natural experiments. For example, if a mutation reducing the ability to metabolize alcohol, and thus the propensity to drink, is randomly distributed in a population, then whether one carries zero, one, or two copies of that gene can *instrument* for drinking in a study of drinking's sequelae. In fact, genes are not distributed with perfect randomness, and can affect health through multiple pathways, so Mendelian randomization does not produce experiments as close to perfect as a conventional randomized trial (Thomas and Conti 2004; Davey Smith and Ebrahim 2003, pp. 13−17). Nevertheless, that these studies have generally failed to find benefit from moderate drinking in comparison to no drinking creates serious doubts about traditional observational studies. A meta-analysis of Mendelian-randomized alcohol studies, aggregating data from 261,991 people of European descent, found that carriers of a particular variant gene for the alcohol dehydrogenase 1B enzyme drank 17% less (95% confidence interval 15.6−18.9%) and had 10% lower odds of coronary heart disease (95% confidence interval 4−14%; Holmes et al. 2014, p. 4). (The gene causes faster conversion of alcohol into toxic aldehyde in the blood stream.) Restricting the analysis to low, moderate, or heavy drinkers did not change the apparent impact (Holmes et al. 2014, p. 7). Even at low levels, more drinking did at least modest harm.

- A study based on a telephone survey of more than 200,000 people in the US checked whether self-reported drinking levels correlated with any of 30 demographic, social, and economic factors thought to contribute to cardiovascular disease, from age to poverty to less exercise to limited access to health care. Among light and moderate drinkers as a group, 27 of the 30 factors were correlated with drinking levels, and in the direction that could create a false appearance of moderate drinking leading to better health. For example, as people enter old age, they drink less and become more likely to die.[38] (Naimi et al. 2005.) Presumably no observational study controls for all 27, and perhaps none needs to go quite that far since the factors are probably inter-correlated. Nevertheless, the study suggests that the endogeneity of moderate drinking to health is hard to eradicate from an observational study.

In light of these facts, Occam's Razor argues for a simple theory: the net marginal impact of drinking on health is negative at all levels; and moderate drinking is a marker for relative youth, affluence, and healthy habits rather than a cause of good health (IOGT-NTO and the Swedish Society of Medicine 2014; Chikritzhs et al. 2015). Pending high-quality evidence to the contrary, alcohol taxes should be presumed to save even more lives in the long run.

## 8. Lives and years of life saved

I worked to translate findings from the highest-powered study settings into more comparable and meaningful estimates of lives saved. (See table below.) For the US-based studies, I relied on my replication regressions since they have larger samples and, in the case of the Cook-Tauchen-style US panel study, are modified to study the same outcome—deaths not just from cirrhosis but also from alcohol poisoning and damage to organs aside from the liver.

In Alaska, directly measured alcohol prices rose some 12% between 2002 and 2003. According to a regression modeled on the Alaska study, the alcohol-caused death rate among those 15 and up fell by 31% in the quarter after the tax increase, producing a mortality-price elasticity of roughly −2.7. I.e., for every 1% increase in alcohol price, deaths fell 2.7%.

Moving to the panel context—studying 30 states at once—we find that a $1/gallon liquor tax increase (in 1967 dollars), or an average of 5% of the retail price, led to a 17% reduction, for an elasticity of −3.4. Note that, for lack of price data for the entire study period, the *price* impact of the representative tax increase is not observed but is inferred from a comparison to the typical $20/gallon liquor price. To the extent that retailers *more* than passed on tax increases to customers (Young and Bielinska-Kwapisz 2002; Kenkel 2005), this calculation underestimates the typical price change, and overestimates the elasticity.

Finally, in Finland, over-30 alcohol-caused mortality rose 16% among men and 31% among women after the alcohol price index plunged 18%, for elasticities of −0.75 and −1.36. (Here the impact numbers are simple changes rather than regression coefficients.)

[38] The analysis excluded respondents reported to be in poor health, in order to prevent unhealthy *former* heavy drinkers from contaminating the sample of non-drinkers.

**Short-term impact of alcohol tax changes in highest-powered study settings**

| Context | Approximate price change | Impact on deaths from alcohol-caused diseases | Price elasticity |
|---|---|---|---|
| Alaska, 2002 | +8% beer, +14% wine, +17% sprits, +12% overall | −31% | −2.7 |
| 30 US states with private liquor retail, 1960−2004 | Benchmark: +$1/gal spirits (1967 $) ≈ +5% retail | −17% | −3.4 |
| Finland, 2004 | −18% alcohol price index | +16% men, +31% women | −0.75 men, −1.36 women |

Sources: *Alaska*: Price changes based on Ponicki (2004, ACCRAdj.xls); overall change an average weighted by 2002 consumption of gallons of pure ethanol (LaVallee, Kim, and Yi 2014). Impact on alcohol-caused deaths from a Wagenaar, Maldonado-Molina, and Wagenaar (2009) log mortality replication regression like those reported earlier but with the dependent variable narrowed to alcohol-*caused* deaths as defined in that study. Price elasticity is −0.31/ln(1.12). *US states*: Population-weighted average price of a bottle of J&B scotch whiskey for the sample was equivalent to about $20/gallon (1967 dollars; Ponicki 2004, ACCRAdj.xls). Impact on alcohol-caused deaths reported in Cook and Tauchen (1982) log mortality replication. Price elasticity is −0.167/ln(1.05). *Finland:* Change in alcohol price index from FRED series CP0210FIM086NEST. Impacts from Herttua, Mäkelä, and Marikainen (2008, Table 1). Elasticities are ln(1.16)/ln(1-.18) and ln(1.31)/ln(1-.18).

We can link this rather wide range of elasticities, 1−3 in absolute value, to tallies of alcohol-caused deaths, which I calculate at 23,000 in the US and 72,000 in all western industrial countries in 2010 (WHO).[39] In the US, raising alcohol prices by 10% would cut the death toll by 9−25% or 2,000−6,000/year.[40] Across these industrial countries, 6,500−18,000 deaths/year would be averted.

How many *years* of life a tax hike would save is a more complicated matter. Perhaps immediately after a tax increase the lives saved would be of people still close to death, so that *years* of life saved would seem modest. But if progression of alcohol-caused diseases slowed at all stages of those diseases, then the lives of (would-be) heavy drinkers of all ages would be lengthened, even if this would not show up in mortality statistics for decades. So it appears reasonable to assume that a permanent tax increase—*if sustained against inflation*—would reduce years of life lost proportionally to deaths. In the US, the CDC estimated that alcohol-caused chronic conditions cost 22.6 years of life per death in 2001 (CDC 2004, Table).[41] Applying that to our figure of 23,000 US deaths in 2010 suggests that a 10% tax-induced alcohol price increase would save 48,000−130,000 years of life for each year it was sustained.

## 9. Conclusion

Despite weaknesses in the majority of studies—the ones not reviewed in depth here because they do not exploit natural experiments—and despite seeming disagreements among the studies given attention here, we can be reasonably confident that taxing alcohol reduces drinking in general and problem drinking in particular. The reasons are several:

1. Common sense and empirical economics say that people usually buy less of a thing when its price goes up. Of course there are caveats: raise the alcohol tax enough and an illicit industry will take root and perhaps *increase* supply as it exploits economies of scale and learning-by-doing. And sometimes tax increase extract monopoly or other rents from suppliers, and are not passed on to consumers. But the assumption that taxes affect sales is a responsible starting point.

[39] The coding definition of "alcohol-caused" is from Maldonado-Molina and Wagenaar (2010, table 2). It excludes "alcohol-related" deaths, which have codes that do not fully attribute alcohol as the underlying cause, yet for which it often is.
[40] Given elasticity bounds of 1 and 3, the calculations are $1-\exp(\ln 1.1 \times (-1)) = 9\%$ and $1-\exp(\ln 1.1 \times (-3)) = 25\%$.
[41] The definition of alcohol-caused deaths used in CDC (2004) results in 788,005 years of potential life lost and 34,833 deaths.

2. Most studies find the expected negative correlation between alcohol prices and total drinking. Even Nelson (2013, 2014a), in his skepticism, agrees. I like his publication bias–adjusted elasticity estimates: −0.29 for beer, −0.46 for wine, −0.54 for spirits, −0.49 for total alcohol (Nelson 2013a).
3. It is hard to construct compelling alternative explanations for this correlation.
4. The evidence suggests that the tax-consumption link carries over to *problem* drinking. Alcohol-related deaths rose in Alaska in 2002 (Wagenaar, Maldonado-Molina, and Wagenaar 2009) and Finland in 2004 (Koski et al. 2007), as did drunk driving in Finland, at least temporarily (Mäkelä and Österberg 2009). Panel studies such as Cook and Tauchen (1982), and my extensions thereof, find similar results across many states and decades. In Switzerland, self-reported drinking problems rose (Mohler-Kuo et al. 2004), about as much in the long run among heavy drinkers as in the general population (Gmel et al. 2007).
5. The evidence for impacts on sexually transmitted diseases and traffic fatalities looks weaker.
6. While some studies of natural experiments in taxation do *not* find impacts on alcohol-related mortality—notably, in Hong Kong and Denmark—and others detect it in a way that leaves room for doubt—in Florida and the U.S. as a whole—one observation can explain *most* of this seeming disagreement. The larger and cleaner the natural experiment, the more apt are studies to detect impacts on drinking. In the context of U.S. state studies, the largest change was in Alaska in 2002, and that is where the identification of impacts is most compelling (Wagenaar, Maldonado-Molina, and Wagenaar 2009). Something similar goes for the tax cuts in Nordic nations, with Finland the unhappy standout. In Hong Kong, where no impacts were found, the tax cut hardly produced changes in the price series that were indistinguishable from long-term trends when relying on surveys 5–6 years apart. Thus the evidence base taken as a whole is most easily explained by a dose-response story. The bigger the tax change, the greater the statistical power to detect it.
7. The idea that moderate drinking is healthier than no drinking is based on observational epidemiological studies, which as a type have fallen somewhat into disrepute in recent decades, and are less persuasive as to causality than the tax impact studies relied on here. It therefore seems unlikely that alcohol tax increases do net harm in the long run by discouraging moderate drinking. Rather, the benefits of slowing the progression of diseases such as cirrhosis accumulate over time, and probably dominate the overall long-term effect.
8. Estimates from the highest-powered study settings suggest an elasticity of mortality with respect to price of −1 to −3. A 10% price increase in the US would, according to this estimate, save 2,000–6,000 lives and 48,000–130,000 years of life each year.